\documentclass[trackchanges,twocolumn]{aastex7} 

\newcommand{\glswithcite}[2]{%
    \ifglsused{#1}{\gls{#1} (#2)}{\glsentrylong{#1} (\glsentryshort{#1}; #2)\glsunset{#1}}
}

\usepackage{makecell}
\usepackage{longtable}
\usepackage[T1]{fontenc}

\newcommand{\todo}[1]{\textcolor{black}{#1}}

\usepackage[acronym,nohypertypes={acronym}]{glossaries} 
\newacronym{LSST}{LSST}{Legacy Survey of Space and Time}
\newacronym{JWST}{JWST}{James Webb Space Telescope}
\newacronym{HST}{HST}{Hubble Space Telescope}
\newacronym{FWHM}{FWHM}{full width at half maximum}
\newacronym{PSF}{PSF}{point spread function}
\newacronym{ATLAS}{ATLAS}{Asteroid Terrestrial-impact Last Alert System}
\newacronym{NEA}{NEA}{near-Earth
asteroids}
\newacronym{ATClean}{\texttt{ATClean}}{ATLAS Clean}
\newacronym{SNR}{SNR}{signal-to-noise ratio}
\newacronym{RSG}{RSG}{red supergiant}
\newacronym{SN}{SN}{supernova}
\newacronym{SN II}{Type II SN}{Type II supernova}
\newacronym{MIR}{MIR}{mid-infrared}
\newacronym{SED}{SED}{spectral energy distribution}
\newacronym{CSM}{CSM}{circumstellar material}
\newacronym{ALMA}{ALMA}{Atacama Large Millimeter/submillimeter Array}
\newacronym{WFPC2}{WFPC2}{Wide Field Planetary Camera 2}
\newacronym{DECam}{DECam}{Dark Energy Camera}
\newacronym{ZTF}{ZTF}{Zwicky Transient Facility}
\newacronym{PTF}{PTF}{Palomar Transient Factory}
\newacronym{ACS}{ACS}{Advanced Camera for Surveys}
\newacronym{WFC}{WFC}{Wide Field Channel} 
\newacronym{CHIPS}{CHIPS}{Complete History of Interaction-Powered Supernovae}
\newacronym{ATCA}{ATCA}{Australia Telescope Compact Array}

\begin{document}


\title{Probing the Nature of the Circumstellar Material Surrounding SN 2024ggi: A Search for Precursor Emission and Constraints from Radio Observations}

\correspondingauthor{Tobias G\'eron}
\email{tobias.geron@utoronto.ca}

\author[0000-0002-6851-9613]{Tobias G\'eron}
\affiliation{Dunlap Institute for Astronomy \& Astrophysics, University of Toronto, 50 St. George Street, Toronto, ON M5S 3H4, Canada}
\email{tobias.geron@utoronto.ca}

\author[0000-0002-9415-3766]{James K. Leung}
\affiliation{Dunlap Institute for Astronomy \& Astrophysics, University of Toronto, 50 St. George Street, Toronto, ON M5S 3H4, Canada}
\affiliation{David A. Dunlap Department of Astronomy \& Astrophysics, University of Toronto, 50 St. George Street, Toronto, ON, M5S 3H4, Canada}
\affiliation{Racah Institute of Physics, The Hebrew University of Jerusalem, Jerusalem 91904, Israel}
\email{-}

\author[0000-0001-7081-0082]{Maria R. Drout}
\affiliation{David A. Dunlap Department of Astronomy \& Astrophysics, University of Toronto, 50 St. George Street, Toronto, ON, M5S 3H4, Canada}
\email{-}

\author[0000-0001-5609-7372]{Rami Z. E. Alsaberi}
\affiliation{Western Sydney University, Locked Bag 1797, Penrith NSW 2751, Australia}
\affiliation{Faculty of Engineering, Gifu University, 1-1 Yanagido, Gifu 501-1193, Japan}
\email{-}

\author[0000-0002-3934-2644]{W.~V.~Jacobson-Gal\'{a}n}
\altaffiliation{NASA Hubble Fellow}
\affiliation{Cahill Center for Astrophysics, California Institute of Technology, MC 249-17, 1216 E California Boulevard, Pasadena, CA, 91125, USA}
\email{-}

\author[0000-0002-5740-7747]{Charles D. Kilpatrick}
\affiliation{Center for Interdisciplinary Exploration and Research in Astrophysics (CIERA), Northwestern University, Evanston, IL 60208, USA}
\email{-}

\author[0000-0002-4410-5387]{A. Rest}
\affiliation{Space Telescope Science Institute, 3700 San Martin Drive, Baltimore, MD 21218, USA}
\affiliation{Department of Physics and Astronomy, The Johns Hopkins University, 3400 North Charles Street, Baltimore, MD 21218, USA}
\email{-}

\author[0000-0002-3825-0553]{S. Rest}
\affiliation{Department of Computer Science, The Johns Hopkins University, Baltimore, MD 21218, USA}
\email{-}

\author[0000-0003-4501-8100]{Stuart D. Ryder}
\affiliation{School of Mathematical and Physical Sciences, Macquarie University, Sydney, NSW 2109, Australia}
\affiliation{Astrophysics and Space Technologies Research Centre, Macquarie University, Sydney, NSW 2109, Australia}
\email{-}

\author[0000-0002-4060-5931]{Toshikazu Shigeyama}
\affiliation{Research Center for the Early Universe (RESCEU), School of Science, The University of Tokyo, 7-3-1 Hongo, Bunkyo-ku, Tokyo
113-0033, Japan}
\affiliation{Department of Astronomy, School of Science, The University of Tokyo, Tokyo, Japan}
\email{-}

\author[0000-0002-8215-5019]{Yuki Takei}
\affiliation{Research Center for the Early Universe (RESCEU), School of Science, The University of Tokyo, 7-3-1 Hongo, Bunkyo-ku, Tokyo
113-0033, Japan}
\affiliation{Yukawa Institute for Theoretical Physics (YITP), Kyoto University, Kitashirakawa-oiwake-cho, Kyoto, Kyoto 606-8502}
\email{-}


\begin{abstract}
We present an analysis of the pre-explosion optical light curve and post-explosion radio observations of the nearby supernova (SN) 2024ggi, a Type II SN discovered in NGC 3621 at a distance of $\sim$7.2 Mpc. Early spectra confirmed that SN~2024ggi showed ``IIn-like'' features, suggesting the presence of dense circumstellar material (CSM). We searched for precursor emission in the light curve up to 8 years pre-explosion from ATLAS using \texttt{ATClean}, which cleaned the light curve and effectively convolved it with rolling Gaussians of several timescales. No significant evidence for precursor variability was found. We find 80\% detection thresholds on the absolute magnitude of $-11.28$ mag and $-8.98$ mag for outbursts with timescales of 2 and 300 days, respectively. Our radio observations taken with the Australia Telescope Compact Array (ATCA) are consistent with a wind-like CSM density profile, with an inferred mass-loss rate of ${\sim}  8 \times10^{-5} M_\odot$ yr$^{-1}$, assuming $v_{\rm wind} = 50$ km s$^{-1}$. However, including CSM density estimates from the literature implies a two-component structure in the overall CSM density profile, with a denser inner structure at smaller radii ($\lesssim 6\times10^{14}$ cm) and an extended wind-like profile at larger radii. We conclude that the dense CSM around SN~2024ggi is likely not deposited there by eruptive outbursts. Instead, we favor a scenario with a lower but more stable mass-loss rate, such as pulsation-driven superwinds, a model with extended chromospheres, or a combination of both.
\end{abstract}





\section{Introduction}

A Type II supernova (SN) is produced following the core collapse of a hydrogen-rich massive star \citep{heger_2003}. A specific subset of these contain ``narrow'' emission lines, and are therefore termed Type IIn \citep{schlegel_1990, filippenko_1997}. These narrow emission lines are typically interpreted as being caused by the interaction between the SN ejecta and the slowly moving, preexisting dense \glswithcite{CSM}{\citealp{gal_yam_2017, smith_2017}}. SNe classified as Type IIn are diverse, but are typically dominated by these narrow emission lines for the entirety of their evolution, which can last a few months to decades (e.g., see \citealp{jencson_2016, ryder_2016, cai_2026}). Type IIn SNe are relatively rare, with around 4 - 9 \% of core collapse SNe belonging to Type IIn \citep{smartt_2009, smith_2011, li_2011} and the origin of the \gls{CSM} is actively debated.

Notably, as wide-field time-domain surveys have expanded and SNe are now regularly discovered within a few days of explosion, the spectra of some Type II SNe have been found to show \emph{transient} narrow emission lines, reminiscent of the aforementioned Type IIn SNe (known as ``IIn-like''), caused by a ``flash-ionized'' \gls{CSM}. However, contrary to Type IIn, these emission lines are short-lived, with a characteristic timescale ($t_{\rm IIn}$) of only a few hours to days. At that point, the fast-moving SN ejecta will break out and sweep up the slow-moving \gls{CSM}, and the Doppler-broadened features of the SN ejecta become visible instead \citep{dessart_2017, yaron_2017, jacobson_galan_2023, jacobson_galan_2024b}. These are often also called ``flash spectroscopy'' SN II, or CSM-interacting SNe II. Interestingly, more than $36$\% of Type II SNe observed within 2 days of explosion show IIn-like features \citep{bruch_2021,bruch_2023}, suggesting that this type might be common, albeit hard to observe due to the short-lived nature of these emission lines.

There is additional evidence for the presence of dense \gls{CSM} in a subset of Type II SNe. For example, the progenitors have cooler effective temperatures \citep{kilpatrick_2018,jencson_2023} and are less luminous than expected \citep{van_dyk_2023,kilpatrick_2023}, suggesting that the progenitor is surrounded and partially obscured by a shell of dense \gls{CSM}. Additionally, light curve modeling of Type II SNe has shown that models that include dense \gls{CSM} tend to fit the light curves better than models without it \citep{morozova_2017,morozova_2018}. Finally, luminous post-explosion radio emission observed in a subset of Type II SNe arises from the interaction of the SN ejecta with an extended and dense \gls{CSM} (e.g., see \citealp{williams_2002, bauer_2008, ryder_2016}).

One mechanism that is often mentioned to explain the dense \gls{CSM} is eruptive outbursts prior to explosion (e.g., \citealt{dessart_2010, fuller_2017, tsang_2022}). These outbursts can manifest themselves in the pre-explosion light curve as faint ``bumps''. Such precursor emission has been found for many Type IIn, such as SN~2009ip \citep{pastorello_2013}, 2010mc \citep{ofek_2013}, 2011ht \citep{fraser_2013}, and 2023vbg \citep{goto_2025}. Other mechanisms have also been proposed to explain the dense surrounding \gls{CSM}, such as binary interactions \citep{matsuoka_2024}, extended atmospheres/chromospheres \citep{dessart_2017, fuller_2024}, and `superwinds' \citep{yoon_2010}.

It is crucial to correctly interpret the origin of the dense \gls{CSM}. For example, many models for creating eruptive pre-SN outbursts involve harnessing energy from the core of the star (e.g., through gravity waves). This can impact the final structure of the progenitor star and therefore both its explodability and the type of compact object it leaves behind \citep{quataert_2012, shiode_2014, meynet_2015,muller_2016,fuller_2017,gossan_2020}. Two particular constraints that can help to distinguish between these different models are (i) searches for precursor emission expected from eruptive outbursts and (ii) probing the larger \gls{CSM} density profile that surrounds the progenitor star, which provides insight into the pre-SN evolution and varies between different models. We elaborate on each of these points below.

Precursor emission is relatively common among Type IIn. For example, \citet{ofek_2014} find that $>50$\% of Type IIn SNe have at least one pre-explosion outburst brighter than $3 \times10^{7} L_\odot$ within $1/3$ yr prior to explosion using \glswithcite{PTF}{\citealp{law_2009}}. Similarly, \citet{strotjohann_2021} find that $25^{+44}_{-20}$\% of Type IIn have a month-long detected precursor emission brighter than $-13$ mag in the $r$-band within the three months before the explosion using the \glswithcite{ZTF}{\citealp{bellm_2019}}. However, it is unclear whether the rest do not have any precursor emission at all, or if it is missed due to insufficient depth and cadence. \citet{ofek_2013} also note that fainter precursor emission is likely even more common, suggesting that the fractions quoted above are lower limits and the true fraction of Type IIn SNe with precursor emission is likely higher.

While precursor emission might be prevalent among Type IIn SNe, it does not seem to be equally common among Type IIP/IIL SNe. \citet{kochanek_2017} and \citet{johnson_2018} both find low or no clear evidence for variability in the pre-explosion light curves of their sample. Interestingly, \citet{jacobson_galan_2022} did find evidence for precursor emission in SN~2020tlf. This was the first reported case of significant precursor emission in a Type II SN with ``IIn-like'' features that indicate the presence of a surrounding \gls{CSM}. This suggests that perhaps only Type II SNe that show clear evidence of \gls{CSM} interaction (i.e., ``IIn-like'') are able to produce detectable precursor emission. However, no precursor emission was found in the very nearby SN~2023ixf, which was also ``IIn-like'' \citep{dong_2023, rest_2025}. Thus, it is still an open question how common precursor emission is among SNe with ``IIn-like'' features. 

As mentioned above, another way to distinguish between the different mass-loss mechanisms is by probing the extended \gls{CSM} density profile that surrounds the progenitor star. This can be done with multi-epoch post-explosion radio observations (e.g., see \citealt{weiler_2007,yaron_2017,iwata_2025,nayana_2025}). As the supernova ejecta expand outward, its interaction with the \gls{CSM} forms a shock front at the interface; synchrotron emission is then produced from electrons accelerated at the shock front. 
The observed synchrotron spectra arising from this interaction can be used to determine the physical properties (i.e., the \gls{CSM} radius and density) of the emitting region \citep{Chevalier1998}. 
By inferring these properties for the synchrotron spectra from multiple epochs, we can trace the mass-loss history of the progenitor back multiple decades pre-explosion. 
This can be used to distinguish between eruptive, short-lived outbursts and long-term, steady mass-loss mechanisms.

In this work, we will investigate the origin of the dense \gls{CSM} surrounding SN~2024ggi \citep{tonry_2024}. It is a Type II SN \citep{hoogendam_2024, zhai_2024} that is remarkably nearby, at a distance of $D = 7.2\pm0.2$ Mpc \citep{saha_2006}. Spectra taken of SN~2024ggi shortly after it was discovered contained prominent narrow emission lines \citep{chen_2024, chen_2024b, jacobson_galan_2024, pessi_2024, shrestha_2024, zhang_2024}. These narrow emission lines remained observable for $3.8 \pm 1.6$ days, which indicates that SN~2024ggi is ``IIn-like'' \citep{jacobson_galan_2024}. Other work confirmed the presence of a dense surrounding \gls{CSM} \citep{chen_2024, chen_2024b, jacobson_galan_2024, ryder_2024, ertini_2025} and an enhanced mass-loss rate prior to explosion \citep{chen_2024b, jacobson_galan_2024, shrestha_2024, zhang_2024, ertini_2025, hu_2025, bostroem_2026}.

SN~2024ggi is one of the closest Type II SNe with clear ``IIn-like'' features in a decade, which makes it an ideal candidate to study the origin of the dense \gls{CSM} that surrounds it. We will use data from \gls{ATLAS} to study the pre-explosion light curve of SN~2024ggi to look for precursor emission. We will place constraints and upper limits on the pre-explosion light curve. We then complement these optical constraints with multi-epoch post-explosion radio observations to probe the extended \gls{CSM} density profile and the long-term mass-loss history of the progenitor star. 

The structure of this paper is as follows. We first summarize previous work on SN~2024ggi in Section \ref{sec:previous_work_ggi}. The optical pre-explosion light curve analysis is presented in Section \ref{sec:optical}, while the radio analysis is shown in Section \ref{sec:radio}. We then discuss our findings from the optical and radio analysis in the context of the different possible mechanisms that can explain the presence of dense \gls{CSM} around SN~2024ggi. Finally, we summarize our conclusions in Section \ref{sec:conclusion}.

\section{SN 2024ggi: a Type II SN in NGC 3621}
\label{sec:previous_work_ggi}

SN~2024ggi was first discovered by the \gls{ATLAS} survey \citep{tonry_2018} on 11 April 2024 (MJD 60411.14) in NGC 3621 with coordinates $\alpha = 11^{h}18^{m}22.09^{s}$, $\delta = -32^{\circ}50'15.27"$ \citep{tonry_2024}. It is classified as a Type II SN \citep{hoogendam_2024,zhai_2024} with a peak absolute magnitude of $M_{g} = -18.1$ mag \citep{jacobson_galan_2024}. SN~2024ggi is particularly interesting due to its proximity; it is an ideal candidate for follow-up studies at a redshift-independent distance of $D = 7.2\pm0.2$ Mpc \citep{saha_2006} and a redshift of $z = 0.002215$, which is based on the Na {\footnotesize I} D absorption line \citep{jacobson_galan_2024}. Indeed, SN~2024ggi has been the target of multiple follow-up observations across the entire wavelength spectrum: radio \citep{chandra_2024,ryder_2024,hu_2025}, optical \citep{chen_2024, jacobson_galan_2024, wyrzykowski_2025}, X-ray \citep{lutovinov_2024,margutti_2024,zhang_2024b,ferdinand_2026}, and $\gamma$-ray \citep{marti_devesa_2024}.

Multiple groups were able to obtain spectra of SN~2024ggi shortly after it was discovered \citep{chen_2024b, jacobson_galan_2024, pessi_2024, shrestha_2024, zhang_2024}. The spectra obtained immediately after the explosion ($+$0.8 days) showed prominent narrow emission lines (such as H {\footnotesize I}, He {\footnotesize I}, C {\footnotesize III}, N {\footnotesize III}). This implies the presence of a dense, optically thick \gls{CSM}. Later spectra ($+$1.5 days) showing the appearance of highly ionized spectral lines (such as N {\footnotesize IV}, N {\footnotesize V}, C {\footnotesize IV}, O {\footnotesize IV}, O {\footnotesize V}) implied a rise in ionization. \citet{jacobson_galan_2024} found that these features were observable for a duration of $t_{\rm IIn} = 3.8 \pm 1.6$ days, suggesting that SN~2024ggi is ``IIn-like''.

The outer radius of the confined \gls{CSM} surrounding SN~2024ggi is estimated to be between $2 \times 10^{14} -  6 \times 10^{14}$ cm \citep{chen_2024b, jacobson_galan_2024, shrestha_2024, zhang_2024}. \citet{ertini_2025} specifically note that the \gls{CSM} likely consists of two distinct components: a compact core and an extended tail. These studies then combine measurements of the \gls{CSM} density with assumed wind velocities (which vary between 10 - 77 km s$^{-1}$) to calculate pre-explosion mass-loss rates for the progenitor of SN~2024ggi between $\dot{M} = 10^{-3} - 10^{-2} M_{\odot} \textrm{yr}^{-1}$ \citep{chen_2024b, jacobson_galan_2024, shrestha_2024, zhang_2024, ertini_2025, hu_2025, bostroem_2026}, which is typical for Type IIn SNe \citep{kiewe_2012}.

Interestingly, estimates of the \gls{CSM} mass around SN~2024ggi show a wide range of values. Estimates obtained from modeling the SN light curve vary between $0.4 - 1.2 M_{\odot}$ \citep{chen_2024, chen_2024b, ertini_2025}. However, recent work by \citet{laplace_2026} has shown that a combination of radial pulsations in \glspl{RSG} and the extended chromospheres predicted by \citet{fuller_2024} can explain the ``early excess'' features in the light curve that mislead \gls{CSM} mass estimates. Indeed, \gls{CSM} mass estimates based on spectroscopy are much lower: $0.02 - 0.04 M_{\odot}$ \citep{jacobson_galan_2024}.

By fitting the stellar spectral models to optical \gls{HST} and \gls{MIR} Spitzer data, \citet{xiang_2024} found that the progenitor of SN~2024ggi has $T_{*} = 3290^{+19}_{-27}$ K, $R_{*} = 887^{+60}_{-51} R_{\odot}$ and $\log(L/L_{\odot}) = 4.92^{+0.05}_{-0.04}$. They mention that the progenitor is consistent with being a solar-metallicity star with an initial mass $M_{*} = 13^{+1}_{-1} M_{\odot}$. The derived dust shell surrounding the progenitor is optically thin ($\tau_{V} = 0.3$). Other work on the progenitor find similar values for the initial mass, ranging from $M_{*} = 10.2 - 17 M_{\odot}$ \citep{chen_2024, hong_2024, ertini_2025}, although other progenitor radius estimates are generally lower $R_{*} = 517 - 550 R_{\odot}$ \citep{chen_2024, ertini_2025}. \citet{xiang_2024} find a progenitor mass-loss rate of $\dot{M} <3 \times 10^{-6} M_{\odot} \textrm{yr}^{-1}$ between 1995-2003 (so ${\sim}21 - 29$ years pre-explosion), consistent with mass-loss rates of typical \glspl{RSG} ($10^{-9} M_{\odot} \textrm{yr}^{-1} < \dot{M} < 10^{-5} M_{\odot} \textrm{yr}^{-1}$) found in \citet{antoniadis_2024}. However, this mass-loss rate is much lower than the rates described above, which implies that the progenitor of SN~2024ggi experienced enhanced mass-loss leading up to its explosion.

All the measurements and parameters noted above are summarized in Table \ref{tab:ggi_properties}.

\renewcommand\cellalign{tl}
\begin{deluxetable}{l l l}
  \tablecaption{Summary of various parameters of SN~2024ggi found in the literature. In the table below, we show the right ascension, declination, distance ($D$), redshift ($z$), peak absolute magnitude in the $g$-band ($M^{\rm peak}_{g}$), the rise time in the $g$-band ($t_{\textrm{rise}, g}$), the duration of the ``IIn-like'' features ($t_{\rm IIn}$), the radius and mass of the surrounding \gls{CSM} ($R_{\rm CSM}$ and $M_{\rm CSM}$, respectively), the mass-loss rate $\dot{M}$, and the explosion energy. We also list properties of the progenitor of SN~2024ggi: temperature ($T_{\rm *, precursor}$), radius ($R_{\rm *, precursor}$), mass ($M_{\rm *, precursor}$), luminosity ($\log(L/L_{\odot})_{\rm precursor}$), and mass-loss rate ($\dot{M}_{\rm precursor}$). A = \citet{tonry_2024}; B = \citet{saha_2006}; C = \citet{jacobson_galan_2024}; D = \citet{chen_2024b}; E = \citet{shrestha_2024}; F = \citet{chen_2024}; G = \citet{ertini_2025}; H = \citet{zhang_2024}; I = \citet{hu_2025}; J = \citet{xiang_2024}; K = \citet{hong_2024}; L = \citet{bostroem_2026}; M = \citet{ferdinand_2026}.}
  \label{tab:ggi_properties}
  \tablehead{
  \colhead{Property} & \colhead{Value} & \colhead{Reference}}
  \decimalcolnumbers
  \startdata
    Host galaxy & NGC 3621 & A\\
    Discovery Date (MJD) & 60411.14 & A\\
    R.A. & $11^{h}18^{m}22.09^{s}$ & A\\
    Dec. & $-32^{\circ}50'15.27"$ & A\\
    $D$ & $7.2 \pm 0.2$ Mpc & B\\
    $z$ & 0.002215 & C\\
    $M^{\rm peak}_{g}$ & $-18.1 \pm 0.06$ mag & C\\
    $t_{\textrm{rise}, g}$ & $6.5 \pm 0.9$ days & C\\
    $t_{\rm IIn}$ & $3.8 \pm 1.6$ days & C\\
    $R_{\rm CSM}$ & $2 \times 10^{14} -  6 \times 10^{14}$ cm & C, D, E, G, H\\
    $M_{\rm CSM}$ & $0.4 - 1.2 M_{\odot}$ (light curve) & D, F, G\\
     & $0.02 - 0.04 M_{\odot}$ (spectra) & C\\
     $\dot{M}$ & $10^{-3} - 10^{-2} M_{\odot} \textrm{yr}^{-1}$ & C, D, E, G, H, I\\
     $E_{\rm exp}$ & $(1.2 - 2) \times 10^{51}$ erg & C, D, G\\
     $T_{\rm *, precursor}$ & $3290^{+19}_{-27}$ K & J\\
     $R_{\rm *, precursor}$ & $517 - 887 R_{\odot}$ & F, G, J\\
     $M_{\rm *, precursor}$ & $10.2 - 17 M_{\odot}$ & F, G, K, J\\
     $\log(L/L_{\odot})_{\rm precursor}$ & $4.92^{+0.05}_{-0.04}$ & J\\
     $\dot{M}_{\rm precursor}$ & $<3 \times 10^{-6} - 9 \times 10^{-5} M_{\odot} \textrm{yr}^{-1}$ & J, L, M
  \enddata
\end{deluxetable}


\section{Constraints on Precursor Emission from ATLAS Observations}
\label{sec:optical}

\subsection{ATLAS Observations}
\label{sec:atlas_observations}

The \gls{ATLAS} survey \citep{tonry_2018} was originally built to find dangerous \gls{NEA}. However, due to its high cadence (${\sim} 2$ days), all-sky coverage, and high sensitivity (m $\sim$ 20~mag), it is also well equipped to find and study transients. \gls{ATLAS} consists of four 0.5m telescopes. Two of the telescopes are located in the northern hemisphere in Hawai`i (one each on Haleakala and Mauna Loa). The other two are located in the southern hemisphere, one in Chile (El Sauce), and the other in South Africa (Sutherland). \gls{ATLAS} has one orange ($o$) and one cyan ($c$) band.

\begin{figure}
	\includegraphics[width=\columnwidth]{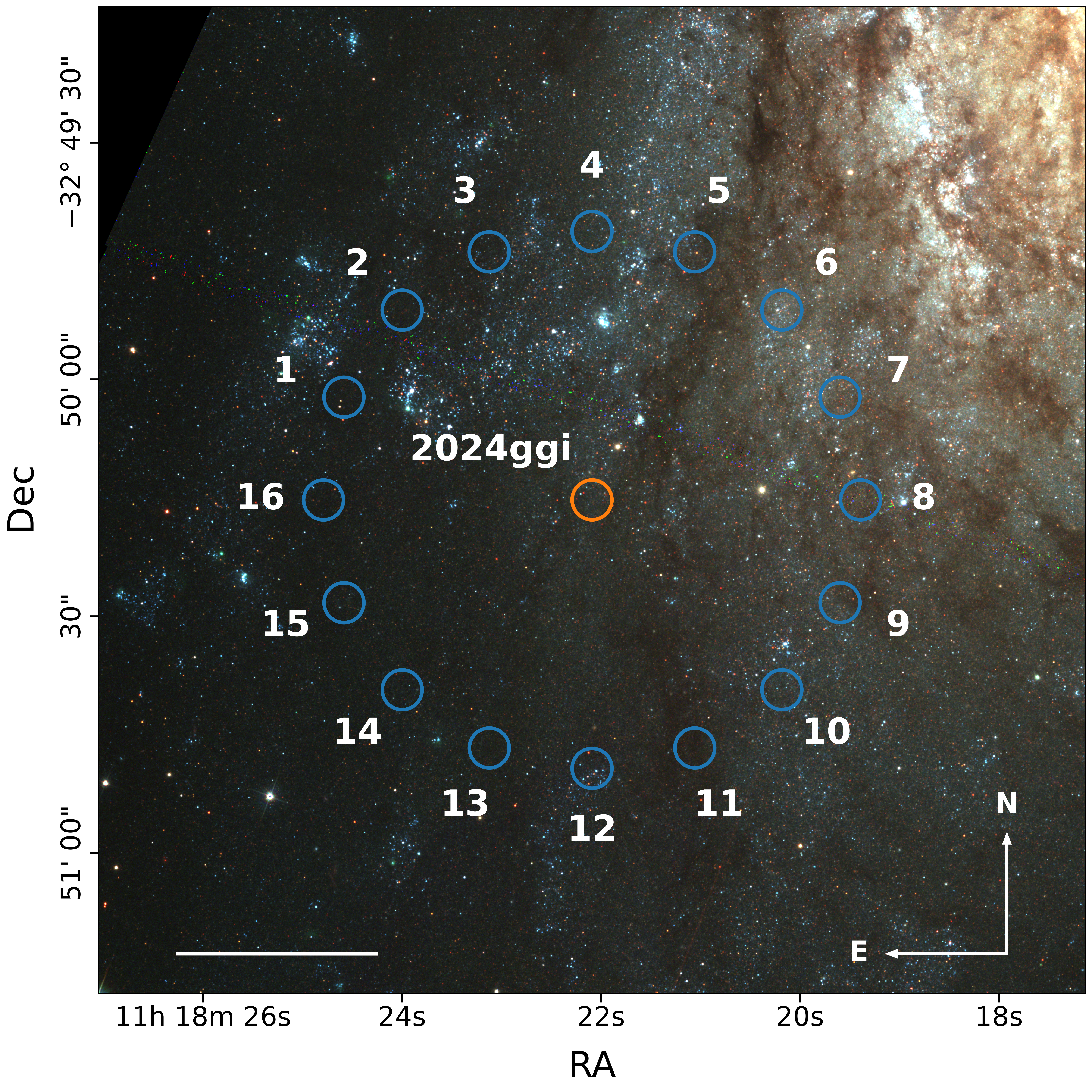}
    \caption{A cutout from the \gls{HST} F435W (blue), F555W (green), and F814W (red) bands taken with the \gls{ACS}. The \gls{HST} image comes from Proposal 9492 in Cycle 11. The image is centered around the location of SN~2024ggi and also shows the locations of the control light curves used in this work. The outer regions of NGC 3621 are visible in the top-right corner. The white line in the bottom-left corner has a length of 30 arcsec.} 
    
    \label{fig:lightcurve_locations}
\end{figure}

The \gls{ATLAS} light curve of SN~2024ggi in both bands was obtained using the forced photometry server \citep{shingles_2021}. After inspecting the light curve, it becomes clear that SN~2024ggi was observed in 10 distinct annual observation seasons during which it was visible from the telescope site, ranging between MJD 57368 and 60719. This includes $\sim$8 years of coverage before explosion and $\sim$300 days after explosion. In this work, we use the pre-explosion data to search for evidence of precursor emission. This is done using \gls{ATClean}\footnote{\url{https://github.com/srest2021/atclean/tree/v5}}, a light curve cleaning and analysis pipeline described in \citet{rest_2025}, which has been previously used to search for precursor emission in SN~2023ixf. We also obtain \todo{16} control light curves placed in a circle around SN~2024ggi with a radius of \todo{37} arcsec. We assume that no real astrophysical transient flux is present in the control light curves. The coordinates of SN~2024ggi and every control light curve are listed in Table~\ref{tab:locations} and visualized in Figure~\ref{fig:lightcurve_locations}.

In the figures below and subsequent analysis, we use the \gls{ATLAS} $o$-band, due to its higher cadence than the \gls{ATLAS} $c$-band \citep{tonry_2018}. This allows us to derive deeper and more accurate constraints with \gls{ATClean} (see Sections \ref{sec:cleaning} and \ref{sec:atclean_lc_analysis}). However, using the $c$-band instead does not significantly change the results or conclusions of this work.

\begin{deluxetable}{c c c}
  \tablecaption{The coordinates of SN~2024ggi and the control light curves used in this work.}
  \label{tab:locations}
  \tablehead{
  \colhead{Name} & \colhead{RA} & \colhead{Dec}}
  \decimalcolnumbers
  \startdata
2024ggi & 169.592030 & -32.837576\\ 
Control 1 & 169.602415 & -32.833961\\ 
Control 2 & 169.599979 & -32.830897\\ 
Control 3 & 169.596332 & -32.828850\\ 
Control 4 & 169.592030 & -32.828131\\ 
Control 5 & 169.587729 & -32.828850\\ 
Control 6 & 169.584082 & -32.830897\\ 
Control 7 & 169.581645 & -32.833961\\ 
Control 8 & 169.580790 & -32.837576\\ 
Control 9 & 169.581645 & -32.841190\\ 
Control 10 & 169.584082 & -32.844254\\ 
Control 11 & 169.587729 & -32.846301\\ 
Control 12 & 169.592030 & -32.847020\\ 
Control 13 & 169.596332 & -32.846301\\ 
Control 14 & 169.599979 & -32.844254\\ 
Control 15 & 169.602415 & -32.841190\\ 
Control 16 & 169.603271 & -32.837576\\ 
    \enddata
\end{deluxetable}

\subsection{Cleaning the ATLAS Light Curves}
\label{sec:cleaning}

We clean the \gls{ATLAS} light curves using \gls{ATClean}. We largely follow the procedure described in \citet{rest_2025}, but elaborate on the main steps below. The first step includes simply removing all measurements whose nominal uncertainty ($\sigma_{f,0}$) exceeds \todo{160} $\mu$Jy. This threshold value was determined in \citet{rest_2025} by calculating the typical uncertainty of bright stars just below the saturation limit. This threshold affected \todo{4.23}\% of measurements in the $o$-band. 

The next step involves updating the uncertainties on the light curve of SN~2024ggi. It is possible that the nominal uncertainties ($\sigma_{f,0}$) underestimate the true uncertainties ($\sigma_{f}$) due to additional sources of systematic uncertainty ($\sigma_{\textrm{sys}}$). The true uncertainty can be calculated by adding the other two in quadrature:

\begin{equation}
    \sigma_{f}^{2} = \sigma_{f,0}^{2} + \sigma_{f,\textrm{sys}}^{2} \; .
    \label{eq:uncertainties}
\end{equation}

We can estimate $\sigma_{f,\textrm{sys}}$ from the control light curves, since they should not contain real flux. We calculate the median of the nominal uncertainties across all light curves ($\tilde{\sigma}_{f,0}$), as well as the 3$\sigma$-clipped standard deviation of the control light curve measurements ($\sigma_{f, \textrm{empirical}}$). We can then calculate $\sigma_{f,\textrm{sys}}$ with:

\begin{equation}
    \sigma_{f,\textrm{sys}}^{2} = 
    \begin{cases}
        0 & \text{if } \sigma_{f, \textrm{empirical}} \leq \tilde{\sigma}_{f,0}\\
        \sigma_{f, \textrm{empirical}}^{2} - \tilde{\sigma}_{f,0}^{2} & \text{otherwise}
    \end{cases}
\end{equation}

We find that, in the $o$-band for SN~2024ggi, \todo{$\sigma_{f,\textrm{sys}} = 12.35\ \mu \textrm{Jy}$}. Applying this additional systematic uncertainty using Equation \ref{eq:uncertainties} increases the uncertainty in the measurements by \todo{16.75\%}. We use $\sigma_{f}^{2}$ instead of $\sigma_{f,0}^{2}$ for the rest of this paper. Note that we only use the control light curve measurements in this step where the reduced chi-square of the PSF fits to our source apertures in the difference imaging ($\chi^{2}_{\rm PSF}$) is less than 20.

In \citet{rest_2025}, they then apply a cut on $\chi^{2}_{\rm PSF}$. However, we find that $\chi^{2}_{\rm PSF}$ significantly increases even for good measurements at very bright epochs. This possibility is also mentioned in the \gls{ATClean} documentation. We therefore opt not to apply this step in this work. Applying this threshold would only affect the epochs around the discovery date. Since we are primarily interested in the pre-explosion light curve, this decision does not significantly change our results.

We then calculate the 3$\sigma$-clipped average of the \todo{16} control light curves for every epoch. This allows us to select unreliable epochs, as the flux in the controls should be consistent with 0. Entire epochs are removed from the light curve if the $\chi^{2}$ associated with the 3$\sigma$-clipped average is greater than 2.5, or if the resultant \gls{SNR} $> 3$. Epochs are also removed if more than 2 measurements in this epoch were clipped or if fewer than 4 measurements were averaged in the 3$\sigma$-clipped average. This step specifically flagged \todo{4.96}\% of measurements in the $o$-band. 

The difference images and the associated light curves in \gls{ATLAS} are created by subtracting templates from the science images \citep{tonry_2018}. However, these template images were updated by \gls{ATLAS} over time. This can cause artificial step discontinuities in the SN and control light curves. These template changes occurred at MJD = \todo{58417} and MJD = \todo{58882}, which means that there are three distinct periods. We correct for these step discontinuities by calculating the $3\sigma$-clipped average of all flux measurements (the SN light curve and all control light curves) that are not flagged as unreliable by one of the steps above for each of the three epochs. These offset values are very low (\todo{-0.14, 0.47, and 0.89} $\mu$Jy for each epoch in the $o$-band), which suggests that the effect of template changes is minimal for the area around SN~2024ggi. Nevertheless, we correct for the template changes by subtracting these offsets from the SN and control light curves in each corresponding epoch. 

Finally, we bin the \gls{ATLAS} light curves by calculating the $3\sigma$-clipped average in every bin with a size of \todo{1} day. \gls{ATLAS} typically takes four exposures every day. Thus, on average, every bin contains $\sim$4 measurements. However, only measurements that were not flagged in any of the steps above were used. We further remove any bins that had fewer than \todo{2} unflagged measurements, bins where more than \todo{1} datapoint was clipped, or where the returned $\chi^{2} > \todo{4.0}$. This removed \todo{6.17}\% of the $o$-band bins.

\begin{figure*}
	\includegraphics[width=\textwidth]{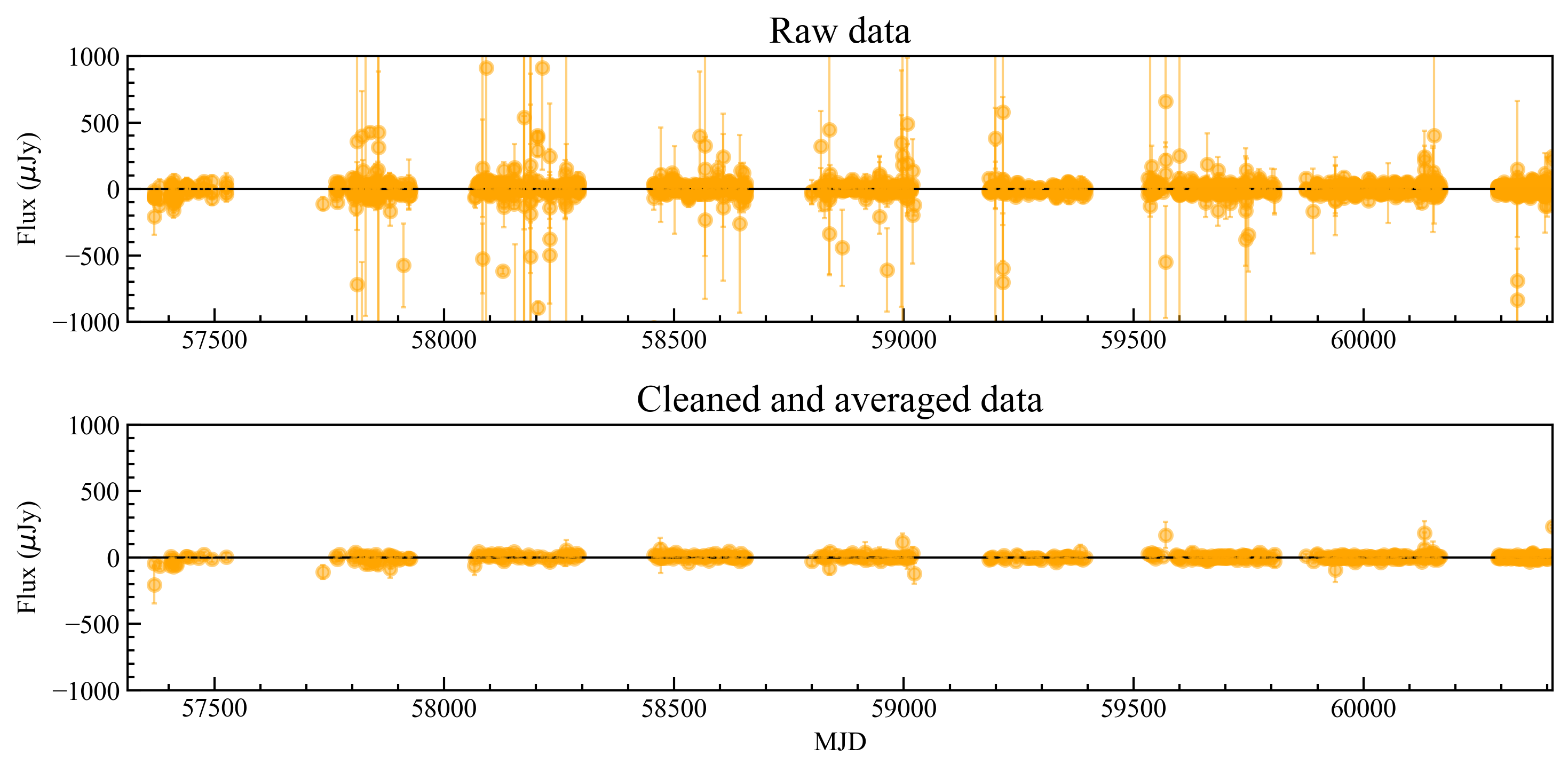}
    \caption{The pre-SN $o$-band light curve of SN~2024ggi before cleaning the data with \gls{ATClean} (top) and after (bottom). The different cleaning steps are described in Section \ref{sec:cleaning}. The breaks in the light curve represent the distinct annual observation seasons during which SN~2024ggi was visible from the telescope site.}
    \label{fig:cleaning}
\end{figure*}
\begin{figure*}
	\includegraphics[width=\textwidth]{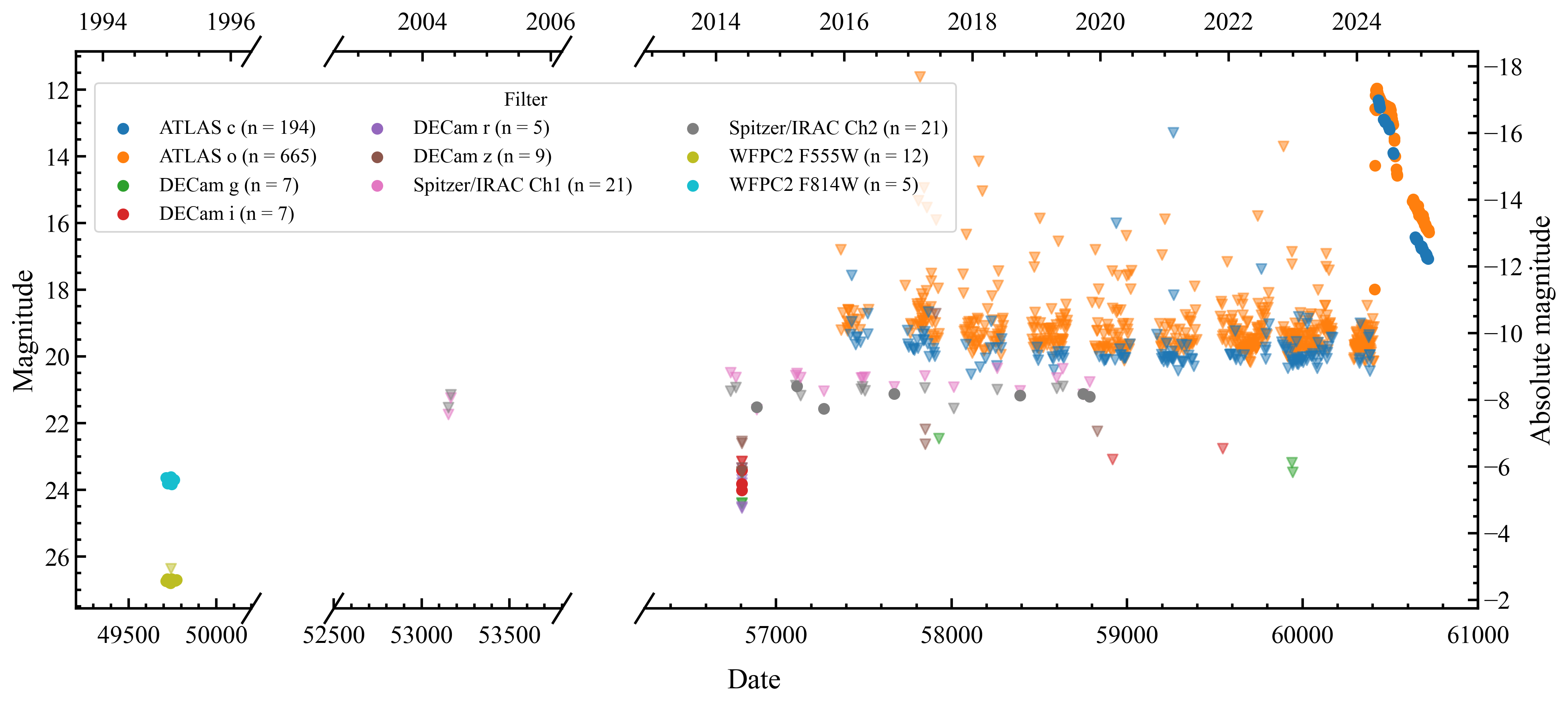}
    \caption{The light curve of SN~2024ggi. The data were obtained from \gls{ATLAS} \citep{tonry_2018} and cleaned with \gls{ATClean}, as described in Section \ref{sec:cleaning}, \gls{DECam} \citep{flaugher_2015}, Spitzer \citep{werner_2004}, and \gls{WFPC2} aboard \gls{HST} \citep{holtzman_1995}. The circles represent detections, while the triangles denote $5\sigma$ upper limits.} 
    \label{fig:lightcurve}
\end{figure*}

The result of these cleaning steps is shown in Figure \ref{fig:cleaning}, where the raw $o$-band pre-SN light curve from \gls{ATLAS} is compared with the cleaned $o$-band pre-SN light curve. All these steps (except for the template change correction) were done with \gls{ATClean}. More details on each of these cleaning steps are provided in Section 3 of \citet{rest_2025}. The full (pre- and post-explosion) cleaned $o$- and $c$-band \gls{ATLAS} light curves of SN~2024ggi are shown in Figure \ref{fig:lightcurve}. We do not find any significant ($>5\sigma$) detections in the averaged and cleaned pre-explosion light curve. This warrants more detailed analysis with \gls{ATClean} (see Sections \ref{sec:atclean_lc_analysis} - \ref{sec:atclean_2024ggi})

In addition to the \gls{ATLAS} photometry, Figure \ref{fig:lightcurve} also includes observations from \gls{DECam} in the $griz$ bands \citep{flaugher_2015}, Spitzer/IRAC ($\text{Ch1}$ and $\text{Ch2}$; \citealp{werner_2004}), and \gls{WFPC2} aboard \gls{HST} (F555W and F814W; \citealp{holtzman_1995}). A few of these epochs show tentative pre-explosion detections of the progenitor star. For example, the \gls{DECam} $i$-band has $>5\sigma$ precursor detections in early 2014 (at MJD = \todo{56805}) of $M_i = -5.5$ mag. Spitzer observed the precursor for multiple epochs up to  $\sim20$ years pre-explosion with $\text{Ch1}$ and $\text{Ch2}$, with absolute magnitudes around $-8$ mag. Finally, the \gls{WFPC2} has detections in the F814W-band that reach absolute magnitudes up to $-5.6$ mag in 1995. We note that these detections are fainter than the \gls{ATLAS} upper limits, and thus consistent with them. Further analysis is needed to confirm whether these precursor detections are actually real, and not false positive detections due to, e.g., image artifacts. However, this is outside the scope of this work. As the data from \gls{DECam}, Spitzer/IRAC and \gls{WFPC2} are very sparse, we decided to focus on the high-cadence observations from \gls{ATLAS} in this work and analyze them in greater detail to search for precursor outbursts.

\subsection{ATLAS Light Curve Analysis Methodology}
\label{sec:atclean_lc_analysis}


After cleaning up the $o$-band pre-SN light curves, we use the \gls{ATClean} detection algorithm to determine whether SN~2024ggi experienced any significant pre-SN outbursts. As noted in \citet{rest_2025}, this detection algorithm works by effectively convolving the pre-SN light curve with a rolling Gaussian $\mathcal{N}$ of different timescales $\sigma_{\rm kernel}$. This will amplify any signal of a similar timescale. In practice, this is done by calculating a figure of merit, $\Sigma_{\rm FOM}$, as follows:

\begin{equation}
    \Sigma_{\rm FOM} = \frac{\int \mathcal{N}(\sigma_{\rm kernel}) * f / \sigma_{f}}{\int \mathcal{N}(\sigma_{\rm kernel}) * \mathcal{B}}\;,
    \label{eq:sigma_fom}
\end{equation}

where $f$ is the flux in each epoch and $\sigma_{f}$ the noise. Finally, $\mathcal{B}$ is a binary function equal to 1 if the epoch contains a measurement, and 0 otherwise. This is used to normalize $\Sigma_{\rm FOM}$ by the number of measurements present. 

Although \gls{ATClean} allows us to probe a wide range of event timescales, there are limits to the timescales we can examine. The \gls{ATLAS} cadence is 1-2 days \citep{tonry_2018}, which means that detecting events with timescales close to this cadence is difficult. On the other hand, the \gls{ATLAS} observing seasons tend to be 200-300 days long, suggesting that looking for events longer than that is equally challenging. Therefore, we limit the kernel sizes ($\sigma_{\rm kernel}$) to values between 5 and 300 days.

The next step in the \gls{ATClean} detection algorithm involves determining a threshold on $\Sigma_{\rm FOM}$ above which measurements are counted as significant detections. This is done by similarly convolving the control light curves, which are assumed to not contain any real astrophysical transient flux, with a weighted rolling Gaussian sum and calculating their distribution of $\Sigma_{\rm FOM}$. The threshold, $\Sigma_{\rm FOM, limit}$, is then found by balancing contamination of false positives in the control light curves while keeping $\Sigma_{\rm FOM, limit}$ as low as possible. In this work, we allow a maximum of \todo{2} false positives in the control light curves over the entire light curve, similar to \citet{rest_2025}. Note that this implies that each $\sigma_{\rm kernel}$ has its own associated $\Sigma_{\rm FOM, limit}$. 

As mentioned in Section \ref{sec:atlas_observations}, there are multiple observation seasons in which SN~2024ggi was observed. In this work, we only calculate $\Sigma_{\rm FOM}$ and look for detections during these observation seasons. We also pad the edges of each observation season with \todo{30} days to avoid edge effects. We also completely exclude the first observation season (\todo{MJD 57368 - 57526}) due to a significantly lower cadence in our field and poorer quality of the controls. Including this season would negatively affect the resultant $\Sigma_{\rm FOM, limit}$ measurement and harm our ability to obtain representative magnitude limits. We therefore opt to work with data from the second observation season onward (\todo{MJD $\geq 57735$}). A similar MJD range was also excluded in \citet{rest_2025} for the same reasons.

\subsection{Calculating Detection Efficiency}
\label{sec:calculate_detection_efficiency}

We want to characterize the detection efficiency of our method in the specific field of our target. This is done with \gls{ATClean} by injecting simulated outbursts into the control light curves and subsequently calculating what fraction of these are recovered by the detection algorithm. The injected outbursts have a Gaussian shape, with a peak apparent magnitude ($m_{\rm peak}$) varying between 16 and 23 mag and a simulated width ($\sigma_{\rm sim}$) between 2 and 300 days. The MJD of the peak is randomly drawn from all available MJDs, and each simulated event is injected into one of the \todo{16} control light curves, also selected randomly. A total of \todo{50,000} simulated outbursts are injected for each combination of $m_{\rm peak}$ and $\sigma_{\rm sim}$.

The control light curves injected with these simulated outbursts are then convolved with a Gaussian with different kernel sizes ($\sigma_{\rm kernel}$) to calculate $\Sigma_{\rm FOM}$ (see Equation \ref{eq:sigma_fom}). This is then compared to $\Sigma_{\rm FOM, limit}$ to see whether the detection algorithm was able to detect the simulated outburst as described in Section \ref{sec:atclean_lc_analysis}. We then generate an efficiency curve for every combination of $\sigma_{\rm kernel}$ and $\sigma_{\rm sim}$. These can be used to calculate a 50\% (or 80\%) magnitude detection threshold, at which 50\% (or 80\%) of the simulated outbursts with a $m_{\rm peak}$ equal to the magnitude detection threshold are detected.


\subsection{Applying \texttt{ATClean} to SN~2024ggi}
\label{sec:atclean_2024ggi}

\begin{figure}
	\includegraphics[width=\columnwidth]{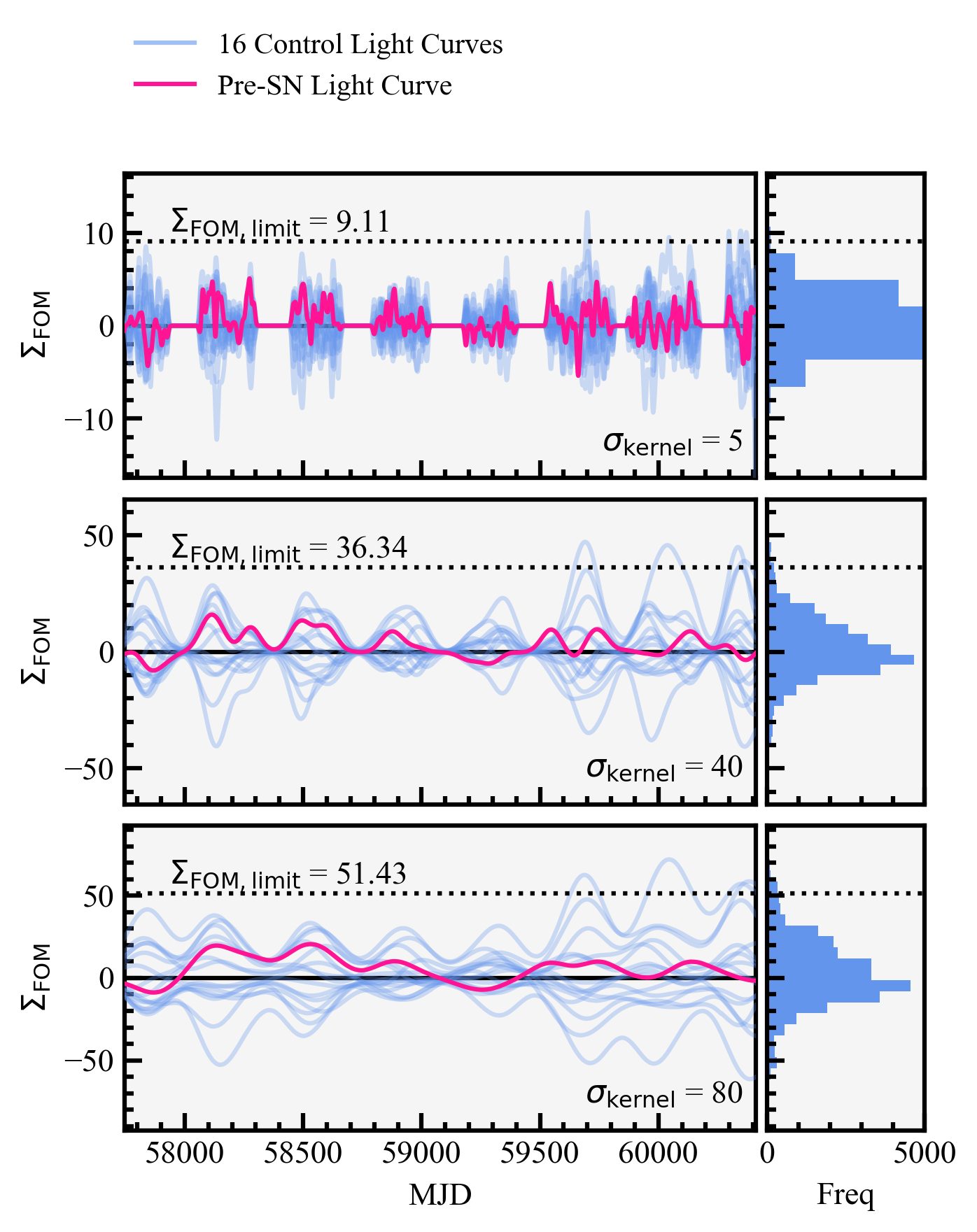}
    \caption{The evolution of $\Sigma_{\rm FOM}$ for the pre-SN light curve (pink) and the control light curves (blue) for a subset of kernel sizes ($\sigma_{\rm kernel}$). The horizontal dashed lines indicate the detection threshold ($\Sigma_{\rm FOM, limit}$). The histograms on the right show the distribution of $\Sigma_{\rm FOM}$ of the control light curves for that specific kernel size. We find no significant detections in the pre-SN light curve for any of the kernel sizes. An extended version of this figure with all values of $\sigma_{\rm kernel}$ tested is shown in Figure \ref{fig:sigma_fom_large} in Appendix \ref{app:optical_appendix}.}
    \label{fig:sigma_fom_small}
\end{figure}

The evolution of $\Sigma_{\rm FOM}$ for the \gls{ATLAS} $o$-band pre-SN light curve and all the controls for $\sigma_{\rm kernel}$ = 5, 40, and 80 is shown in Figure \ref{fig:sigma_fom_small}. The $\Sigma_{\rm FOM}$ evolution for all $\sigma_{\rm kernel}$ tested can be found in Figure \ref{fig:sigma_fom_large} in Appendix \ref{app:optical_appendix}. We find that the pre-SN light curve does not cross the $\Sigma_{\rm FOM, limit}$, suggesting that there is no significant precursor emission detected in any of the kernel sizes tested. This is in agreement with the findings of \citet{shrestha_2024}, who also find no evidence for precursor emission in SN~2024ggi.

\begin{figure}
	\includegraphics[width=\columnwidth]{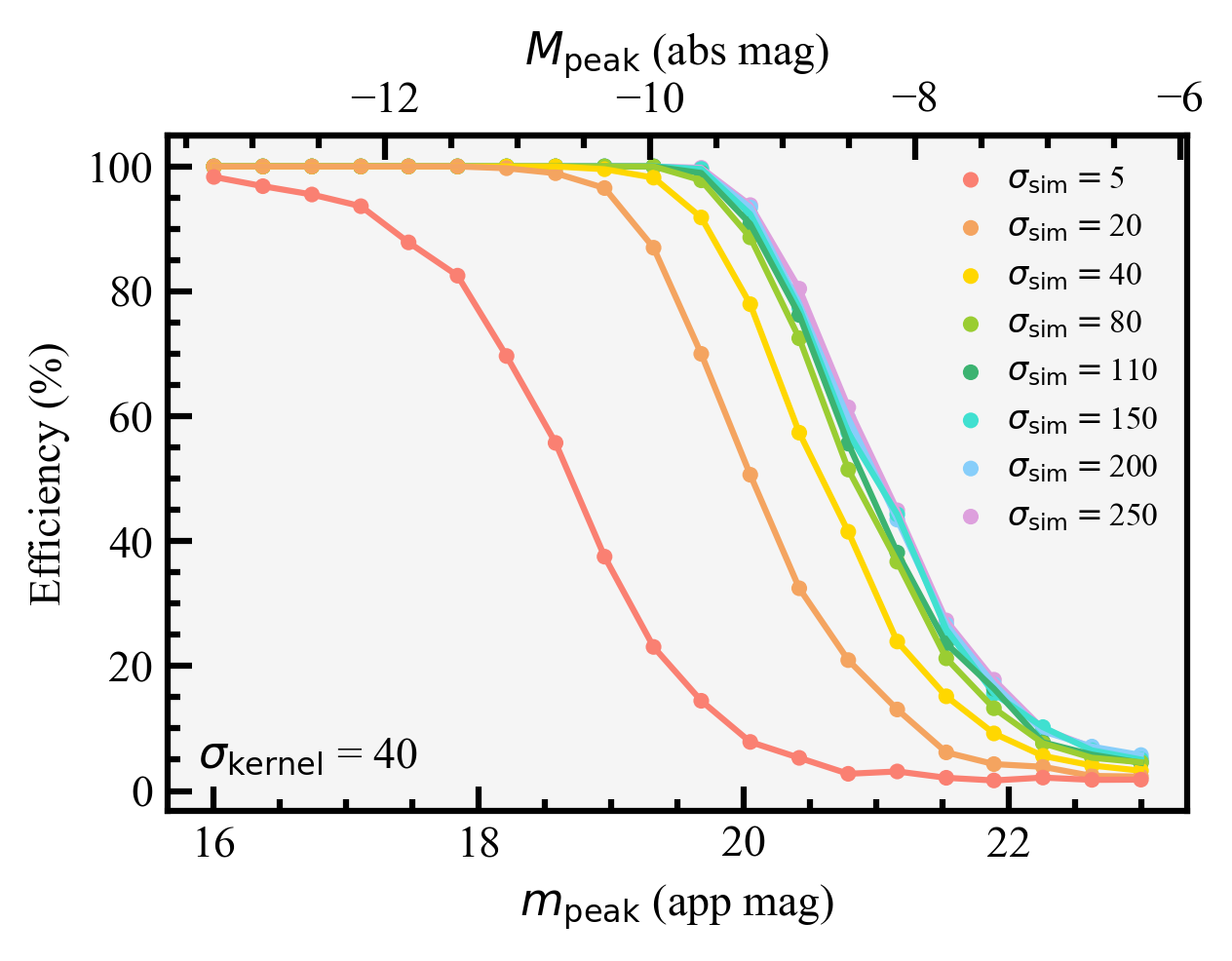}
    \caption{The detection efficiency for simulated events of various timescales over apparent magnitude (bottom axis) and absolute magnitude (top axis) for a kernel size ($\sigma_{\rm kernel}$) of 40 days. An extended version of this figure for all kernel sizes is shown in Figure \ref{fig:efficiency_large} in Appendix \ref{app:optical_appendix}.}
    \label{fig:efficiency_small}
\end{figure}

Despite finding no detections, we can characterize the detection efficiency of \gls{ATClean} and calculate detection thresholds. These detection thresholds will act as upper limits that can still be used to infer information about the physical processes that occurred in the precursor of SN~2024ggi (see Section \ref{sec:calculate_detection_efficiency} for more details). The detection efficiency depends both on the kernel size ($\sigma_{\rm kernel}$) and the timescale of the simulated event ($\sigma_{\rm sim}$). Figure~\ref{fig:efficiency_small} shows the detection efficiency curves for \todo{$\sigma_{\rm kernel} = 40$ days} for all simulated event timescales. The detection efficiency goes from 100\% to 0\% as we consider fainter sources, as expected. The detection efficiency tends to be higher for higher $\sigma_{\rm sim}$ (i.e., we can detect fainter events if they last longer). The detection efficiency curves for all $\sigma_{\rm kernel}$ are shown in Figure \ref{fig:efficiency_large} in Appendix \ref{app:optical_appendix}.

\begin{figure}
	\includegraphics[width=\columnwidth]{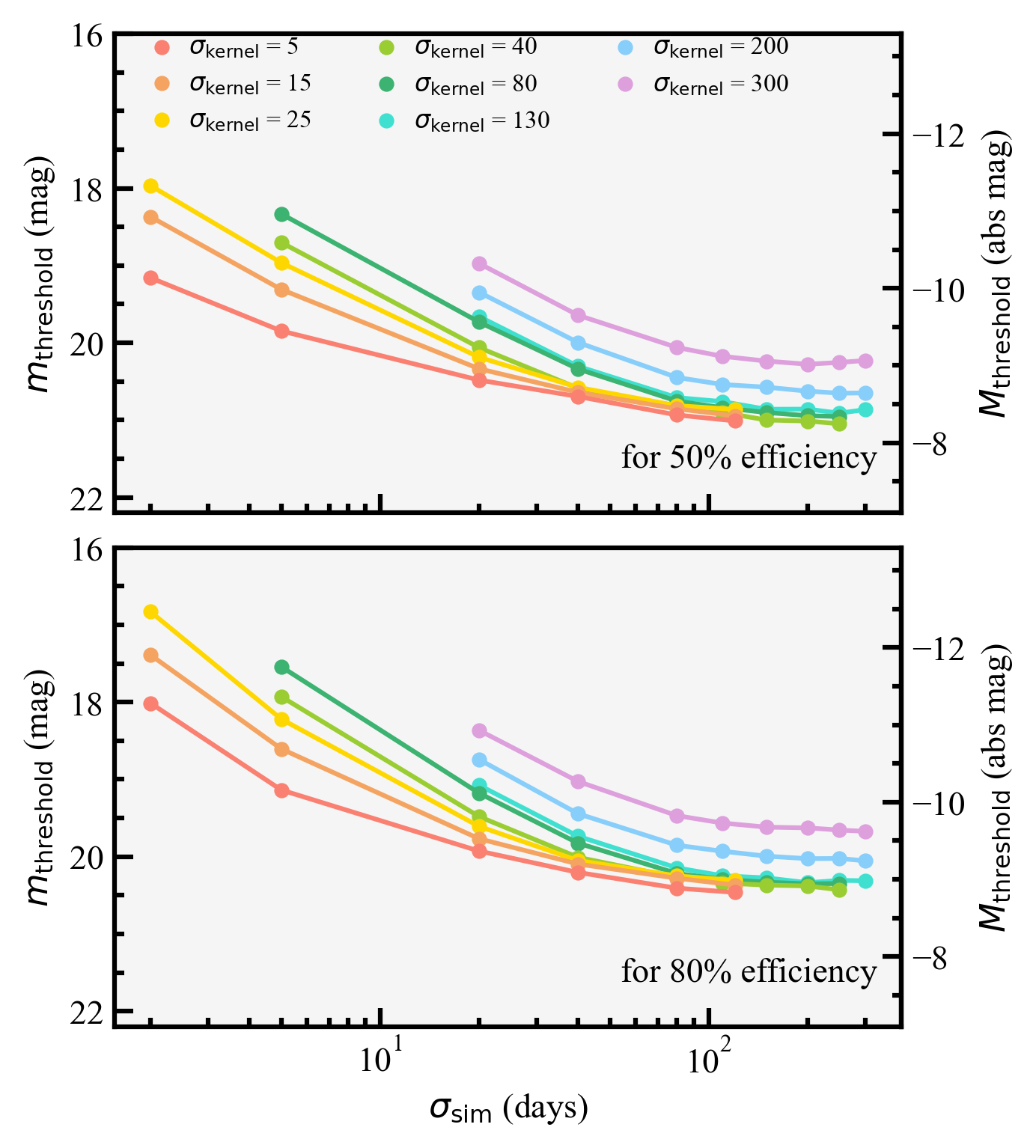}
    \caption{The top panel shows the 50\% apparent magnitude efficiency threshold (left y-axis) and  50\% absolute magnitude efficiency threshold (right y-axis) against the simulated event timescale ($\sigma_{\rm sim}$) for different kernel sizes ($\sigma_{\rm kernel}$). The bottom panel is similar, but for the 80\% efficiency threshold instead.}
    \label{fig:mag_thresholds}
\end{figure}

These efficiency curves are then used to calculate the 50\% and 80\% detection thresholds for every combination of $\sigma_{\rm kernel}$ and $\sigma_{\rm sim}$. These are shown in the top and bottom panels of Figure \ref{fig:mag_thresholds}, respectively. We find that, for a given $\sigma_{\rm kernel}$, the detection threshold becomes deeper as $\sigma_{\rm sim}$ increases, in agreement with the results presented above. This is true for any kernel size used. The deepest thresholds were found for the largest simulated event timescales (\todo{$\sigma_{\rm sim} > 200$ days}) and intermediate kernel sizes (\todo{$40 < \sigma_{\rm kernel} < 80$ days}). The deepest 50\% and 80\% apparent and absolute magnitude thresholds for every simulated timescale $\sigma_{\rm sim}$ are shown in Table \ref{tab:best_mag_thresholds}. The magnitude thresholds for every combination of $\sigma_{\rm kernel}$ and $\sigma_{\rm sim}$ is shown in Table \ref{tab:mag_thresholds} in Appendix \ref{app:optical_appendix}.

\subsection{Comparison of Upper Limits to Other SNe}
\label{sec:comparison_limits}

We did not find any significant evidence for precursor emission in SN~2024ggi using \gls{ATClean}. However, using the calculated detection thresholds, we are able to put constraints on the physical processes that occurred in the precursor of SN~2024ggi. These constraints can be used to shed light on different mechanisms that can explain the dense surrounding \gls{CSM} observed in SN~2024ggi as inferred from other work (e.g., see \citealp{jacobson_galan_2024}).

\begin{figure}
    \includegraphics[width=\columnwidth]{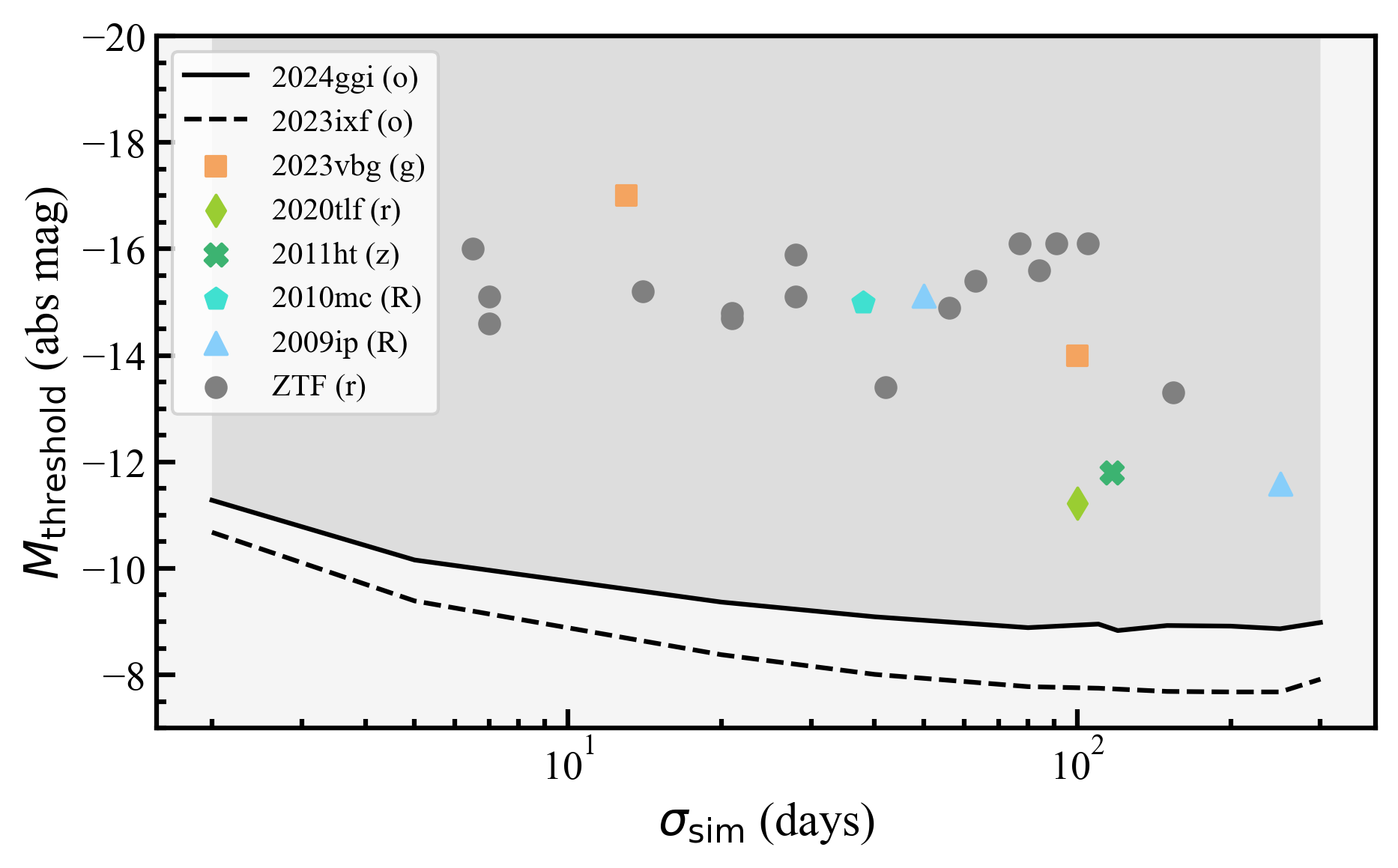}
    \caption{The deepest 80\% absolute magnitude thresholds obtained for every simulated timescale are shown by the full line. We can exclude the entire parameter space above this line (gray background) for SN~2024ggi based on this analysis. The dashed line shows a similar limit obtained for 2023ixf in \citet{rest_2025}. We show the median absolute magnitude values of the pre-SN variability of the 2011 and 2012a events of the SN impostor SN~2009ip in the $R$-band \citep{pastorello_2013} to facilitate comparison. We also show other known SNe in this parameter space: 2010mc \citep{ofek_2013}, 2011ht \citep{fraser_2013}, 2020tlf \citep{jacobson_galan_2022, jacobson_galan_2025}, 2023vbg \citep{goto_2025}, and multiple Type IIn SNe found in \citet{strotjohann_2021} using \gls{ZTF} \citep{bellm_2019}.}
    \label{fig:best_absmag_thresholds}
\end{figure}

We show the deepest 80\% absolute magnitude thresholds obtained for the simulated events of every timescale in Figure \ref{fig:best_absmag_thresholds}. This allows us to visualize which regions of this parameter space are excluded for SN~2024ggi (grey background), and to compare with other SNe with known pre-SN variability. We also show the median absolute magnitudes of the pre-SN variability of the 2011 and 2012a events of SN impostor SN~2009ip in the $R$-band, which lasted ${\sim}250$ and ${\sim} 50$ days, respectively \citep{pastorello_2013}. We additionally include other known Type IIn SNe with precursor emission: 2010mc \citep{ofek_2013}, 2011ht \citep{fraser_2013}, 2023vbg \citep{goto_2025}, and multiple Type IIn SNe found in \gls{ZTF} \citep{bellm_2019, strotjohann_2021}. We also show the precursor outburst of 2020tlf, a SN with ``IIn-like'' features \citep{jacobson_galan_2022, jacobson_galan_2025}. The absolute magnitude limits for SN~2024ggi found in this work confidently exclude the region of the parameter space where these other Type IIn SNe with detected precursor emission are located, as well as any outbursts similar to the outburst observed in SN~2020tlf. The upper limits for SN~2023ixf found by \citet{rest_2025} are also shown in Figure \ref{fig:best_absmag_thresholds}.

\begin{deluxetable}{c c c c c}
  \tablecaption{The best 50\% and 80\% apparent magnitude thresholds ($m_{\rm threshold, 50}$ and $m_{\rm threshold, 80}$) and the 50\% and 80\% absolute magnitude thresholds ($M_{\rm threshold, 50}$ and $M_{\rm threshold, 80}$) for any $\sigma_{\rm sim}$. The complete table with all combinations of $\sigma_{\rm sim}$ and $\sigma_{\rm kernel}$ is found in Table \ref{tab:mag_thresholds} in Appendix \ref{app:optical_appendix}.}
  \label{tab:best_mag_thresholds}
  \tablehead{
  \colhead{$\sigma_{\rm sim}$ [days]} & \colhead{$m_{\rm threshold, 50}$ [mag]} & \colhead{$m_{\rm threshold, 80}$ [mag]} & \colhead{$M_{\rm threshold, 50}$ [mag]} & \colhead{$M_{\rm threshold, 80}$ [mag]}}
  \decimalcolnumbers
  \startdata
    2 & 19.16 & 18.01 & -10.14 & -11.28 \\
    5 & 19.85 & 19.14 & -9.45 & -10.16 \\
    20 & 20.48 & 19.93 & -8.81 & -9.37 \\
    40 & 20.70 & 20.20 & -8.60 & -9.09 \\
    80 & 20.93 & 20.41 & -8.36 & -8.89 \\
    110 & 20.90 & 20.34 & -8.39 & -8.95 \\
    120 & 21.01 & 20.46 & -8.28 & -8.83 \\
    150 & 21.00 & 20.37 & -8.29 & -8.93 \\
    200 & 21.01 & 20.38 & -8.28 & -8.91 \\
    250 & 21.05 & 20.43 & -8.25 & -8.86 \\
    300 & 20.86 & 20.31 & -8.43 & -8.98 \\
  \enddata
\end{deluxetable}

\subsection{Implications for Mass of CSM Ejected in Pre-SN Eruptions}
\label{sec:chips_csm_mass}

We aim to better constrain whether eruptive mass-loss events can explain the surrounding \gls{CSM} of SN~2024ggi. This is done by converting our magnitude thresholds to more physically meaningful thresholds on the \gls{CSM} mass using simulations generated by \texttt{CHIPS} \citep{takei_2022,takei_2024}. This code performs one-dimensional radiation hydrodynamical simulations that are able to reproduce both the \gls{CSM} resulting from mass eruptions around the star prior to core collapse, and its associated light curve. We simulate light curves of a range of progenitor masses: 11, 13, 14, and 17 $M_{\odot}$, and we also use the preexisting light curve from \citet{kuriyama_2020} that corresponds to a progenitor mass of 20 $M_{\odot}$. This covers the range of stellar masses predicted for the progenitor of SN~2024ggi (see Table \ref{tab:ggi_properties}). For each simulation, we measure the peak luminosity from each light curve, as well as the corresponding \gls{CSM} mass, which allows us to convert our magnitude limits ($m_{\rm threshold, 80}$), each associated with a specific $\sigma_{\rm sim}$, to limits on the ejected mass. 

\begin{figure}
    \includegraphics[width=\columnwidth]{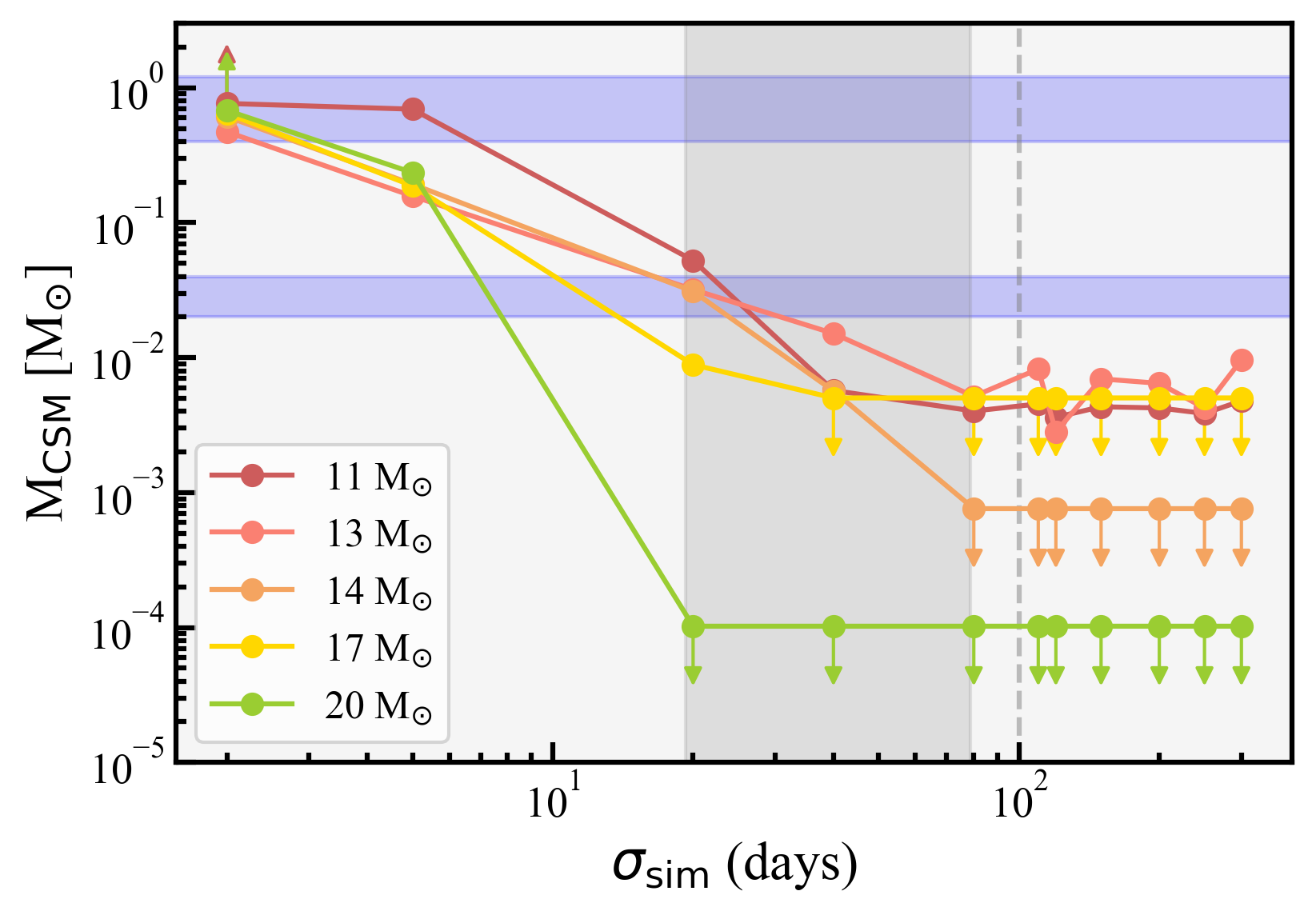}
    \caption{Constraints on the \gls{CSM} mass around SN~2024ggi as a function of outburst timescale ($\sigma_{\rm sim}$), obtained from comparing the detection thresholds obtained with \gls{ATClean} to simulations performed with \texttt{CHIPS} for progenitor masses ranging between 11 - 20 $M_{\odot}$. The blue horizontal bands represent the \gls{CSM} mass estimates inferred from light curve modeling (0.4 - 1.2 $M_{\odot}$; top blue band) and spectroscopy (0.02 - 0.04 $M_{\odot}$; bottom blue band). The gray vertical band indicates the 25$^{\rm th}$ - 75$^{\rm th}$ percentile range of precursor outburst timescales found for Type IIn SNe in \gls{ZTF} found by \citet{strotjohann_2021}. Finally, the vertical dashed line indicates the outburst timescale found for SN~2020tlf \citep{jacobson_galan_2025}. } 
    \label{fig:chips_sim}
\end{figure}

This is visualized in Figure \ref{fig:chips_sim}, where we plot $\sigma_{\rm sim}$ against \gls{CSM} mass for each of the models with a different progenitor mass. It is clear that, as expected, the deeper luminosity limits for larger $\sigma_{\rm sim}$ result in lower (i.e., more constraining) limits on the \gls{CSM} mass. There is also a dependency on the progenitor stellar mass: the CSM mass limit is typically lower for higher progenitor masses at any given $\sigma_{\rm sim}$. Figure \ref{fig:chips_sim} also highlights the \gls{CSM} mass estimates found for SN~2024ggi based on spectroscopy (0.02 - 0.04 $M_{\odot}$; \citealp{jacobson_galan_2024}) and light curve modeling (0.4 - 1.2 $M_{\odot}$; \citealp{chen_2024, chen_2024b, ertini_2025}) with blue horizontal bands.

Our \gls{CSM} mass limits are below the estimated \gls{CSM} mass range based on light curve modeling for most of the $\sigma_{\rm sim}$ range, except for $\sigma_{\rm sim} = 2$ days and $\sigma_{\rm sim} = 2-5$ days for the model with the lowest progenitor mass ($M_{*} = 11 M_{\odot}$). In other words, if the progenitor of SN~2024ggi experienced an outburst that resulted in a \gls{CSM} mass of 0.4 - 1.2 $M_{\odot}$ that lasted $>5$ days, we would have detected it. We can perform a similar analysis for the estimated \gls{CSM} masses based on spectroscopy. The different models cross that region at different values for $\sigma_{\rm sim}$, depending on their progenitor mass. However, this is around $\sigma_{\rm sim} \sim 10 - 30$ days for most models. Thus, if the progenitor of SN~2024ggi experienced an outburst that resulted in a \gls{CSM} mass of 0.02 - 0.04 $M_{\odot}$ that lasted $>10-30$ days, we would have detected it. 

Figure \ref{fig:chips_sim} clearly shows that the thresholds on the \gls{CSM} mass depend on $\sigma_{\rm sim}$. It implies that, in principle, it would still be possible for SN~2024ggi to have experienced a short precursor outburst that could explain the reported \gls{CSM} mass. However, we can look at the precursor outburst timescales of previously observed Type II SNe to see whether such a short timescale is realistic. This is shown with the gray vertical shaded region in Figure \ref{fig:chips_sim}, which represents the 25$^{\rm th}$ - 75$^{\rm th}$ percentile range of precursor outburst timescales found for Type IIn SNe in \gls{ZTF} by \citet{strotjohann_2021}. The vertical dashed line indicates the outburst timescale found for SN~2020tlf \citep{jacobson_galan_2025}. The \gls{CSM} mass limits at those timescales are below any of the \gls{CSM} mass values estimated for SN~2024ggi. Thus, if SN~2024ggi had an outburst with a timescale similar to what was observed for Type IIn SNe in \gls{ZTF} or for SN~2020tlf that produced the reported \gls{CSM} mass, then we would have detected it. Since we did not observe any pre-SN variability for SN~2024ggi, it seems unlikely that the surrounding \gls{CSM} was deposited by eruptive outbursts by the progenitor of SN~2024ggi.

One important caveat to highlight is that each simulated light curve also has its own associated outburst timescale, which we have not taken into account here. For the simulated light curves in this work, this typically ranged between 5 and 50 days. It is reasonable to assume that a similar amount of \gls{CSM} mass can be deposited by an outburst that is fainter and longer, compared to a brighter and shorter outburst. However, this is beyond the scope of this work, and we leave this analysis for future work.

\section{Constraints on Large-Scale CSM Profile from Radio Observations}
\label{sec:radio}

\subsection{Radio Observations}
\label{sec:radio_observations}

\label{ssec:atca}
The \gls{ATCA} is a compact radio interferometer consisting of six 22-meter dishes located near Narrabri, Australia. 
We observed the field of SN~2024ggi with the \gls{ATCA} under two separate projects: C3618 (PI: Leung) and C1473 (PI: Ryder). 
The observations were taken over 10 epochs, using a selection of the 2.1, 5.5/9, and 16.7/21.2\,GHz receivers, with the first taking place on 2024-04-25 11:49:46 UT (14 days post-discovery) and the last taking place on 2025-02-22 16:23:28 UT (318 days post-discovery). 

\begin{figure*}
    \includegraphics[width=0.8\linewidth]{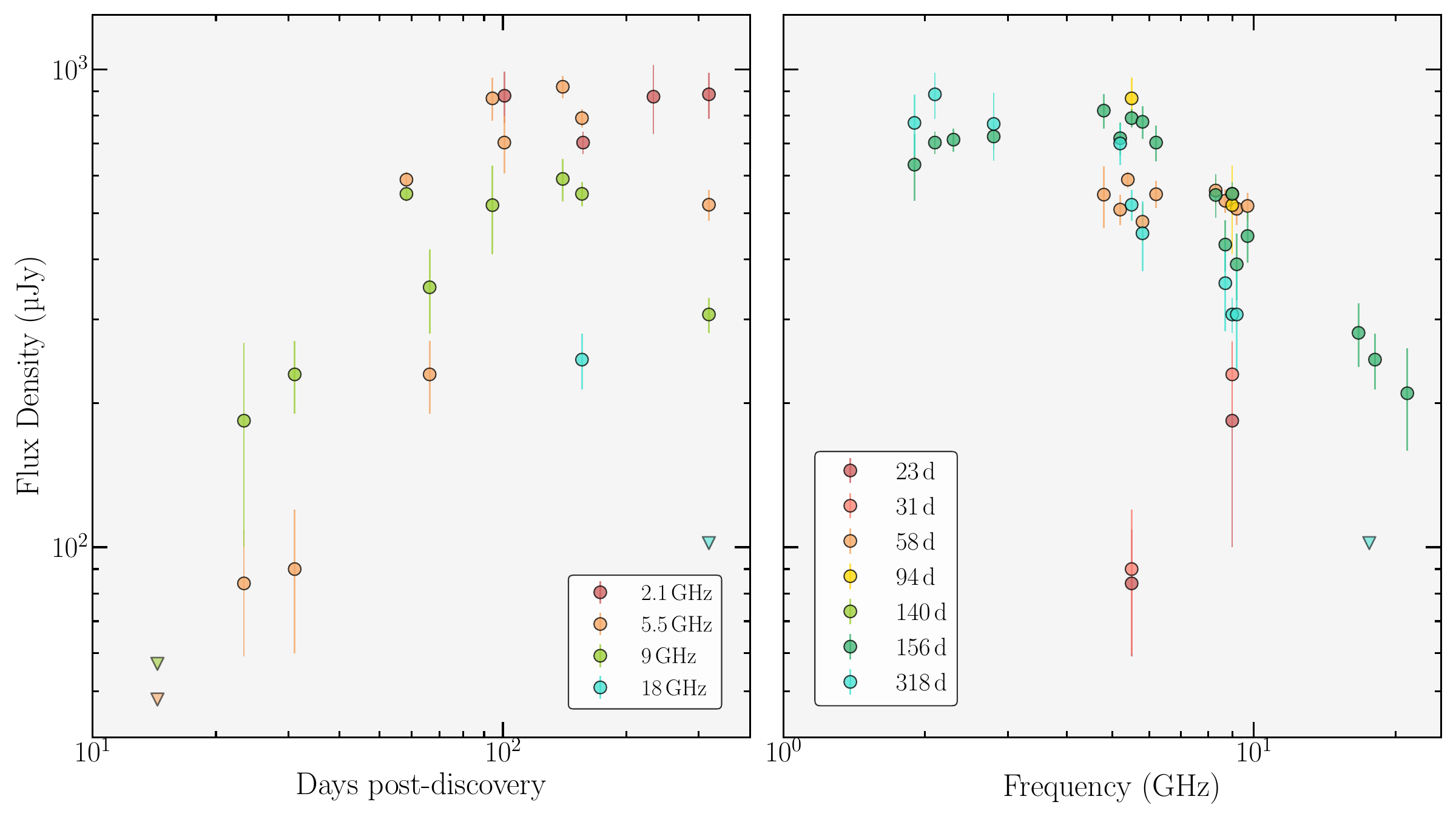}
    \centering
    \caption{The left panel shows the radio light curve of SN~2024ggi at 4 different frequencies indicated in the legend. The right panel shows the radio spectrum of SN~2024ggi at 7 different epochs indicated in the legend.
    The radio light curve and spectral evolution can approximately be described by the passage of the synchrotron self-absorption frequency through the observed frequency bands, starting from optically thick and ending up optically thin. 
    }
    \label{fig:radio_lcsed}
\end{figure*}

We reduced the visibility data following standard procedures with {\sc Miriad} \citep{Sault1995}. 
We used B1934$-$638 to set the flux-density scale and calibrate the bandpass response, and both B1048$-$313 and B1101$-$325 to calibrate the complex gains (depending on observing frequency and array configuration). 
At 16.7/21.2\,GHz, a separate bright, compact source selected from the standard set of delay calibration sources\footnote{\url{https://www.narrabri.atnf.csiro.au/observing/users_guide/html/atug.htm\#calibration-delays}} was used to calibrate the bandpass, as B1934$-$638 is too radio-faint for this purpose at these frequencies. 
If B1934$-$638 was not visible or suffered from problems such as low elevation during the course of the observation, B0823$-$500 (or in some cases the gain calibrator) was used to calibrate the flux-density scale instead. 
In order to do this, we bootstrapped the flux-density scale of B0823$-$500 (or the gain calibrator) to that of B1934$-$638 via the regular observatory-led calibrator monitoring program C007 (PI: Stevens) following standard flux-density calibration procedures. 
As shown in Table~\ref{tab:radiodata}, most observations were taken with a 6\,km array configuration.\footnote{\url{https://www.narrabri.atnf.csiro.au/operations/array_configurations/configurations.html}} 
However, for those taken in the H214 configuration, where the $(u,v)$-spacing between short and long baselines created imaging problems due to a highly irregular dirty beam shape, we imaged the data using only long baselines (with antenna CA06) as the shorter baselines were affected by sidelobes of a nearby bright source. 
We performed the image deconvolution using the multi-frequency synthesis CLEAN algorithm \citep{Hogbom1974, Clark1980, Sault1994} and fit a point source in the image plane for our flux-density measurements. 
We report our flux densities (and $3\sigma$ upper limits for non-detections) along with the array configurations of each observation in  Table~\ref{tab:radiodata}. 
The radio light curves at different frequencies and radio spectra at different epochs are shown in Figure~\ref{fig:radio_lcsed}. 
We note that there was an apparent discrepancy between the flux-density measurements at 5.5 and 9\,GHz for the epochs on 2024-06-08 UT and 2024-06-16 UT. We have checked the flux density of the secondary calibrator and that of other sources in the target field in both epochs and find no discrepancy in the flux-density scale and/or any gain calibration problems. We conclude this flux-density variation, while extreme, is likely physical, possibly due to density variations in the circumburst environment.

\begin{longtable*}{ccccccc}
\caption{ATCA radio flux-density measurements for SN~2024ggi. 
Columns 1 through 7 show the date/time of the observation, the number of days since discovery (03:22:36 UT on 2024 April 11), the project code, the array configuration, the central frequency of the radio image produced from the observation, the bandwidth of the data used to produce the radio image, and the flux-density measurements for SN~2024ggi extracted from the radio image. 
We report the $3\sigma$ limit for all non-detections. 
The reported uncertainties are purely statistical: these were the uncertainties used for our analysis and modeling. 
The systematic errors arising from flux-density scale calibration (not factored into our quoted numbers) are $\lesssim 5$ per cent for the \gls{ATCA} -- the uncertainties are therefore expected to be dominated by the statistical component. 
\label{tab:radiodata}}\\
\hline\hline
Date (UTC) & $\delta t$ (d) & Project & Config & $\nu$ (GHz) & BW (MHz) & $F_\nu$ (\textmu Jy) \\
$(1)$ & $(2)$ & $(3)$ & $(4)$ & $(5)$ & $(6)$ & $(7)$  \\
\hline
2024-04-25 11:49:46 & 14.4 & C3618 & 6A & 5.5 & 2048 & $< 50$ \\
2024-04-25 11:49:46 & 14.4 & C3618 & 6A  & 9.0 & 2048 & $< 60$ \\

2024-05-04 11:49:27 & 23.4 & C1473 & 6A & 5.5 & 2048 & $\phantom{0}80 \pm 30$ \\
2024-05-04 11:49:27 & 23.4 & C1473 & 6A & 9.0 & 2048 & $180 \pm 80$ \\

2024-05-12 06:27:39 & 31.1 & C1473 & 6A & 5.5 & 2048 & $\phantom{0}90 \pm 30$ \\
2024-05-12 06:27:39 & 31.1 & C1473 & 6A & 9.0 & 2048 & $230 \pm 40$ \\

2024-06-08 07:44:09 & 58.2 & C3618 & 6D & 4.8 & 512 & $550 \pm 80$ \\
2024-06-08 07:44:09 & 58.2 & C3618 & 6D & 5.2 & 512 & $510 \pm 40$ \\
2024-06-08 07:44:09 & 58.2 & C3618 & 6D & 5.4 & 2048 & $590 \pm 20$ \\
2024-06-08 07:44:09 & 58.2 & C3618 & 6D & 5.8 & 512 & $480 \pm 40$ \\
2024-06-08 07:44:09 & 58.2 & C3618 & 6D & 6.2 & 512 & $550 \pm 40$ \\
2024-06-08 07:44:12 & 58.2 & C3618 & 6D & 8.3 & 512 & $560 \pm 40$ \\
2024-06-08 07:44:12 & 58.2 & C3618 & 6D & 8.7 & 512 & $530 \pm 30$ \\
2024-06-08 07:44:12 & 58.2 & C3618 & 6D & 9.0 & 2048 & $550 \pm 20$ \\
2024-06-08 07:44:12 & 58.2 & C3618 & 6D & 9.2 & 512 & $510 \pm 40$ \\
2024-06-08 07:44:12 & 58.2 & C3618 & 6D & 9.7 & 512 & $520 \pm 30$ \\

2024-06-16 09:57:17 & 66.3 & C1473 & 6D & 5.5 & 2048 & $230 \pm 40$ \\
2024-06-16 09:57:17 & 66.3 & C1473 & 6D & 9.0 & 2048 & $350 \pm 70$ \\

2024-07-14 10:58:36 & 94.3 & C1473 & 6D & 5.5 & 2048 & $870 \pm 90$ \\
2024-07-14 10:58:36 & 94.3 & C1473 & 6D & 9.0 & 2048 & $\phantom{0}520 \pm 110$ \\

2024-07-21 00:44:23 & 100.9 & C3618 & H214 & 2.1 & 2048 & $\phantom{0}880 \pm 110$ \\
2024-07-20 23:42:26 & 100.8 & C3618 & H214 & 5.4 & 2048 & $\phantom{0}700 \pm 100$ \\

2024-08-29 00:12:18 & 139.9 & C1473 & 6A & 5.5 & 2048 & $920 \pm 50$ \\
2024-08-29 00:12:18 & 139.9 & C1473 & 6A & 9.0 & 2048 & $590 \pm 60$ \\

2024-09-14 22:07:00 & 156.8 & C3618 & 6A & 1.9 & 512 & $\phantom{0}630 \pm 100$ \\
2024-09-14 22:07:00 & 156.8 & C3618 & 6A & 2.1 & 2048 & $700 \pm 40$ \\
2024-09-14 22:07:00 & 156.8 & C3618 & 6A & 2.3 & 512 & $710 \pm 40$ \\
2024-09-14 22:07:00 & 156.8 & C3618 & 6A & 2.8 & 512 & $720 \pm 50$ \\
2024-09-13 22:01:34 & 155.8 & C3618 & 6A & 4.8 & 512 & $820 \pm 70$ \\
2024-09-13 22:01:34 & 155.8 & C3618 & 6A & 5.2 & 512 & $720 \pm 60$ \\
2024-09-13 22:01:34 & 155.8 & C3618 & 6A & 5.5 & 2048 & $790 \pm 30$ \\
2024-09-13 22:01:34 & 155.8 & C3618 & 6A & 5.8 & 512 & $780 \pm 60$ \\
2024-09-13 22:01:34 & 155.8 & C3618 & 6A & 6.2 & 512 & $700 \pm 60$ \\
2024-09-13 22:01:34 & 155.8 & C3618 & 6A & 8.3 & 512 & $550 \pm 60$ \\
2024-09-13 22:01:34 & 155.8 & C3618 & 6A & 8.7 & 512 & $430 \pm 50$ \\
2024-09-13 22:01:34 & 155.8 & C3618 & 6A & 9.0 & 2048 & $550 \pm 30$ \\
2024-09-13 22:01:34 & 155.8 & C3618 & 6A & 9.2 & 512 & $390 \pm 60$ \\
2024-09-13 22:01:34 & 155.8 & C3618 & 6A & 9.7 & 512 & $450 \pm 50$ \\
2024-09-13 22:37:25 & 155.8 & C3618 & 6A & 16.7\phantom{0} & 2048 & $280 \pm 40$ \\
2024-09-13 22:37:25 & 155.8 & C3618 & 6A & 18.1\phantom{0} & 4096 & $250 \pm 30$ \\
2024-09-13 22:37:25 & 155.8 & C3618 & 6A & 21.2\phantom{0} & 2048 & $210 \pm 50$ \\

2024-11-29 19:36:13 & 232.7 & C3618 & 6A & 2.0 & 2048 & $\phantom{0}880 \pm 150$ \\

2025-02-22 17:19:30 & 317.6 & C3618 & 6D & 1.9 & 512 & $\phantom{0}770 \pm 110$ \\
2025-02-22 17:19:30 & 317.6 & C3618 & 6D & 2.1 & 2048 & $\phantom{0}890 \pm 100$ \\
2025-02-22 17:19:30 & 317.6 & C3618 & 6D & 2.8 & 512 & $\phantom{0}770 \pm 130$ \\
2025-02-22 17:29:00 & 317.6 & C3618 & 6D & 5.2 & 512 & $700 \pm 70$ \\
2025-02-22 17:29:00 & 317.6 & C3618 & 6D & 5.5 & 2048 & $520 \pm 40$ \\
2025-02-22 17:29:00 & 317.6 & C3618 & 6D & 5.8 & 512 & $450 \pm 80$ \\
2025-02-22 17:29:00 & 317.6 & C3618 & 6D & 8.7 & 512 & $360 \pm 70$ \\
2025-02-22 17:29:00 & 317.6 & C3618 & 6D & 9.0 & 2048 & $310 \pm 30$ \\
2025-02-22 17:29:00 & 317.6 & C3618 & 6D & 9.2 & 512 & $310 \pm 70$ \\
2025-02-22 16:23:28 & 317.5 & C3618 & 6D & 17.6\phantom{0} & 4096 & $< 100$ \\
\hline
\end{longtable*}

\subsection{Radio Synchrotron Modeling}
\label{sec:radio_modelling}

\citet{Chevalier1998} presents a model for interpreting the radio supernova light curves by relating the radio-emitting source's physical properties to its observed synchrotron self-absorption properties. 
In particular, expressions for the outer shock radius $R_p$ and the magnetic field at the shock front $B_p$ \citep[equations 11 and 12 in][]{Chevalier1998} were obtained by intersecting the equations that describe the flux density at the optically thick and thin limits \citep[equations 2 and 3 in][]{Chevalier1998} on either side of the self-absorption peak. 
In the specific scenario where the electron spectral index $p=3$, where $p$ is the power-law index of the relativistic electron energy distribution $N(E) \propto E^{-p}$, the expressions for $R_p$ and $B_p$ simplify to equations 13 and 14 in \citet{Chevalier1998}, reproduced below for clarity.

The outer shock radius is given by:
\begin{equation}
\begin{split}
    & R_p = 8.8 \times 10^{15} \left( \frac{\epsilon_e}{\epsilon_B}\right)^{-1/19}  \left( \frac{f}{0.5} \right)^{-1/19} \left( \frac{F_{\nu_p}}{\textrm{Jy}} \right)^{9/19} \\
    & \indent\indent\indent \cdot \left( \frac{D}{\textrm{Mpc}} \right)^{18/19} \left( \frac{\nu_p}{\textrm{5 GHz}} \right)^{-1}\,\textrm{cm}, 
\end{split}
    \label{eq:radius}
\end{equation}
and the magnetic field at the shock front is given by:
\begin{equation}
\begin{split}
    & B_p = 0.58 \left( \frac{\epsilon_e}{\epsilon_B}\right)^{-4/19}  \left( \frac{f}{0.5} \right)^{-4/19} \left( \frac{F_{\nu_p}}{\textrm{Jy}} \right)^{-2/19} 
    \\
    & \indent\indent\indent \cdot \left(\frac{D}{\textrm{Mpc}} \right)^{-4/19} \left( \frac{\nu_p}{\textrm{5 GHz}} \right)\,\textrm{G}, 
\end{split}
    \label{eq:magnetic_field}
\end{equation}
where $\epsilon_e$ and $\epsilon_B$ are the fractional shock energies in the relativistic electrons and in the magnetic fields, respectively, $f$ is the emission filling factor (while the emitting region is approximated as a circular planar region with thickness $s$ and radius $R$ such that the emitting volume is $\pi R^2s$, the emission filling factor $f$ provides an alternative approximation as a spherical emitting volume such that $V = \pi R^2s = 4\pi R^3f/3$), $\nu_p$ is the frequency where the optical depth is unity, and $F_{\nu_p}$ is the flux density at the extrapolated meeting point between the optically thin and optically thick power-law asymptotes.

By considering the Rankine–Hugoniot jump conditions in the strong shock limit for a monatomic gas, and also assuming that a fixed fraction $\epsilon_B$ of the energy density in the shocked ejecta is converted into the magnetic energy at the shock front (given by Equation~\ref{eq:magnetic_field}), we obtain an expression for the density of the ambient CSM: 
\begin{equation}
\begin{split}
    & \rho_\textrm{csm} = 1.15 \times 10^{-24} \left( \frac{1}{\epsilon_B}\right)\left( \frac{\epsilon_e}{\epsilon_B}\right)^{-6/19}  \left( \frac{f}{0.5} \right)^{-6/19} \\
    & \indent\indent\indent \cdot \left( \frac{F_{\nu_p}}{\textrm{Jy}} \right)^{-22/19} \left( \frac{D}{\textrm{Mpc}} \right)^{-44/19} \\
    & \indent\indent\indent \cdot \left( \frac{\nu_p}{\textrm{5 GHz}} \right)^{4} \left( \frac{t_p}{\textrm{1 day}} \right)^{2}\,\textrm{g\,cm}^{-3}, 
\end{split}
    \label{eq:density}
\end{equation}
where $t_p$ is the time post-explosion in the observer frame. 

By using Equations~\ref{eq:radius} and \ref{eq:density}, we obtain a measurement (or limit) on $R_p$ and $\rho_\textrm{csm}$ at the time post-explosion corresponding to each epoch of radio data (see Figure~\ref{fig:radio_lcsed}). 
For these equations, we use $D = 7.2$\,Mpc, and assume $f = 0.1$, $\epsilon_e = 0.1$, and $\epsilon_B = 0.01$; we note that the dependence of $R_p$ and $\rho_\textrm{csm}$ on most of these assumed values are quite low, with the exception of $\rho_\textrm{csm} \propto \epsilon_B^{-13/19}$. 

For epochs where the spectral turnover appears within the observed frequency range and is well sampled (58, 156, and 318 days post-explosion), we obtain a measurement of $\nu_p$ and $F_{\nu_p}$ by fitting the radio spectrum with a smoothly broken power law of the form:
\begin{equation}
    F_\nu(t=t_p) = F_{\nu_p}(t_p)\,\left[\left(\frac{\nu}{\nu_p}\right)^{-\alpha_1s} + \left(\frac{\nu}{\nu_p}\right)^{-\alpha_2s}\right]^{-1/s}, 
    \label{eq:sa_break}
\end{equation}
where $\alpha_1 = 5/2$ and $\alpha_2 = -(p-1)/2$ (corresponding to the optically thick and thin spectral slopes, respectively), and the smoothness parameter is set to $s=1$. 
The observed peak of the radio spectrum can be recovered with $F_{\nu_p} \cdot 2^{-1/s}$.  
To perform the fit, we fixed $\alpha_2$ (or equivalently, $p$) to be the same across all three epochs, i.e., performing the fit simultaneously across the three epochs where the spectral turnover appears within the observed frequency. 
We fit the data to Equation~\ref{eq:sa_break} using the nested sampler {\sc Dynesty} \citep{Speagle2020} as implemented in the Bayesian inference software {\sc Bilby} \citep{Ashton2019}. 
We performed the nested sampling using uniform priors (sampled uniformly over linear space), with 2\,000 live points, and a stopping criterion on the change in the estimated Bayesian evidence $\hat{\mathcal{Z}}$ (i.e., the normalization integral of the posterior probability function) from one iteration to the next of $\Delta \mathrm{ln}(\hat{\mathcal{Z}}) = 0.05$. 
Table~\ref{tab:radiosed_fits} and Figure~\ref{fig:cornerplot} in Appendix \ref{app:radio_appendix} (corner plot) show the results of the fit to $\nu_p$ and $F_{\nu_p}$ at each epoch. 
This fit also verified that Equations~\ref{eq:radius} and \ref{eq:density}, which assume $p = 3$, are appropriate for us to use for analyzing SN~2024ggi as our nested sampling procedure constrained $p = 2.90^{+0.13}_{-0.13}$. 

In the case where the spectral turnover did not clearly appear within the observed frequency range (for the epochs at $23$, $31$, $94$, $140$ days post-explosion), we could obtain limits on $\nu_p$ and $F_{\nu_p}$. 
Other epochs that are not listed above were discarded for this analysis as they were suspected to be undersampled near the spectral turnover (i.e., when it was uncertain whether the spectral turnover lay within or outside the observed frequency range).

\begin{figure*}
    \includegraphics[width=0.8\linewidth]{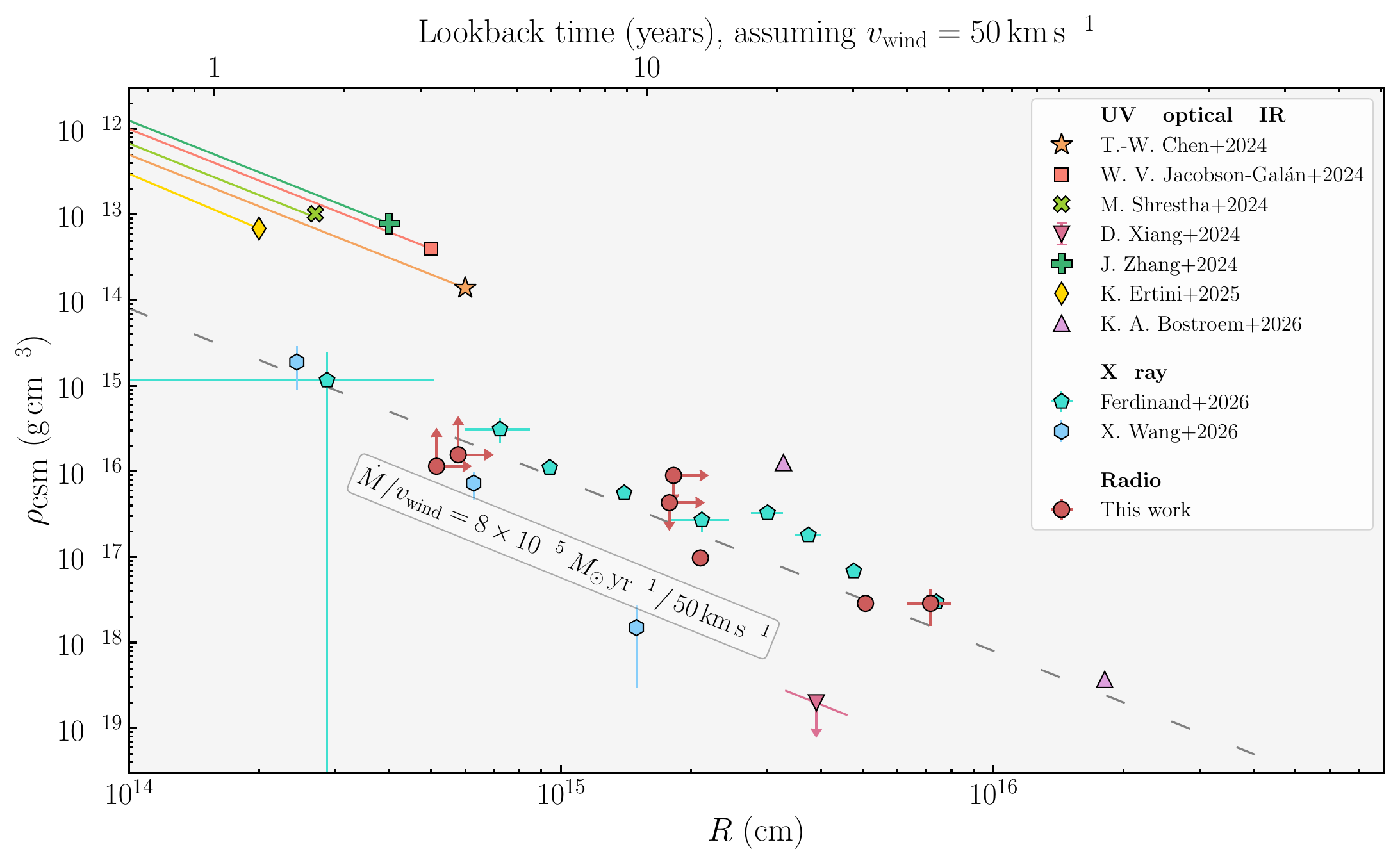}
    \centering
    \caption{Ambient CSM density $\rho_\mathrm{csm}$ as a function of distance $R$ from the explosion site of SN~2024ggi. 
    The errors on the density and radius measurements from the radio data points are purely statistical, from the nested sampling fits of the synchrotron self-absorption turnover. 
    For the radio epochs where the synchrotron self-absorption turnover is not sampled within the observed frequency range, limits are provided and are indicated on the plot using L-shaped markers with their arrows showing the direction of the limits. 
    The equivalent lookback time $(=R/v_\mathrm{wind})$ is given on the top axis, assuming a wind velocity of $v_\mathrm{wind} = 50\,\mathrm{km}\,\mathrm{s}^{-1}$. 
    The gray dashed line indicates an approximate mass-loss rate, normalized to this wind velocity, which could reproduce the ambient CSM density profile inferred from the radio data. We also include data from other work that used UV-optical-IR data \citep{chen_2024b, jacobson_galan_2024, shrestha_2024, xiang_2024, zhang_2024, ertini_2025, bostroem_2026} and X-ray data  \citep{ferdinand_2026, wang_2026}. If the \gls{CSM} density was not explicitly stated in these studies, then we calculated the inferred density using their quoted mass-loss rates and assumed wind velocities using $\dot{M} = 4\pi R^2v_\mathrm{wind}\rho_\mathrm{csm}$.}
    \label{fig:density_vs_radius}
\end{figure*}

Table~\ref{tab:fig:r+d_table} shows a summary of the measurements and limits on $\nu_p$ and $F_{\nu_p}$ as well as the inferred outer shock radius $R_p$ and ambient CSM density $\rho_\textrm{csm}$ at each epoch. 
The ambient CSM density as a function of the outer shock radius as inferred from the radio data is also shown in Figure~\ref{fig:density_vs_radius}. 

\begin{deluxetable}{lccccccc}
    \tablecaption{Nested sampling results from the broken power-law model fit to selected radio spectra. 
    Column 1 describes the information for each parameter given in each row, while columns 2 through 8 are the model parameters that we performed the nested-sampling procedure to constrain. 
    The units for the peak frequencies and the flux densities at the corresponding peak frequencies are in GHz and $\mu$Jy, respectively. 
    The rows of the table show the prior range we used (all of which assumed a uniform prior) and a result summary, which shows the median and the 16$^{\text{th}}$/84$^{\text{th}}$ percentile of the marginalized posterior distribution, for each parameter. 
    The full corner plot can be found in Appendix \ref{app:radio_appendix} in Figure~\ref{fig:cornerplot}.}
    \label{tab:radiosed_fits}
    \tablehead{\colhead{Information} & \colhead{$p$} & \colhead{$\nu_{p}$ (58\,d)} & \colhead{$\nu_{p}$ (156\,d)} & \colhead{$\nu_{p}$ (318\,d)} & \colhead{$F_{\nu_{p}}$ (58\,d)} & \colhead{$F_{\nu_{p}}$ (156\,d)} & \colhead{$F_{\nu_{p}}$ (318\,d)}
    }
    \colnumbers
    \startdata
    Prior Range &
    $[2.5, 3.5]$ & 
    $[3, 10]$ & 
    $[1, 7]$ & 
    $[1, 7]$ & 
    $[500, 1600]$ & 
    $[800, 2400]$ & 
    $[800, 2400]$ \\
    Result &
    $2.90^{+0.13}_{-0.13}$ & 
    $5.05^{+0.22}_{-0.23}$ & 
    $2.63^{+0.11}_{-0.11}$ & 
    $1.77^{+0.17}_{-0.19}$ & 
    $1052.67^{+26.51}_{-27.56}$ & 
    $1625.93^{+62.47}_{-63.50}$ & 
    $1561.18^{+108.28}_{-101.48}$ \\
    \hline
    \enddata
\end{deluxetable}

\begin{deluxetable}{c c c c c}
    \tablecaption{The table shows the physical properties derived from the \citet{Chevalier1998} model describing synchrotron self-absorption in radio supernovae. 
    Columns 1 through 5 show the time post-explosion that the derived physical properties correspond to, the peak frequency, which is assumed to be the self-absorption frequency, the flux density at this peak frequency, the derived outer shock radius, and the derived density of the ambient CSM. 
    Measurements come from the nested sampling results presented in Table~\ref{tab:radiosed_fits} (where the errors represent the $1\sigma$ credible interval), while the limits come from radio spectra taken at times where the peak frequency and flux density at this peak frequency are unconstrained. 
    }
    \label{tab:fig:r+d_table}
    \tablehead{\colhead{$\delta t$ (d)} & \colhead{$\nu_p$ (GHz)} & \colhead{$F_{\nu_p}$ (\textmu Jy)} & \colhead{$R$ (cm)} & \colhead{$\rho_\textrm{csm}$ (g\,cm$^{-3}$)}}
    \decimalcolnumbers
    \startdata
    23 & $>9$ & $>180$ & $>5.15 \times 10^{14}$  & $>1.15 \times 10^{-16}$ \\
    31 & $>9$ & $>230$ & $>5.78 \times 10^{14}$  & $>1.57 \times 10^{-16}$ \\
    58 & $5.2 \pm 0.2$  & $1053 \pm 28$ & $(2.10 \pm 0.09) \times 10^{15}$ & $(9.75 \pm 1.56) \times 10^{-18}$ \\
    94 & $<5.5$ & $>870$ & $>1.78 \times 10^{15}$  & $<4.32 \times 10^{-17}$ \\
    140 & $<5.5$ & $>920$ & $>1.82 \times 10^{15}$  & $<8.99 \times 10^{-17}$ \\
    156 & $2.7 \pm 0.1$  & $1626 \pm 64$ & $(5.06 \pm 0.22) \times 10^{15}$ & $(2.88 \pm 4.62) \times 10^{-18}$ \\
    318 & $1.8 \pm 0.2$ & $1561 \pm 109$ & $(7.16 \pm 0.83) \times 10^{15}$ & $(2.88 \pm 1.30) \times 10^{-18}$ \\
    \enddata
\end{deluxetable}

\subsection{The Overall CSM Profile surrounding SN~2024ggi}
\label{sec:radio_comparison}

The \gls{CSM} density profile based on our radio observations presented in Figure \ref{fig:density_vs_radius} is consistent with being a wind-like ($\rho \propto r^{-2}$) density profile. Assuming a wind velocity of $v_\mathrm{wind} = 50\,\mathrm{km}\,\mathrm{s}^{-1}$, the approximate mass-loss rate that could reproduce this ambient \gls{CSM} density profile, assumed to be shaped by steady mass-loss from the progenitor's stellar winds, is $\dot{M} = 4\pi R^2v_\mathrm{wind}\rho_\mathrm{csm}=8\times10^{-5} M_\odot\,\mathrm{yr}^{-1}$. Our radio observations cover ${\sim} 3 - 50$ years pre-explosion (with the actual detections covering ${\sim}10 - 50$ years pre-explosion), assuming the same wind velocity $v_\mathrm{wind} = 50\,\mathrm{km}\,\mathrm{s}^{-1}$. Figure \ref{fig:density_vs_radius} also includes the \gls{CSM} density profile from multiple other studies found in the literature: \citet{chen_2024b, jacobson_galan_2024, shrestha_2024, xiang_2024, zhang_2024, ertini_2025, bostroem_2026, ferdinand_2026, wang_2026}. If the \gls{CSM} density was not explicitly stated in these studies, then we calculated the inferred density using their quoted mass-loss rates and assumed wind velocities. 

The density profile implied by our radio observations agrees well with the profile inferred from the X-ray observations from \citet{ferdinand_2026}, which are also shown in Figure \ref{fig:density_vs_radius}. They quote a slightly lower progenitor mass-loss rate of $(6.2 \pm 0.2) \times 10^{-5}$ $M_\odot$ yr$^{-1}$, but also assume a lower wind velocity ($v_\mathrm{wind} = 20\,\mathrm{km}\,\mathrm{s}^{-1}$). We disagree more significantly with \citet{xiang_2024}, who find a progenitor mass-loss rate of $\dot{M} <3 \times 10^{-6} M_{\odot} \textrm{yr}^{-1}$ (assuming $v_\mathrm{wind} = 50\,\mathrm{km}\,\mathrm{s}^{-1}$) by fitting stellar spectral models to optical \gls{HST} and \gls{MIR} Spitzer data. Their data cover ${\sim}21 - 29$ years pre-explosion, which overlaps with our range. This disagreement might be because of systemic issues or assumptions in e.g., the gas-to-dust ratio in \citet{xiang_2024} or in the adopted $\epsilon_e$ and $\epsilon_B$ values for our radio analysis. Interestingly, using ultraviolet spectra from \gls{HST}, \citet{bostroem_2026} find a higher mass-loss rate of $1\times10^{-3} M_{\odot} \textrm{yr}^{-1}$ at 30 years before explosion and a mass-loss rate of $9 \times 10^{-5} M_{\odot} \textrm{yr}^{-1}$ at 150 years before explosion, possibly due to systematics or differences in the assumed shock or wind velocities. However, their inferred \gls{CSM} densities are broadly consistent with our radio observations.

Interestingly, the densities inferred by multiple other early-time spectroscopic and photometric studies consistently show increased densities at lower radii \citep{chen_2024b, jacobson_galan_2024, shrestha_2024, zhang_2024, ertini_2025}, which are also visualized in Figure \ref{fig:density_vs_radius}. These studies typically compare the light curve or spectra to a grid of models with a confined \gls{CSM} component up to a certain radius.

Thus, there is some disagreement in the overall \gls{CSM} density profile between the different studies shown in Figure \ref{fig:density_vs_radius}. This is likely because they all employ different methods, which all come with their own uncertainties and systematics. A full systematic comparison of the different methods is outside the scope of this paper. However, it appears that the different studies broadly agree that the overall \gls{CSM} density profile of SN~2024ggi has a two-component structure, with a dense inner structure at radii $\lesssim6 \times 10^{14}$ cm and an extended broadly wind-like profile at larger radii. \citet{ertini_2025} noted that their analysis also preferred a two-component structure with a compact core and an extended tail for the \gls{CSM} distribution of SN~2024ggi.

A similar story emerges when looking at the reported mass-loss rates from different studies. The aforementioned early-time studies that probe ${\sim}1-4$ years pre-explosion all quote mass-loss rates between $10^{-3}-10^{-2}$ $M_{\odot} \textrm{yr}^{-1}$ \citep{chen_2024b, jacobson_galan_2024, shrestha_2024, zhang_2024, ertini_2025}. In contrast, as mentioned above, studies that probe further into the progenitor's past (${\sim}3 - 50$ years pre-explosion; this work, but also \citealt{xiang_2024}, \citealt{ferdinand_2026}, and \citealt{bostroem_2026}) imply lower mass-loss rates between $<3 \times 10^{-6}$ and $10^{-4} M_{\odot} \textrm{yr}^{-1}$

This all points towards an evolution in the mass-loss rate; a stable mass-loss rate at earlier times (${\sim}10 - 50$ years pre-explosion), and a substantial increase in the mass-loss rate in the last ${\sim}1-4$ years pre-explosion. Alternatively, the increase in density at lower radii could be explained by an extended chromosphere (see \citealt{fuller_2024}). The full implications of this two-component structure for different physical mechanisms that could explain the \gls{CSM} around SN~2024ggi are discussed in more detail in Section \ref{sec:discussion}.


\section{Discussion}
\label{sec:discussion}

We will go over different proposed mechanisms that can explain the presence of the dense \gls{CSM} surrounding SN~2024ggi: eruptive outbursts in Section \ref{sec:eruptive_outbursts}, binary interactions in Section \ref{sec:binary_interactions}, pulsation-driven superwinds in Section \ref{sec:superwinds}, and extended chromospheres in Section \ref{sec:chromospheres}. We will examine the viability of each mechanism for the case of SN~2024ggi based on the optical pre-SN light-curve analysis presented in Section \ref{sec:optical} and the radio observations presented in Section \ref{sec:radio}.

\subsection{Eruptive Outbursts and Wave-Driven Mass-Loss}
\label{sec:eruptive_outbursts}

Eruptive outbursts prior to explosion (e.g., \citealt{dessart_2010, fuller_2017, tsang_2022}) are often invoked to explain the dense \gls{CSM} that surrounds many SNe. The associated precursor emission has been found for many Type IIn SNe, such as SN~2009ip \citep{pastorello_2013}, 2010mc \citep{ofek_2013}, 2011ht \citep{fraser_2013}, and 2023vbg \citep{goto_2025}, which makes it a promising candidate explanation for the \gls{CSM} around SN~2024ggi as well. 

Many simulations (e.g., see Section \ref{sec:chips_csm_mass} and \citealt{kuriyama_2020}) are agnostic to the actual physical mechanism that results in eruptive outbursts. Instead, they simply inject a certain amount of energy at the bottom of the stellar envelope. One possibility for the physical mechanism is wave heating (or wave-driven mass-loss). During the core neon- and oxygen-burning phases in the last years before core collapse of \gls{RSG} SN progenitors, vigorous convection in the core can generate waves that can transport energy from the core to the envelope. The large decrease in density at the bottom of the hydrogen envelope damps these waves, which heats up the surrounding envelope. This subsequently results in pressure waves that move outwards to the surface, which can result in eruptive outbursts and mass-loss \citep{quataert_2012, shiode_2014, fuller_2017}.

\citet{fuller_2017} note that wave-driven mass-loss is unlikely to produce Type IIn SNe, but it may result in SNe with ``IIn-like'' features, such as SN~2024ggi. This mechanism is also predicted to cause eruptive outbursts that can form a \gls{CSM} with masses between $10^{-3} - 1 M_{\odot}$ \citep{shiode_2014, fuller_2017}, which covers the entire estimated \gls{CSM} mass range of SN~2024ggi (see Table \ref{tab:ggi_properties}). Furthermore, we noted a higher density in the \gls{CSM} density profile at lower radii ($1 - 6 \times 10^{14}$ cm) in Figure \ref{fig:density_vs_radius}. Assuming a wind velocity of \todo{50} km s$^{-1}$, this corresponds roughly to a lookback time of \todo{1-4} years pre-explosion. This is broadly consistent with the timescales of core neon- and oxygen-burning phases in progenitors with similar masses to the progenitor of SN~2024ggi \citep{woosley_2002}. As mentioned above, these core burning stages are exactly when core convection vigorously drives gravity waves and thus when we expect eruptive outbursts to happen \citep{shiode_2014, fuller_2017, woosley_2002}.

However, as presented in Section \ref{sec:atclean_2024ggi}, we did not find any observational evidence for eruptive outbursts in the pre-explosion light curve of SN~2024ggi. Furthermore, as shown in Section \ref{sec:comparison_limits} and specifically Figure \ref{fig:best_absmag_thresholds}, the inferred upper limits confidently exclude eruptions previously observed for many known Type IIn SNe with precursor emission: 2010mc \citep{ofek_2013}, 2011ht \citep{fraser_2013}, 2023vbg \citep{goto_2025}, and multiple Type IIn SNe found in \gls{ZTF} \citep{bellm_2019, strotjohann_2021}. We also exclude the precursor outburst observed for 2020tlf, a SN with ``IIn-like'' features \citep{jacobson_galan_2022, jacobson_galan_2025}. In addition, \citet{strotjohann_2021} note that the typical absolute magnitude of the precursor emission of Type IIn is brighter than -13 mag. Our detection thresholds are much fainter than that, suggesting that the precursor to SN~2024ggi behaved differently compared to these other Type IIn and SN~2020tlf. Thus, if the progenitor of SN~2024ggi experienced eruptive outbursts that were similar to any of the aforementioned previously observed outbursts, then we would have detected them. The upper limits found in this work are converted to limits on the \gls{CSM} mass in Section \ref{sec:chips_csm_mass}, where we came to a similar conclusion. If SN~2024ggi had an outburst with a similar timescale to what was observed for Type IIn SNe in \gls{ZTF} or for SN~2020tlf that produced the reported \gls{CSM} mass, then we would have detected it. 

Nevertheless, it is certainly still possible that faint eruptions occurred with magnitudes below our detection threshold. Perhaps multiple of these smaller and shorter eruptions could lead to the build-up of a significant amount of \gls{CSM} mass. However, these eruptions would have to be uncharacteristically short and/or faint, and unlike any of the previously observed precursor outbursts.

Instead, it appears more likely that SN~2024ggi is similar to SN~2023ixf. \citet{rest_2025} also found no precursor emission for SN~2023ixf and note that an eruptive mass-loss mechanism is unlikely to be the source of the confined \gls{CSM} around SN~2023ixf. 

We do note that there is a possibility that the aforementioned wave heating does not result in full eruptions, but instead only inflates the stellar envelope \citep{fuller_2017, fuller_2018}. However, the simulations of \citet{fuller_2018} found that the envelope only fully unbinds in hydrogen-poor stars, whereas the precursor of SN~2024ggi was hydrogen-rich. Thus, such an inflation would likely need to be coupled with another mechanism (such as binary interactions, see Section \ref{sec:binary_interactions}) to unbind the envelope.

In conclusion, although wave-driven mass-loss is a good model to explain the presence of \gls{CSM} for SNe with ``IIn-like'' features in general, we do not find evidence for eruptive outbursts for SN~2024ggi specifically. Thus, it is unlikely that this is the physical mechanism that resulted in the dense \gls{CSM} around SN~2024ggi.

\subsection{Binary Interactions}
\label{sec:binary_interactions}

Another possibility that we want to discuss here is mass-loss associated with binary interactions, as more than half of massive ($M \gtrsim 8 M_\odot$) stars have a binary companion \citep{sana_2012, sana_2013, rizzuto_2013, kobulnicky_2014, offner_2023, sana_2025, shenar_2026} and such companions can remain undetected in the pre-explosion photometry. 

For example, common envelope ejections can result in enhanced mass-loss (e.g., see \citealt{chevalier_2012, smith_2014}). However, such common envelope ejections are typically associated with faint transients (e.g., see \citealt{dong_2024}). As noted in Section \ref{sec:atclean_2024ggi}, we do not find any evidence for outbursts in the optical pre-SN light curve of SN~2024ggi, with upper limits going down to absolute magnitudes between \todo{$-11.28$} and \todo{$-8.98$} mag. 

The simulations of \citet{matsuoka_2024} show that binary interactions can also cause an enhanced mass-loss rate due to Roche lobe overflow. They find that \gls{CSM} masses that cover the entire range of estimated \gls{CSM} masses of SN~2024ggi ($0.02 - 1.2 M_{\odot}$; see Table \ref{tab:ggi_properties}) can be produced by binary interactions, depending on the period, the stellar masses, and the assumed outflowing gas velocity.

However, the Roche lobe overflows in these simulations occur $>1000$ years pre-explosion, and result in so-called shell-like or cliff-like structures in the \gls{CSM} density profile (see Figure 2 in \citealt{matsuoka_2024}). These structures appear at radii of ${\sim}10^{17}$ cm, though the exact radius depends on the orbital period of the binary. This radius range is not probed by our radio observations, which go up to ${\sim}7 \times 10^{15}$ cm (see Table \ref{tab:fig:r+d_table}). 
Continued radio monitoring of SN~2024ggi could probe this radius range in the future and help us probe whether this type of binary interaction occurred prior to explosion.

Notably, as shown in Figure \ref{fig:density_vs_radius} in Section \ref{sec:radio_comparison}, the \gls{CSM} density profile has two distinct components: a denser inner region and an extended profile at larger radii. Perhaps Roche lobe overflows caused the increased density structure at lower radii, although then it must have been a very late-stage interaction (${\sim}1-4$ years pre-explosion). This could have, in principle, been triggered by the inflation of the stellar envelope due to the wave heating in the core neon- and oxygen-burning phases during the last few years pre-explosion (see Section \ref{sec:eruptive_outbursts} and \citealt{fuller_2017,fuller_2018}). However, this does not necessarily explain the relatively dense but wind-like extended \gls{CSM} tail at larger radii (which could be reproduced by invoking an additional mechanism such as pulsation-driven superwinds from the progenitor; see Section \ref{sec:superwinds}). Alternatively, if non-conservative binary interactions cause more consistent and regular mass-loss, and are thus responsible for the more diffuse extended tail in the \gls{CSM} profile, then explaining the denser inner regions via the same mechanism would likely require a late-stage change in the mass transfer rate and/or efficiency. Thus, the two-component structure of the density profile makes explaining the presence of the dense \gls{CSM} through binary interactions alone more challenging.

Furthermore, \citet{jiang_2025} show that progenitor envelope stripping by a companion could result in a significantly shorter duration of the plateau phase in the light curve. For example, they find a short plateau phase that lasts ${\sim}60$ days for KSP-SN-2022c. In contrast, \citet{ertini_2025} found that the plateau phase for SN~2024ggi lasted ${\sim}90$ days. This plateau length is consistent with expectations for explosions of single stars with masses similar to those estimated from the progenitor detection of SN~2024ggi \citep{eldridge_2018}. However, given that the estimates for the mass of the dense \gls{CSM} around SN~2024ggi based on spectroscopy are $0.02 - 0.04 M_\odot$ \citep{jacobson_galan_2024}, and the lower end of the \gls{CSM} mass estimates based on light curve modeling are ${\sim}0.4 M\odot$ \citep{chen_2024b}, it is unclear whether that is enough to significantly change the length of the plateau phase, as the \gls{CSM} mass in \citet{jiang_2025} was equal to $0.73 M_\odot$.

In summary, it remains possible that the dense \gls{CSM} around SN~2024ggi was deposited there by binary interactions (through e.g., Roche lobe overflow), although we do not have direct observational evidence for it.

\subsection{Pulsation-Driven Superwinds}
\label{sec:superwinds}

A moderately high, but more stable, mass-loss rate can be achieved through pulsation-driven superwinds. \citet{yoon_2010} find that superwinds can cause a buildup of dense \gls{CSM} that will eventually result in a Type IIn SN. \citet{rest_2025} also favor a steady, enhanced mass-loss mechanism, such as superwinds, to explain the \gls{CSM} around SN~2023ixf.

As noted in Figure \ref{fig:density_vs_radius}, the multi-epoch radio observations presented in this work are consistent with a wind-like density profile ($\rho \propto r^{-2}$). Furthermore, these radio observations imply a mass-loss rate of ${\sim} 8\times10^{-5}$ $M_\odot$ yr$^{-1}$ at a lookback time of ${\sim}10-50$ years, assuming a wind velocity of 50 km s$^{-1}$ (see Section \ref{sec:radio_comparison}). This is consistent with these pulsation-driven superwinds, as they can support mass-loss rates up to $10^{-2} M_{\odot} \textrm{yr}^{-1}$ \citep{yoon_2010}. Even the measured enhanced mass-loss rates right before explosion (${\sim}1-4$ years pre-explosion) for the progenitor of SN~2024ggi ($\dot{M} = 10^{-3} - 10^{-2} M_{\odot} \textrm{yr}^{-1}$; \citealp{chen_2024b, jacobson_galan_2024, shrestha_2024, zhang_2024, ertini_2025}) are consistent with this threshold. 

Crucially, unlike the eruptive outbursts (see Section \ref{sec:eruptive_outbursts}) or common envelope ejections (see Section \ref{sec:binary_interactions}), pulsation-driven superwinds do not predict precursor emission. The absence of any detections in the pre-SN light curve (see Section \ref{sec:atclean_2024ggi}) is therefore consistent with pulsation-driven superwinds.

Furthermore, \citet{xiang_2024} found that the progenitor of SN~2024ggi pulsated with a period of 378.5 $\pm$ 29.4 days using data from the Spitzer Space Telescope and \gls{HST}, which adds additional credibility that significant pulsation-driven superwinds occurred in the progenitor of SN~2024ggi. However, \citet{laplace_2026} argue that this apparent periodicity might be caused by the periodic observation schedule of Spitzer.

Interestingly, \citet{davies_2022} argue against a superwind model, as they show that this model implies that the progenitor would be heavily obscured for many decades before explosion. However, the progenitors of many Type IIn SNe and SNe with ``IIn-like'' features, including the one for SN~2024ggi \citep{xiang_2024}, have been observed before explosion. 

The resultant picture is a continuous and accelerating superwind, which started at ${\sim}8 \times 10^{-5}$ $M_\odot$ yr$^{-1}$ at ${\sim}10-50$ years pre-explosion (the extended wind-like tail inferred by our radio observations in Figure \ref{fig:density_vs_radius}), which then increased up to $10^{-3} - 10^{-2}$ $M_\odot$ yr$^{-1}$ at $\sim1-4$ years pre-explosion (the denser inner region in Figure \ref{fig:density_vs_radius}). Alternatively, pulsation-driven superwinds can account for the lower mass-loss inferred by our radio observations at larger radii, while the denser inner region is caused by another mechanism. The denser inner region can be caused by pulsations that change the progenitor's structure prior to explosion \citep{laplace_2026}, binary interactions (see Section \ref{sec:binary_interactions}), or extended chromospheres (see Section \ref{sec:chromospheres}).

\subsection{Extended Chromospheres}
\label{sec:chromospheres}

\citet{fuller_2024} argue that the narrow flash-ionized emission lines found in SNe with ``IIn-like'' features and the implied \gls{CSM} are the consequence of a dense, extended chromosphere between the stellar surface and the dust formation radius. This dense chromosphere is supported by outgoing shock waves generated by turbulent convection near the photosphere of the star. Interestingly, they argue that this material is not really mass that is ejected from the star prior to explosion, but is instead better interpreted as the chromosphere of the \gls{RSG}. This model provides a promising explanation for our observations of SN~2024ggi.

First of all, the extended chromosphere model is consistent with our optical pre-SN light curve analysis (see Section \ref{sec:atclean_2024ggi}), as it does not predict any precursor emission. Furthermore, the expected chromosphere extends to the dust formation radius, which is at $\sim5$ stellar radii \citep{fuller_2024}. The progenitor of SN~2024ggi had a radius of $517-887 R_\odot$ (see Table \ref{tab:ggi_properties}; \citealt{chen_2024, ertini_2025, xiang_2024}). This implies that the dust formation radius is at ${\sim}1.8 - 3.1 \times 10^{14}$ cm. As shown in Figure \ref{fig:density_vs_radius}, this is roughly consistent with the radii where increased densities are observed relative to our radio observations. This model is also broadly consistent with estimates of the \gls{CSM} mass around SN~2024ggi. \citet{fuller_2024} predict a chromospheric mass of ${\sim}10^{-2} - 10^{-1} M_{\odot}$ (see their Figure 3). This is consistent with the \gls{CSM} mass estimates based on spectroscopy ($0.02 - 0.04 M_{\odot}$; \citealp{jacobson_galan_2024}), although it is below the \gls{CSM} mass estimates based on the light curve of SN~2024ggi ($0.4 - 1.2 M_{\odot}$; \citealp{chen_2024,ertini_2025}).

However, a potential issue arises when considering the different mass-loss rates measured for the progenitor of SN~2024ggi. The predicted mass-loss rates from the extended chromosphere model are $\lesssim10^{-5} M_{\odot} \textrm{yr}^{-1}$ for models around the stellar mass range of the progenitor of SN~2024ggi ($M_{*} = 12 - 15 M_{\odot}$). While this is consistent with the progenitor mass-loss rate measured by \citet{xiang_2024} of  $\dot{M} <3 \times 10^{-6} M_{\odot} \textrm{yr}^{-1}$, it is lower than what was measured from the radio observations presented in Section \ref{sec:radio_modelling}. We find a mass-loss rate of ${\sim}8 \times 10^{-5} M_{\odot} \textrm{yr}^{-1}$, assuming a wind velocity of 50 km s$^{-1}$. However, \citet{fuller_2024} argue that such discrepancies arise because the actual outflow velocity is much lower than what is typically assumed, which will result in exaggerated mass-loss rates. Thus, the true pre-explosion mass-loss rate of the progenitor of SN~2024ggi might be much lower than what we reported. For example, if we instead assume a wind velocity of 10 km s$^{-1}$, then the inferred mass-loss rate would decrease to $1.6\times10^{-5} M_{\odot} \textrm{yr}^{-1}$, which is much closer to the aforementioned values quoted by \citet{fuller_2024}. Alternatively, as the mass-loss rates in \citet{fuller_2024} increase with the progenitor radius, the mass-loss rates inferred from our radio observations become more consistent if we assume that the progenitor of SN~2024ggi is more massive (e.g., the pre-SN mass-loss rate for a 20 $M_{\odot}$ progenitor is ${\sim}4\times10^{-5} M_{\odot} \textrm{yr}^{-1}$; see their Figure 6). Thus, the mass-loss rates found in this work can be consistent with the rates found in \citet{fuller_2024} if we assume a lower wind velocity, a more massive progenitor for SN~2024ggi, or a combination of both. 

In conclusion, the extended chromosphere model provides a promising explanation for the presence of the dense \gls{CSM} around SN~2024ggi. It is especially well suited to explain the denser inner region of the two-component \gls{CSM} density profile. A similar conclusion was also reached by \citet{laplace_2026}, who suggest that radial pulsations and an extended chromosphere can in principle explain the flash ionization ``IIn-like'' features observed for many Type II SNe without invoking other mass-loss mechanisms.

\section{Conclusion}
\label{sec:conclusion}

SN~2024ggi was discovered on \todo{11 April 2024} by the \gls{ATLAS} survey. SN~2024ggi is located in a nearby galaxy, NGC 3621, at a distance of $\sim$7.2 Mpc. It was classified as a Type II SN, though spectra obtained immediately after the explosion (+0.8 days) revealed that it exhibited ``IIn-like'' features. This implies the presence of a dense \gls{CSM} around the progenitor of SN~2024ggi. Possible explanations for the presence of this dense \gls{CSM} include: eruptive outbursts, binary interactions, pulsation-driven superwinds, and extended chromospheres. In this work, we constrain the physical processes that can result in the presence of this dense \gls{CSM}. This is done with two complementary approaches: i) we analyzed the optical pre-explosion light curve using data from \gls{ATLAS} to look for evidence of precursor emission and ii) we probed the extended \gls{CSM} density profile with post-explosion radio observations from \gls{ATCA}.

The optical pre-explosion light curve was analyzed with \gls{ATClean}, a tool previously designed by \citet{rest_2025}. We do not find any evidence for precursor emission at any of the timescales tested ($2 < \sigma_{\rm sim} < 300$ days). The detection thresholds tend to be deeper at longer timescales. Assuming the 80\% detection threshold as an upper limit, we can exclude outbursts up to \todo{18.01} mag for $\sigma_{\rm sim} = 2$ days, which increases to \todo{$20.31$} mag for $\sigma_{\rm sim} = 300$ days. This corresponds to absolute magnitude limits of \todo{$-11.28$} mag and \todo{$-8.98$} mag for SN~2024ggi, respectively. The detection thresholds found for SN~2024ggi confidently exclude outbursts that are similar in timescale and absolute magnitude to the precursor variability found in other SNe, such as SN~2009ip and 2020tlf, as well as those found in multiple other Type IIn SNe. A similar conclusion was reached after converting our thresholds to limits on the \gls{CSM} mass with \texttt{CHIPS}, a code to perform one-dimensional radiation hydrodynamical simulations. In principle, faint outbursts with very short timescales could have evaded detection, but comparison to the timescales of outbursts found among other SNe suggests that this is unlikely. In other words, if the progenitor of SN~2024ggi experienced an outburst that was similar to any previously known outburst, then we would have detected it.

The radio observations were analyzed using a standard synchrotron model \citep{Chevalier1998} relating the observed radio spectra to the physical properties of the radio-emitting source. 
We found that our radio observations are consistent with an extended wind-like ($\rho \propto r^{-2}$) density profile. We infer a mass-loss rate of ${\sim} 8\times10^{-5} M_\odot\,\mathrm{yr}^{-1}$, assuming a wind velocity of $v_\mathrm{wind} = 50\,\mathrm{km}\,\mathrm{s}^{-1}$. Our radio observations cover radii between $\sim5 \times 10^{14} - 7 \times10^{15}$ cm, which correspond to $\sim 3 - 50$ years pre-explosion. Interestingly, comparison of our observations with \gls{CSM} density estimates found in the literature revealed a two-component structure to the overall \gls{CSM} density profile. It appears that there is a denser inner structure at smaller radii ($\lesssim6 \times 10^{14}$ cm) and an extended wind-like profile at larger radii.

We then interpreted our optical and radio observations in the context of different mechanisms that have been put forward to explain the presence of dense \gls{CSM} around Type IIn SNe and SNe with ``IIn-like'' features. We conclude that the progenitor of SN~2024ggi is not likely to have experienced significant eruptive outbursts. Instead, the denser inner regions of the \gls{CSM} density profile are best explained with the extended chromosphere model (or possibly a very late-stage binary interaction), while the extended tail in the \gls{CSM} density profile is most consistent with pulsation-driven superwinds. Perhaps a combination of these mechanisms is needed to fully explain the complete \gls{CSM} density profile of SN~2024ggi.

In future work, we aim to search for precursor variability in a more systematic approach for multiple targets. This will be possible with the Vera C. Rubin Observatory \glswithcite{LSST}{\citealp{ivezic_2019}} using software such as \texttt{DETECT}, presented in \citet{geron_2026}. Although its cadence is not as high as \gls{ATLAS} (e.g., see \citealp{bianco_2022}), the detection thresholds are markedly deeper. Using \gls{ATClean} on the $o$-band, we were able to probe magnitudes up to ${\sim}\todo{20}$ mag, while the 5$\sigma$ point source depths of the \gls{LSST} $r$-band for the single-visit images will be 24.7 mag, and the 10-year coadded depth is projected to be 27.5 mag \citep{bianco_2022}. In addition, Rubin \gls{LSST} is expected to find several million alerts per night \citep{ridgway_2014}. \citet{gagliano_2025} estimate that Rubin will find ${\sim}40-130$ SN IIP/IIL and ${\sim}110$ SN IIn with precursor emission per year in single-epoch photometry. We will be able to obtain a more consistent picture of the rate of precursor emission across SN types and better constrain the physical mechanisms that cause the presence of the dense \gls{CSM} that surrounds Type IIn SNe and SNe with IIn-like features as the number of core collapse SNe with detected precursor emission keeps growing.

\begin{acknowledgments}
We thank Fabio de Colle for fruitful discussions about synchrotron self-absorption in radio supernovae. 

TG is a Canadian Rubin Fellow at the Dunlap Institute. The Dunlap Institute is funded through an endowment established by the David Dunlap family and the University of Toronto. TG is also supported through the LSST-DA Catalyst Fellowship; this publication was thus made possible through the support of grant 62192 from the John Templeton Foundation to LSST-DA.

JKL and MRD acknowledge support from the University of Toronto and Hebrew University of Jerusalem through the University of Toronto - Hebrew University of Jerusalem Research and Training Alliance program. 

YT is supported by JSPS KAKENHI grant No. 23H04900.

This work has made use of data from the Asteroid Terrestrial-impact Last Alert System (ATLAS) project. The Asteroid Terrestrial-impact Last Alert System (ATLAS) project is primarily funded to search for near earth asteroids through NASA grants NN12AR55G, 80NSSC18K0284, and 80NSSC18K1575; byproducts of the NEO search include images and catalogs from the survey area. This work was partially funded by Kepler/K2 grant J1944/80NSSC19K0112 and HST GO-15889, and STFC grants ST/T000198/1 and ST/S006109/1. The ATLAS science products have been made possible through the contributions of the University of Hawaii Institute for Astronomy, the Queen’s University Belfast, the Space Telescope Science Institute, the South African Astronomical Observatory, and The Millennium Institute of Astrophysics (MAS), Chile.


The Australia Telescope Compact Array is part of the Australia Telescope National Facility (\url{https://ror.org/05qajvd42}) which is funded by the Australian Government for operation as a National Facility managed by CSIRO. We acknowledge the Gomeroi people as the Traditional Owners of the Observatory site.

\end{acknowledgments}





%

\facilities{ATLAS \citep{tonry_2018}, ATCA}

\software{\texttt{Astropy} \citep{astropy_2013, astropy_2018, astropy_2022}, \texttt{ATClean} \citep{rest_2025}, \texttt{Bilby} \citep{Ashton2019}, \texttt{CHIPS} \citep{takei_2022,takei_2024}, \texttt{Dynesty} \citep{Speagle2020}, \texttt{Matplotlib} \citep{matplotlib_2007}, \texttt{Miriad} \citep{Sault1995}, \texttt{NumPy} \citep{numpy_2020}, \texttt{SciPy} \citep{scipy_2020}}



\appendix

\section{Figure-of-merit and detection efficiency for all kernel sizes}
\label{app:optical_appendix}

\begin{figure}
    \includegraphics[width=\columnwidth]{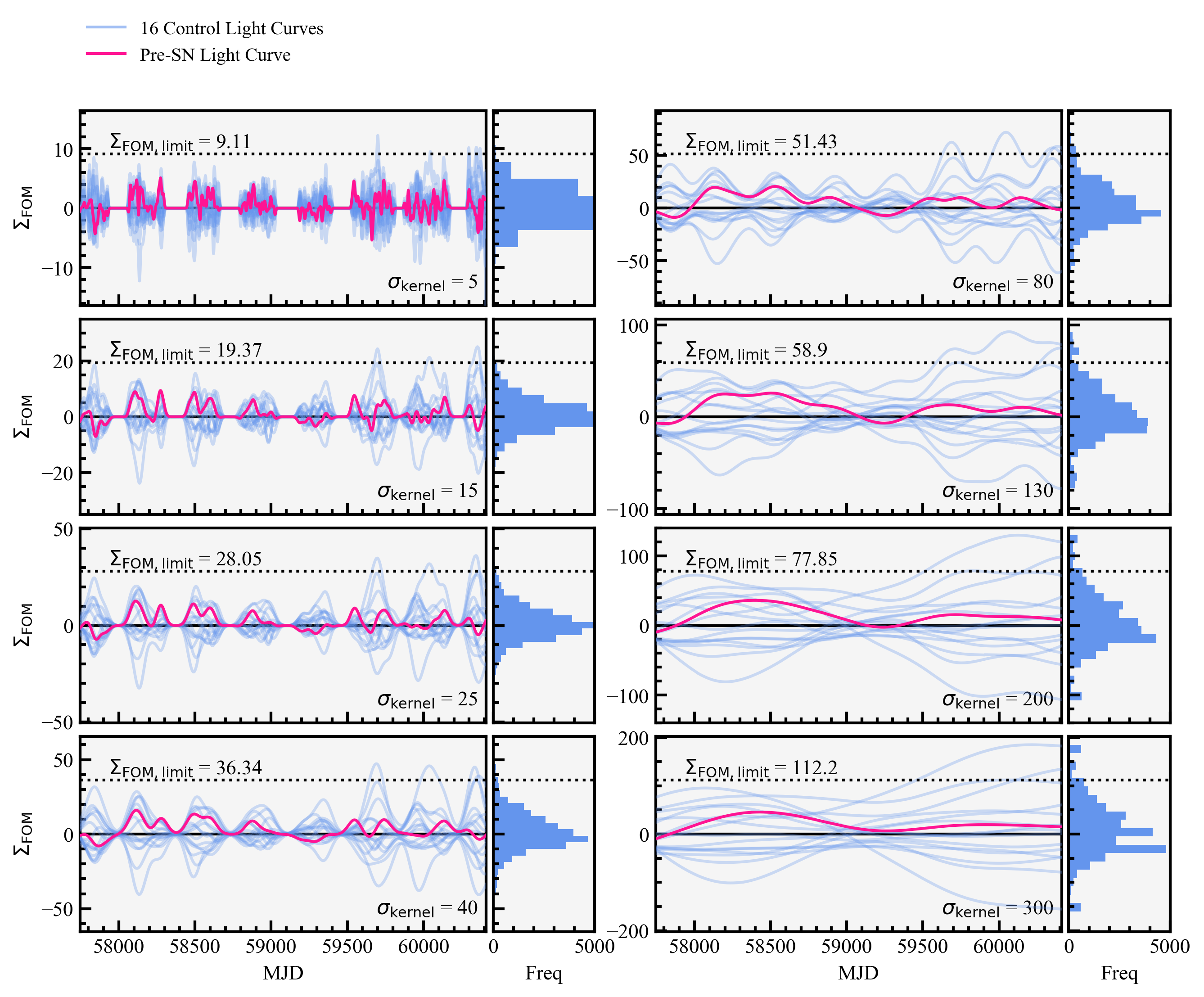}
    \centering
    \caption{The evolution of $\Sigma_{\rm FOM}$ for the pre-SN light curve (pink) and the control light curves (blue) for all kernel sizes tested ($\sigma_{\rm kernel}$). The horizontal dashed lines indicate the detection threshold. The histograms on the right side of each panel show the distribution of $\Sigma_{\rm FOM}$ of the control light curves for that specific kernel size. We find no significant detections in the pre-SN light curve for any of the kernel sizes.}
    \label{fig:sigma_fom_large}
\end{figure}

In this work, we ran \gls{ATClean} on the pre-expolosion optical light curve of SN~2024ggi to look for precursor emission. As noted in Section \ref{sec:atclean_lc_analysis}, this is done by calculating $\Sigma_{\rm FOM}$, which is effectively done by convolving the pre-explosion light curves with Gaussians of different sizes ($\sigma_{\rm kernel}$). In the main text of this paper, we showed the evolution of $\Sigma_{\rm FOM}$ for a select few kernel sizes ($\sigma_{\rm kernel}$ of 5, 40, and 80 days; see Figure \ref{fig:sigma_fom_small}). However, we calculated this for more kernel sizes, ranging from 5 to 300 days. The evolution of $\Sigma_{\rm FOM}$ for all kernel sizes is shown here in Figure \ref{fig:sigma_fom_large}. No significant detections are found in the pre-explosion light curve with any of the kernel sizes used. 

\begin{figure}
	\includegraphics[width=\columnwidth]{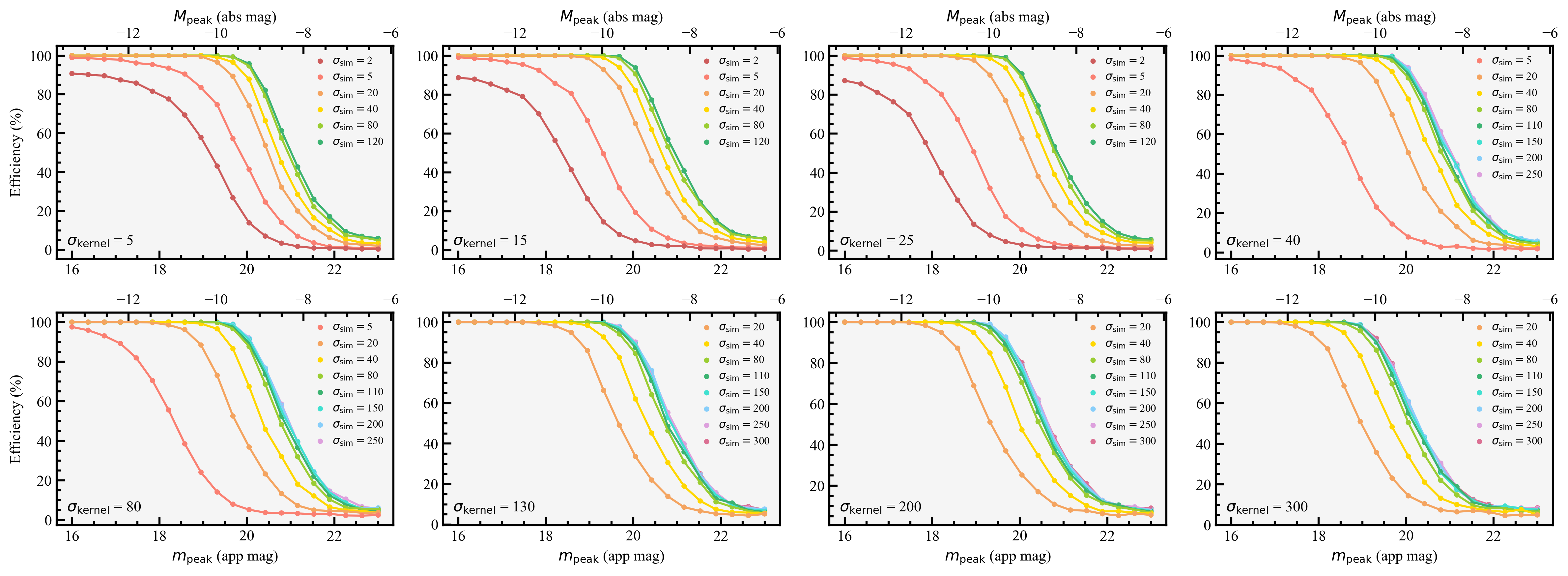}
    \caption{The detection efficiency for simulated events of various timescales over apparent magnitude (bottom axis) and absolute magnitude (top axis) for different kernel sizes. The kernel size is indicated in the bottom left corner of each panel.}
    \label{fig:efficiency_large}
\end{figure}

Since no significant detections were found, we quantified the detection efficiency with \gls{ATClean} (see Section \ref{sec:calculate_detection_efficiency}). Similar to before, we only showed the detection efficiency curves for one kernel size in the main text of the paper ($\sigma_{\rm kernel}$ = 40 days; see Figure \ref{fig:efficiency_small}). However, these efficiency curves were calculated for all values of $\sigma_{\rm kernel}$ that were used. These are shown in Figure \ref{fig:efficiency_large}. Furthermore, the 50\% and 80\% apparent magnitude thresholds ($m_{\rm threshold, 50}$ and $m_{\rm threshold, 80}$) and the corresponding 50\% and 80\% absolute magnitude thresholds ($M_{\rm threshold, 50}$ and $M_{\rm threshold, 80}$) for any combination of $\sigma_{\rm kernel}$ and $\sigma_{\rm sim}$ are summarized in Table \ref{tab:mag_thresholds}.

\begin{deluxetable*}{c c c c c c c}
    \tabletypesize{\scriptsize}
  \tablecaption{The 50\% and 80\% apparent magnitude thresholds ($m_{\rm threshold, 50}$ and $m_{\rm threshold, 80}$) and the 50\% and 80\% absolute magnitude thresholds ($M_{\rm threshold, 50}$ and $M_{\rm threshold, 80}$) for any combination of $\sigma_{\rm kernel}$ and $\sigma_{\rm sim}$.}
  \label{tab:mag_thresholds}
  \tablehead{
  \colhead{$\sigma_{\rm kernel}$ [days]} & \colhead{$\sigma_{\rm sim}$ [days]} & \colhead{$\Sigma_{\rm FOM, limit}$} & \colhead{$m_{\rm threshold, 50}$ [mag]} & \colhead{$m_{\rm threshold, 80}$ [mag]} & \colhead{$M_{\rm threshold, 50}$ [mag]} & \colhead{$M_{\rm threshold, 80}$ [mag]}}
  \decimalcolnumbers
  \startdata
        5 & 2 & 9.11 & 19.16 & 18.01 & -10.14 & -11.28 \\
        5 & 5 & 9.11 & 19.85 & 19.14 & -9.45 & -10.16 \\
        5 & 20 & 9.11 & 20.48 & 19.93 & -8.81 & -9.37 \\
        5 & 40 & 9.11 & 20.70 & 20.20 & -8.60 & -9.09 \\
        5 & 80 & 9.11 & 20.93 & 20.41 & -8.36 & -8.89 \\
        5 & 120 & 9.11 & 21.01 & 20.46 & -8.28 & -8.83 \\
        15 & 2 & 19.37 & 18.37 & 17.39 & -10.93 & -11.90 \\
        15 & 5 & 19.37 & 19.31 & 18.60 & -9.98 & -10.69 \\
        15 & 20 & 19.37 & 20.34 & 19.77 & -8.96 & -9.53 \\
        15 & 40 & 19.37 & 20.64 & 20.10 & -8.65 & -9.20 \\
        15 & 80 & 19.37 & 20.85 & 20.28 & -8.44 & -9.02 \\
        15 & 120 & 19.37 & 20.94 & 20.37 & -8.35 & -8.93 \\
        25 & 2 & 28.05 & 17.97 & 16.83 & -11.33 & -12.46 \\
        25 & 5 & 28.05 & 18.96 & 18.22 & -10.33 & -11.07 \\
        25 & 20 & 28.05 & 20.19 & 19.61 & -9.11 & -9.69 \\
        25 & 40 & 28.05 & 20.58 & 20.05 & -8.72 & -9.24 \\
        25 & 80 & 28.05 & 20.81 & 20.25 & -8.48 & -9.05 \\
        25 & 120 & 28.05 & 20.86 & 20.31 & -8.43 & -8.99 \\
        40 & 5 & 36.34 & 18.70 & 17.93 & -10.59 & -11.36 \\
        40 & 20 & 36.34 & 20.06 & 19.48 & -9.23 & -9.81 \\
        40 & 40 & 36.34 & 20.59 & 20.01 & -8.71 & -9.29 \\
        40 & 80 & 36.34 & 20.82 & 20.27 & -8.47 & -9.02 \\
        40 & 110 & 36.34 & 20.90 & 20.34 & -8.39 & -8.95 \\
        40 & 150 & 36.34 & 21.00 & 20.37 & -8.29 & -8.93 \\
        40 & 200 & 36.34 & 21.01 & 20.38 & -8.28 & -8.91 \\
        40 & 250 & 36.34 & 21.05 & 20.43 & -8.25 & -8.86 \\
        80 & 5 & 51.43 & 18.33 & 17.54 & -10.96 & -11.75 \\
        80 & 20 & 51.43 & 19.73 & 19.18 & -9.56 & -10.12 \\
        80 & 40 & 51.43 & 20.34 & 19.82 & -8.96 & -9.47 \\
        80 & 80 & 51.43 & 20.75 & 20.22 & -8.54 & -9.08 \\
        80 & 110 & 51.43 & 20.84 & 20.29 & -8.45 & -9.00 \\
        80 & 150 & 51.43 & 20.90 & 20.33 & -8.39 & -8.96 \\
        80 & 200 & 51.43 & 20.94 & 20.35 & -8.35 & -8.94 \\
        80 & 250 & 51.43 & 20.95 & 20.35 & -8.34 & -8.95 \\
        130 & 20 & 58.90 & 19.66 & 19.08 & -9.63 & -10.22 \\
        130 & 40 & 58.90 & 20.30 & 19.74 & -8.99 & -9.56 \\
        130 & 80 & 58.90 & 20.71 & 20.14 & -8.59 & -9.15 \\
        130 & 110 & 58.90 & 20.76 & 20.25 & -8.53 & -9.05 \\
        130 & 150 & 58.90 & 20.86 & 20.27 & -8.43 & -9.02 \\
        130 & 200 & 58.90 & 20.86 & 20.34 & -8.44 & -8.96 \\
        130 & 250 & 58.90 & 20.91 & 20.31 & -8.38 & -8.99 \\
        130 & 300 & 58.90 & 20.86 & 20.31 & -8.43 & -8.98 \\
        200 & 20 & 77.85 & 19.35 & 18.74 & -9.95 & -10.55 \\
        200 & 40 & 77.85 & 20.00 & 19.44 & -9.30 & -9.85 \\
        200 & 80 & 77.85 & 20.45 & 19.85 & -8.85 & -9.44 \\
        200 & 110 & 77.85 & 20.54 & 19.93 & -8.75 & -9.36 \\
        200 & 150 & 77.85 & 20.57 & 19.99 & -8.72 & -9.30 \\
        200 & 200 & 77.85 & 20.63 & 20.03 & -8.67 & -9.27 \\
        200 & 250 & 77.85 & 20.65 & 20.02 & -8.64 & -9.27 \\
        200 & 300 & 77.85 & 20.65 & 20.05 & -8.65 & -9.24 \\
        300 & 20 & 112.20 & 18.97 & 18.36 & -10.32 & -10.93 \\
        300 & 40 & 112.20 & 19.64 & 19.03 & -9.65 & -10.27 \\
        300 & 80 & 112.20 & 20.06 & 19.47 & -9.23 & -9.82 \\
        300 & 110 & 112.20 & 20.17 & 19.56 & -9.12 & -9.73 \\
        300 & 150 & 112.20 & 20.24 & 19.62 & -9.06 & -9.68 \\
        300 & 200 & 112.20 & 20.28 & 19.63 & -9.02 & -9.67 \\
        300 & 250 & 112.20 & 20.25 & 19.66 & -9.04 & -9.64 \\
        300 & 300 & 112.20 & 20.23 & 19.67 & -9.06 & -9.62 \\
    \enddata
\end{deluxetable*}


\onecolumngrid 
\section{Posterior Distribution of broken power-law model}
\label{app:radio_appendix}

\begin{figure}
    \includegraphics[width=\columnwidth]{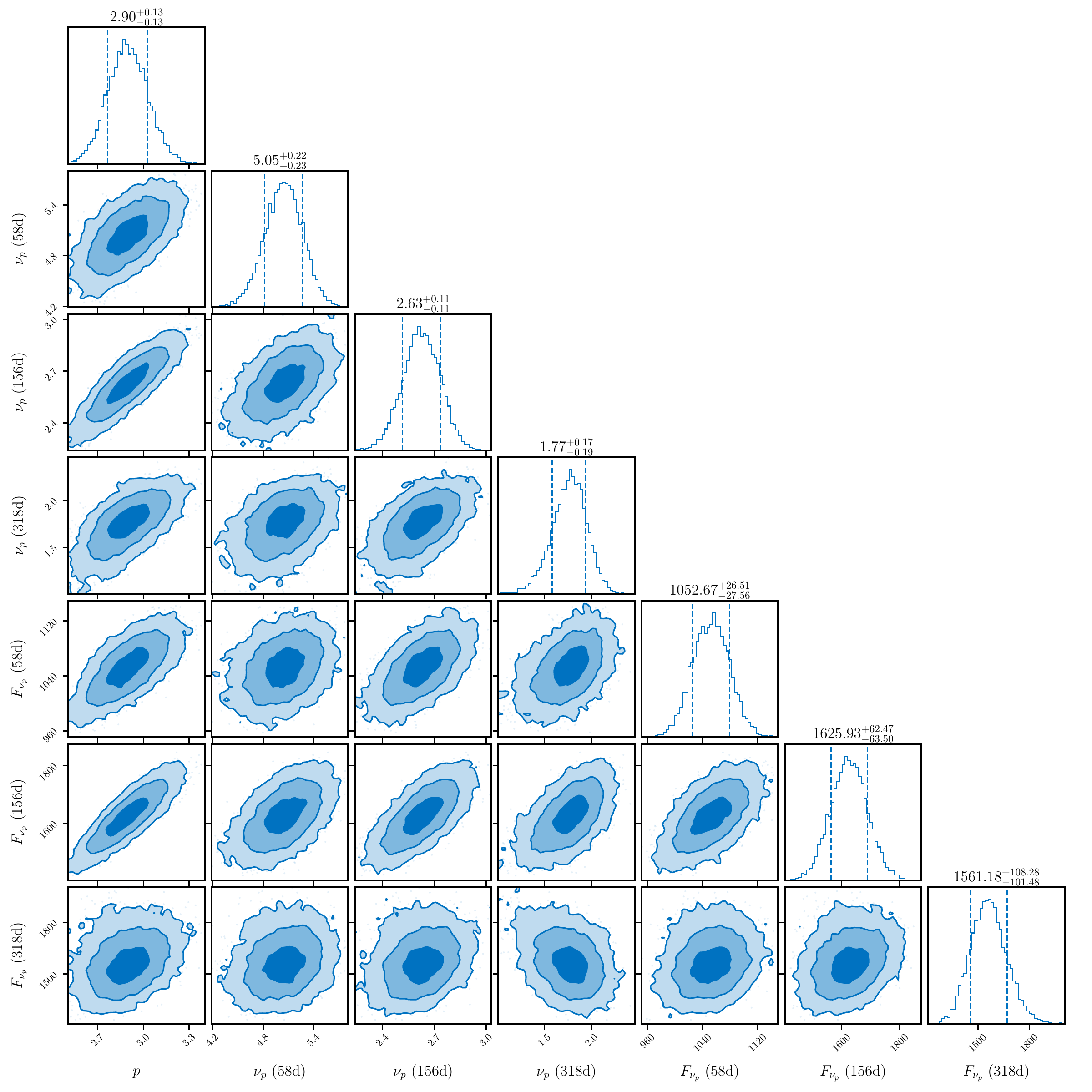}
    \centering
    \caption{Posterior distributions of the parameters in the broken power-law model fit of the radio spectra (see Table~\ref{tab:radiosed_fits}). The dashed vertical lines on the histograms represent the 68\% credible interval of the corresponding marginalized posterior distribution.}
    \label{fig:cornerplot}
\end{figure}

As discussed in Section \ref{sec:radio_modelling}, we fit a smoothly broken power-law model to the radio spectra using the nested sampler \texttt{Dynesty} \citep{Speagle2020}. The full posterior distribution of the model parameters are shown in Figure \ref{fig:cornerplot}.

\bibliography{bibtex}{}

@ARTICLE{astropy_2022,
       author = {{Astropy Collaboration} and {Price-Whelan}, Adrian M. and {Lim}, Pey Lian and {Earl}, Nicholas and {Starkman}, Nathaniel and {Bradley}, Larry and {Shupe}, David L. and {Patil}, Aarya A. and {Corrales}, Lia and {Brasseur}, C.~E. and {N{\"o}the}, Maximilian and {Donath}, Axel and {Tollerud}, Erik and {Morris}, Brett M. and {Ginsburg}, Adam and {Vaher}, Eero and {Weaver}, Benjamin A. and {Tocknell}, James and {Jamieson}, William and {van Kerkwijk}, Marten H. and {Robitaille}, Thomas P. and {Merry}, Bruce and {Bachetti}, Matteo and {G{\"u}nther}, H. Moritz and {Aldcroft}, Thomas L. and {Alvarado-Montes}, Jaime A. and {Archibald}, Anne M. and {B{\'o}di}, Attila and {Bapat}, Shreyas and {Barentsen}, Geert and {Baz{\'a}n}, Juanjo and {Biswas}, Manish and {Boquien}, M{\'e}d{\'e}ric and {Burke}, D.~J. and {Cara}, Daria and {Cara}, Mihai and {Conroy}, Kyle E. and {Conseil}, Simon and {Craig}, Matthew W. and {Cross}, Robert M. and {Cruz}, Kelle L. and {D'Eugenio}, Francesco and {Dencheva}, Nadia and {Devillepoix}, Hadrien A.~R. and {Dietrich}, J{\"o}rg P. and {Eigenbrot}, Arthur Davis and {Erben}, Thomas and {Ferreira}, Leonardo and {Foreman-Mackey}, Daniel and {Fox}, Ryan and {Freij}, Nabil and {Garg}, Suyog and {Geda}, Robel and {Glattly}, Lauren and {Gondhalekar}, Yash and {Gordon}, Karl D. and {Grant}, David and {Greenfield}, Perry and {Groener}, Austen M. and {Guest}, Steve and {Gurovich}, Sebastian and {Handberg}, Rasmus and {Hart}, Akeem and {Hatfield-Dodds}, Zac and {Homeier}, Derek and {Hosseinzadeh}, Griffin and {Jenness}, Tim and {Jones}, Craig K. and {Joseph}, Prajwel and {Kalmbach}, J. Bryce and {Karamehmetoglu}, Emir and {Ka{\l}uszy{\'n}ski}, Miko{\l}aj and {Kelley}, Michael S.~P. and {Kern}, Nicholas and {Kerzendorf}, Wolfgang E. and {Koch}, Eric W. and {Kulumani}, Shankar and {Lee}, Antony and {Ly}, Chun and {Ma}, Zhiyuan and {MacBride}, Conor and {Maljaars}, Jakob M. and {Muna}, Demitri and {Murphy}, N.~A. and {Norman}, Henrik and {O'Steen}, Richard and {Oman}, Kyle A. and {Pacifici}, Camilla and {Pascual}, Sergio and {Pascual-Granado}, J. and {Patil}, Rohit R. and {Perren}, Gabriel I. and {Pickering}, Timothy E. and {Rastogi}, Tanuj and {Roulston}, Benjamin R. and {Ryan}, Daniel F. and {Rykoff}, Eli S. and {Sabater}, Jose and {Sakurikar}, Parikshit and {Salgado}, Jes{\'u}s and {Sanghi}, Aniket and {Saunders}, Nicholas and {Savchenko}, Volodymyr and {Schwardt}, Ludwig and {Seifert-Eckert}, Michael and {Shih}, Albert Y. and {Jain}, Anany Shrey and {Shukla}, Gyanendra and {Sick}, Jonathan and {Simpson}, Chris and {Singanamalla}, Sudheesh and {Singer}, Leo P. and {Singhal}, Jaladh and {Sinha}, Manodeep and {Sip{\H{o}}cz}, Brigitta M. and {Spitler}, Lee R. and {Stansby}, David and {Streicher}, Ole and {{\v{S}}umak}, Jani and {Swinbank}, John D. and {Taranu}, Dan S. and {Tewary}, Nikita and {Tremblay}, Grant R. and {de Val-Borro}, Miguel and {Van Kooten}, Samuel J. and {Vasovi{\'c}}, Zlatan and {Verma}, Shresth and {de Miranda Cardoso}, Jos{\'e} Vin{\'\i}cius and {Williams}, Peter K.~G. and {Wilson}, Tom J. and {Winkel}, Benjamin and {Wood-Vasey}, W.~M. and {Xue}, Rui and {Yoachim}, Peter and {Zhang}, Chen and {Zonca}, Andrea and {Astropy Project Contributors}},
        title = "{The Astropy Project: Sustaining and Growing a Community-oriented Open-source Project and the Latest Major Release (v5.0) of the Core Package}",
      journal = {\apj},
         year = 2022,
        month = aug,
       volume = {935},
       number = {2},
          eid = {167},
        pages = {167},
          doi = {10.3847/1538-4357/ac7c74},
archivePrefix = {arXiv},
       eprint = {2206.14220},
 primaryClass = {astro-ph.IM},
       adsurl = {https://ui.adsabs.harvard.edu/abs/2022ApJ...935..167A}
}

@ARTICLE{ridgway_2014,
       author = {{Ridgway}, Stephen T. and {Matheson}, Thomas and {Mighell}, Kenneth J. and {Olsen}, Knut A. and {Howell}, Steve B.},
        title = "{The Variable Sky of Deep Synoptic Surveys}",
      journal = {\apj},
         year = 2014,
        month = nov,
       volume = {796},
       number = {1},
          eid = {53},
        pages = {53},
          doi = {10.1088/0004-637X/796/1/53},
archivePrefix = {arXiv},
       eprint = {1409.3265},
 primaryClass = {astro-ph.SR},
       adsurl = {https://ui.adsabs.harvard.edu/abs/2014ApJ...796...53R}
}

@ARTICLE{pastorello_2013,
       author = {{Pastorello}, A. and {Cappellaro}, E. and {Inserra}, C. and {Smartt}, S.~J. and {Pignata}, G. and {Benetti}, S. and {Valenti}, S. and {Fraser}, M. and {Tak{\'a}ts}, K. and {Benitez}, S. and {Botticella}, M.~T. and {Brimacombe}, J. and {Bufano}, F. and {Cellier-Holzem}, F. and {Costado}, M.~T. and {Cupani}, G. and {Curtis}, I. and {Elias-Rosa}, N. and {Ergon}, M. and {Fynbo}, J.~P.~U. and {Hambsch}, F. -J. and {Hamuy}, M. and {Harutyunyan}, A. and {Ivarson}, K.~M. and {Kankare}, E. and {Martin}, J.~C. and {Kotak}, R. and {LaCluyze}, A.~P. and {Maguire}, K. and {Mattila}, S. and {Maza}, J. and {McCrum}, M. and {Miluzio}, M. and {Norgaard-Nielsen}, H.~U. and {Nysewander}, M.~C. and {Ochner}, P. and {Pan}, Y. -C. and {Pumo}, M.~L. and {Reichart}, D.~E. and {Tan}, T.~G. and {Taubenberger}, S. and {Tomasella}, L. and {Turatto}, M. and {Wright}, D.},
        title = "{Interacting Supernovae and Supernova Impostors: SN 2009ip, is this the End?}",
      journal = {\apj},
         year = 2013,
        month = apr,
       volume = {767},
       number = {1},
          eid = {1},
        pages = {1},
          doi = {10.1088/0004-637X/767/1/1},
archivePrefix = {arXiv},
       eprint = {1210.3568},
 primaryClass = {astro-ph.SR},
       adsurl = {https://ui.adsabs.harvard.edu/abs/2013ApJ...767....1P}
}

@ARTICLE{kobulnicky_2014,
       author = {{Kobulnicky}, Henry A. and {Kiminki}, Daniel C. and {Lundquist}, Michael J. and {Burke}, Jamison and {Chapman}, James and {Keller}, Erica and {Lester}, Kathryn and {Rolen}, Emily K. and {Topel}, Eric and {Bhattacharjee}, Anirban and {Smullen}, Rachel A. and {Vargas {\'A}lvarez}, Carlos A. and {Runnoe}, Jessie C. and {Dale}, Daniel A. and {Brotherton}, Michael M.},
        title = "{Toward Complete Statistics of Massive Binary Stars: Penultimate Results from the Cygnus OB2 Radial Velocity Survey}",
      journal = {\apjs},
         year = 2014,
        month = aug,
       volume = {213},
       number = {2},
          eid = {34},
        pages = {34},
          doi = {10.1088/0067-0049/213/2/34},
archivePrefix = {arXiv},
       eprint = {1406.6655},
 primaryClass = {astro-ph.SR},
       adsurl = {https://ui.adsabs.harvard.edu/abs/2014ApJS..213...34K}
}

@ARTICLE{rizzuto_2013,
       author = {{Rizzuto}, A.~C. and {Ireland}, M.~J. and {Robertson}, J.~G. and {Kok}, Y. and {Tuthill}, P.~G. and {Warrington}, B.~A. and {Haubois}, X. and {Tango}, W.~J. and {Norris}, B. and {ten Brummelaar}, T. and {Kraus}, A.~L. and {Jacob}, A. and {Laliberte-Houdeville}, C.},
        title = "{Long-baseline interferometric multiplicity survey of the Sco-Cen OB association}",
      journal = {\mnras},
         year = 2013,
        month = dec,
       volume = {436},
       number = {2},
        pages = {1694-1707},
          doi = {10.1093/mnras/stt1690},
archivePrefix = {arXiv},
       eprint = {1309.3811},
 primaryClass = {astro-ph.SR},
       adsurl = {https://ui.adsabs.harvard.edu/abs/2013MNRAS.436.1694R}
}

@ARTICLE{sana_2012,
       author = {{Sana}, H. and {de Mink}, S.~E. and {de Koter}, A. and {Langer}, N. and {Evans}, C.~J. and {Gieles}, M. and {Gosset}, E. and {Izzard}, R.~G. and {Le Bouquin}, J.-B. and {Schneider}, F.~R.~N.},
        title = "{Binary Interaction Dominates the Evolution of Massive Stars}",
      journal = {Science},
         year = 2012,
        month = jul,
       volume = {337},
       number = {6093},
        pages = {444},
          doi = {10.1126/science.1223344},
archivePrefix = {arXiv},
       eprint = {1207.6397},
 primaryClass = {astro-ph.SR},
       adsurl = {https://ui.adsabs.harvard.edu/abs/2012Sci...337..444S}
}

@ARTICLE{flaugher_2015,
       author = {{Flaugher}, B. and {Diehl}, H.~T. and {Honscheid}, K. and {Abbott}, T.~M.~C. and {Alvarez}, O. and {Angstadt}, R. and {Annis}, J.~T. and {Antonik}, M. and {Ballester}, O. and {Beaufore}, L. and {Bernstein}, G.~M. and {Bernstein}, R.~A. and {Bigelow}, B. and {Bonati}, M. and {Boprie}, D. and {Brooks}, D. and {Buckley-Geer}, E.~J. and {Campa}, J. and {Cardiel-Sas}, L. and {Castander}, F.~J. and {Castilla}, J. and {Cease}, H. and {Cela-Ruiz}, J.~M. and {Chappa}, S. and {Chi}, E. and {Cooper}, C. and {da Costa}, L.~N. and {Dede}, E. and {Derylo}, G. and {DePoy}, D.~L. and {de Vicente}, J. and {Doel}, P. and {Drlica-Wagner}, A. and {Eiting}, J. and {Elliott}, A.~E. and {Emes}, J. and {Estrada}, J. and {Fausti Neto}, A. and {Finley}, D.~A. and {Flores}, R. and {Frieman}, J. and {Gerdes}, D. and {Gladders}, M.~D. and {Gregory}, B. and {Gutierrez}, G.~R. and {Hao}, J. and {Holland}, S.~E. and {Holm}, S. and {Huffman}, D. and {Jackson}, C. and {James}, D.~J. and {Jonas}, M. and {Karcher}, A. and {Karliner}, I. and {Kent}, S. and {Kessler}, R. and {Kozlovsky}, M. and {Kron}, R.~G. and {Kubik}, D. and {Kuehn}, K. and {Kuhlmann}, S. and {Kuk}, K. and {Lahav}, O. and {Lathrop}, A. and {Lee}, J. and {Levi}, M.~E. and {Lewis}, P. and {Li}, T.~S. and {Mandrichenko}, I. and {Marshall}, J.~L. and {Martinez}, G. and {Merritt}, K.~W. and {Miquel}, R. and {Mu{\~n}oz}, F. and {Neilsen}, E.~H. and {Nichol}, R.~C. and {Nord}, B. and {Ogando}, R. and {Olsen}, J. and {Palaio}, N. and {Patton}, K. and {Peoples}, J. and {Plazas}, A.~A. and {Rauch}, J. and {Reil}, K. and {Rheault}, J. -P. and {Roe}, N.~A. and {Rogers}, H. and {Roodman}, A. and {Sanchez}, E. and {Scarpine}, V. and {Schindler}, R.~H. and {Schmidt}, R. and {Schmitt}, R. and {Schubnell}, M. and {Schultz}, K. and {Schurter}, P. and {Scott}, L. and {Serrano}, S. and {Shaw}, T.~M. and {Smith}, R.~C. and {Soares-Santos}, M. and {Stefanik}, A. and {Stuermer}, W. and {Suchyta}, E. and {Sypniewski}, A. and {Tarle}, G. and {Thaler}, J. and {Tighe}, R. and {Tran}, C. and {Tucker}, D. and {Walker}, A.~R. and {Wang}, G. and {Watson}, M. and {Weaverdyck}, C. and {Wester}, W. and {Woods}, R. and {Yanny}, B. and {DES Collaboration}},
        title = "{The Dark Energy Camera}",
      journal = {\aj},
         year = 2015,
        month = nov,
       volume = {150},
       number = {5},
          eid = {150},
        pages = {150},
          doi = {10.1088/0004-6256/150/5/150},
archivePrefix = {arXiv},
       eprint = {1504.02900},
 primaryClass = {astro-ph.IM},
       adsurl = {https://ui.adsabs.harvard.edu/abs/2015AJ....150..150F}
}

@ARTICLE{astropy_2013,
       author = {{Astropy Collaboration} and {Robitaille}, Thomas P. and {Tollerud}, Erik J. and {Greenfield}, Perry and {Droettboom}, Michael and {Bray}, Erik and {Aldcroft}, Tom and {Davis}, Matt and {Ginsburg}, Adam and {Price-Whelan}, Adrian M. and {Kerzendorf}, Wolfgang E. and {Conley}, Alexander and {Crighton}, Neil and {Barbary}, Kyle and {Muna}, Demitri and {Ferguson}, Henry and {Grollier}, Fr{\'e}d{\'e}ric and {Parikh}, Madhura M. and {Nair}, Prasanth H. and {Unther}, Hans M. and {Deil}, Christoph and {Woillez}, Julien and {Conseil}, Simon and {Kramer}, Roban and {Turner}, James E.~H. and {Singer}, Leo and {Fox}, Ryan and {Weaver}, Benjamin A. and {Zabalza}, Victor and {Edwards}, Zachary I. and {Azalee Bostroem}, K. and {Burke}, D.~J. and {Casey}, Andrew R. and {Crawford}, Steven M. and {Dencheva}, Nadia and {Ely}, Justin and {Jenness}, Tim and {Labrie}, Kathleen and {Lim}, Pey Lian and {Pierfederici}, Francesco and {Pontzen}, Andrew and {Ptak}, Andy and {Refsdal}, Brian and {Servillat}, Mathieu and {Streicher}, Ole},
        title = "{Astropy: A community Python package for astronomy}",
      journal = {\aap},
         year = 2013,
        month = oct,
       volume = {558},
          eid = {A33},
        pages = {A33},
          doi = {10.1051/0004-6361/201322068},
archivePrefix = {arXiv},
       eprint = {1307.6212},
 primaryClass = {astro-ph.IM},
       adsurl = {https://ui.adsabs.harvard.edu/abs/2013A&A...558A..33A}
}

@ARTICLE{astropy_2018,
       author = {{Astropy Collaboration} and {Price-Whelan}, A.~M. and {Sip{\H{o}}cz}, B.~M. and {G{\"u}nther}, H.~M. and {Lim}, P.~L. and {Crawford}, S.~M. and {Conseil}, S. and {Shupe}, D.~L. and {Craig}, M.~W. and {Dencheva}, N. and {Ginsburg}, A. and {VanderPlas}, J.~T. and {Bradley}, L.~D. and {P{\'e}rez-Su{\'a}rez}, D. and {de Val-Borro}, M. and {Aldcroft}, T.~L. and {Cruz}, K.~L. and {Robitaille}, T.~P. and {Tollerud}, E.~J. and {Ardelean}, C. and {Babej}, T. and {Bach}, Y.~P. and {Bachetti}, M. and {Bakanov}, A.~V. and {Bamford}, S.~P. and {Barentsen}, G. and {Barmby}, P. and {Baumbach}, A. and {Berry}, K.~L. and {Biscani}, F. and {Boquien}, M. and {Bostroem}, K.~A. and {Bouma}, L.~G. and {Brammer}, G.~B. and {Bray}, E.~M. and {Breytenbach}, H. and {Buddelmeijer}, H. and {Burke}, D.~J. and {Calderone}, G. and {Cano Rodr{\'\i}guez}, J.~L. and {Cara}, M. and {Cardoso}, J.~V.~M. and {Cheedella}, S. and {Copin}, Y. and {Corrales}, L. and {Crichton}, D. and {D'Avella}, D. and {Deil}, C. and {Depagne}, {\'E}. and {Dietrich}, J.~P. and {Donath}, A. and {Droettboom}, M. and {Earl}, N. and {Erben}, T. and {Fabbro}, S. and {Ferreira}, L.~A. and {Finethy}, T. and {Fox}, R.~T. and {Garrison}, L.~H. and {Gibbons}, S.~L.~J. and {Goldstein}, D.~A. and {Gommers}, R. and {Greco}, J.~P. and {Greenfield}, P. and {Groener}, A.~M. and {Grollier}, F. and {Hagen}, A. and {Hirst}, P. and {Homeier}, D. and {Horton}, A.~J. and {Hosseinzadeh}, G. and {Hu}, L. and {Hunkeler}, J.~S. and {Ivezi{\'c}}, {\v{Z}}. and {Jain}, A. and {Jenness}, T. and {Kanarek}, G. and {Kendrew}, S. and {Kern}, N.~S. and {Kerzendorf}, W.~E. and {Khvalko}, A. and {King}, J. and {Kirkby}, D. and {Kulkarni}, A.~M. and {Kumar}, A. and {Lee}, A. and {Lenz}, D. and {Littlefair}, S.~P. and {Ma}, Z. and {Macleod}, D.~M. and {Mastropietro}, M. and {McCully}, C. and {Montagnac}, S. and {Morris}, B.~M. and {Mueller}, M. and {Mumford}, S.~J. and {Muna}, D. and {Murphy}, N.~A. and {Nelson}, S. and {Nguyen}, G.~H. and {Ninan}, J.~P. and {N{\"o}the}, M. and {Ogaz}, S. and {Oh}, S. and {Parejko}, J.~K. and {Parley}, N. and {Pascual}, S. and {Patil}, R. and {Patil}, A.~A. and {Plunkett}, A.~L. and {Prochaska}, J.~X. and {Rastogi}, T. and {Reddy Janga}, V. and {Sabater}, J. and {Sakurikar}, P. and {Seifert}, M. and {Sherbert}, L.~E. and {Sherwood-Taylor}, H. and {Shih}, A.~Y. and {Sick}, J. and {Silbiger}, M.~T. and {Singanamalla}, S. and {Singer}, L.~P. and {Sladen}, P.~H. and {Sooley}, K.~A. and {Sornarajah}, S. and {Streicher}, O. and {Teuben}, P. and {Thomas}, S.~W. and {Tremblay}, G.~R. and {Turner}, J.~E.~H. and {Terr{\'o}n}, V. and {van Kerkwijk}, M.~H. and {de la Vega}, A. and {Watkins}, L.~L. and {Weaver}, B.~A. and {Whitmore}, J.~B. and {Woillez}, J. and {Zabalza}, V. and {Astropy Contributors}},
        title = "{The Astropy Project: Building an Open-science Project and Status of the v2.0 Core Package}",
      journal = {\aj},
         year = 2018,
        month = sep,
       volume = {156},
       number = {3},
          eid = {123},
        pages = {123},
          doi = {10.3847/1538-3881/aabc4f},
archivePrefix = {arXiv},
       eprint = {1801.02634},
 primaryClass = {astro-ph.IM},
       adsurl = {https://ui.adsabs.harvard.edu/abs/2018AJ....156..123A}
}

@ARTICLE{ivezic_2019,
       author = {{Ivezi{\'c}}, {\v{Z}}eljko and {Kahn}, Steven M. and {Tyson}, J. Anthony and {Abel}, Bob and {Acosta}, Emily and {Allsman}, Robyn and {Alonso}, David and {AlSayyad}, Yusra and {Anderson}, Scott F. and {Andrew}, John and {Angel}, James Roger P. and {Angeli}, George Z. and {Ansari}, Reza and {Antilogus}, Pierre and {Araujo}, Constanza and {Armstrong}, Robert and {Arndt}, Kirk T. and {Astier}, Pierre and {Aubourg}, {\'E}ric and {Auza}, Nicole and {Axelrod}, Tim S. and {Bard}, Deborah J. and {Barr}, Jeff D. and {Barrau}, Aurelian and {Bartlett}, James G. and {Bauer}, Amanda E. and {Bauman}, Brian J. and {Baumont}, Sylvain and {Bechtol}, Ellen and {Bechtol}, Keith and {Becker}, Andrew C. and {Becla}, Jacek and {Beldica}, Cristina and {Bellavia}, Steve and {Bianco}, Federica B. and {Biswas}, Rahul and {Blanc}, Guillaume and {Blazek}, Jonathan and {Blandford}, Roger D. and {Bloom}, Josh S. and {Bogart}, Joanne and {Bond}, Tim W. and {Booth}, Michael T. and {Borgland}, Anders W. and {Borne}, Kirk and {Bosch}, James F. and {Boutigny}, Dominique and {Brackett}, Craig A. and {Bradshaw}, Andrew and {Brandt}, William Nielsen and {Brown}, Michael E. and {Bullock}, James S. and {Burchat}, Patricia and {Burke}, David L. and {Cagnoli}, Gianpietro and {Calabrese}, Daniel and {Callahan}, Shawn and {Callen}, Alice L. and {Carlin}, Jeffrey L. and {Carlson}, Erin L. and {Chandrasekharan}, Srinivasan and {Charles-Emerson}, Glenaver and {Chesley}, Steve and {Cheu}, Elliott C. and {Chiang}, Hsin-Fang and {Chiang}, James and {Chirino}, Carol and {Chow}, Derek and {Ciardi}, David R. and {Claver}, Charles F. and {Cohen-Tanugi}, Johann and {Cockrum}, Joseph J. and {Coles}, Rebecca and {Connolly}, Andrew J. and {Cook}, Kem H. and {Cooray}, Asantha and {Covey}, Kevin R. and {Cribbs}, Chris and {Cui}, Wei and {Cutri}, Roc and {Daly}, Philip N. and {Daniel}, Scott F. and {Daruich}, Felipe and {Daubard}, Guillaume and {Daues}, Greg and {Dawson}, William and {Delgado}, Francisco and {Dellapenna}, Alfred and {de Peyster}, Robert and {de Val-Borro}, Miguel and {Digel}, Seth W. and {Doherty}, Peter and {Dubois}, Richard and {Dubois-Felsmann}, Gregory P. and {Durech}, Josef and {Economou}, Frossie and {Eifler}, Tim and {Eracleous}, Michael and {Emmons}, Benjamin L. and {Fausti Neto}, Angelo and {Ferguson}, Henry and {Figueroa}, Enrique and {Fisher-Levine}, Merlin and {Focke}, Warren and {Foss}, Michael D. and {Frank}, James and {Freemon}, Michael D. and {Gangler}, Emmanuel and {Gawiser}, Eric and {Geary}, John C. and {Gee}, Perry and {Geha}, Marla and {Gessner}, Charles J.~B. and {Gibson}, Robert R. and {Gilmore}, D. Kirk and {Glanzman}, Thomas and {Glick}, William and {Goldina}, Tatiana and {Goldstein}, Daniel A. and {Goodenow}, Iain and {Graham}, Melissa L. and {Gressler}, William J. and {Gris}, Philippe and {Guy}, Leanne P. and {Guyonnet}, Augustin and {Haller}, Gunther and {Harris}, Ron and {Hascall}, Patrick A. and {Haupt}, Justine and {Hernandez}, Fabio and {Herrmann}, Sven and {Hileman}, Edward and {Hoblitt}, Joshua and {Hodgson}, John A. and {Hogan}, Craig and {Howard}, James D. and {Huang}, Dajun and {Huffer}, Michael E. and {Ingraham}, Patrick and {Innes}, Walter R. and {Jacoby}, Suzanne H. and {Jain}, Bhuvnesh and {Jammes}, Fabrice and {Jee}, M. James and {Jenness}, Tim and {Jernigan}, Garrett and {Jevremovi{\'c}}, Darko and {Johns}, Kenneth and {Johnson}, Anthony S. and {Johnson}, Margaret W.~G. and {Jones}, R. Lynne and {Juramy-Gilles}, Claire and {Juri{\'c}}, Mario and {Kalirai}, Jason S. and {Kallivayalil}, Nitya J. and {Kalmbach}, Bryce and {Kantor}, Jeffrey P. and {Karst}, Pierre and {Kasliwal}, Mansi M. and {Kelly}, Heather and {Kessler}, Richard and {Kinnison}, Veronica and {Kirkby}, David and {Knox}, Lloyd and {Kotov}, Ivan V. and {Krabbendam}, Victor L. and {Krughoff}, K. Simon and {Kub{\'a}nek}, Petr and {Kuczewski}, John and {Kulkarni}, Shri and {Ku}, John and {Kurita}, Nadine R. and {Lage}, Craig S. and {Lambert}, Ron and {Lange}, Travis and {Langton}, J. Brian and {Le Guillou}, Laurent and {Levine}, Deborah and {Liang}, Ming and {Lim}, Kian-Tat and {Lintott}, Chris J. and {Long}, Kevin E. and {Lopez}, Margaux and {Lotz}, Paul J. and {Lupton}, Robert H. and {Lust}, Nate B. and {MacArthur}, Lauren A. and {Mahabal}, Ashish and {Mandelbaum}, Rachel and {Markiewicz}, Thomas W. and {Marsh}, Darren S. and {Marshall}, Philip J. and {Marshall}, Stuart and {May}, Morgan and {McKercher}, Robert and {McQueen}, Michelle and {Meyers}, Joshua and {Migliore}, Myriam and {Miller}, Michelle and {Mills}, David J. and {Miraval}, Connor and {Moeyens}, Joachim and {Moolekamp}, Fred E. and {Monet}, David G. and {Moniez}, Marc and {Monkewitz}, Serge and {Montgomery}, Christopher and {Morrison}, Christopher B. and {Mueller}, Fritz and {Muller}, Gary P. and {Mu{\~n}oz Arancibia}, Freddy and {Neill}, Douglas R. and {Newbry}, Scott P. and {Nief}, Jean-Yves and {Nomerotski}, Andrei and {Nordby}, Martin and {O'Connor}, Paul and {Oliver}, John and {Olivier}, Scot S. and {Olsen}, Knut and {O'Mullane}, William and {Ortiz}, Sandra and {Osier}, Shawn and {Owen}, Russell E. and {Pain}, Reynald and {Palecek}, Paul E. and {Parejko}, John K. and {Parsons}, James B. and {Pease}, Nathan M. and {Peterson}, J. Matt and {Peterson}, John R. and {Petravick}, Donald L. and {Libby Petrick}, M.~E. and {Petry}, Cathy E. and {Pierfederici}, Francesco and {Pietrowicz}, Stephen and {Pike}, Rob and {Pinto}, Philip A. and {Plante}, Raymond and {Plate}, Stephen and {Plutchak}, Joel P. and {Price}, Paul A. and {Prouza}, Michael and {Radeka}, Veljko and {Rajagopal}, Jayadev and {Rasmussen}, Andrew P. and {Regnault}, Nicolas and {Reil}, Kevin A. and {Reiss}, David J. and {Reuter}, Michael A. and {Ridgway}, Stephen T. and {Riot}, Vincent J. and {Ritz}, Steve and {Robinson}, Sean and {Roby}, William and {Roodman}, Aaron and {Rosing}, Wayne and {Roucelle}, Cecille and {Rumore}, Matthew R. and {Russo}, Stefano and {Saha}, Abhijit and {Sassolas}, Benoit and {Schalk}, Terry L. and {Schellart}, Pim and {Schindler}, Rafe H. and {Schmidt}, Samuel and {Schneider}, Donald P. and {Schneider}, Michael D. and {Schoening}, William and {Schumacher}, German and {Schwamb}, Megan E. and {Sebag}, Jacques and {Selvy}, Brian and {Sembroski}, Glenn H. and {Seppala}, Lynn G. and {Serio}, Andrew and {Serrano}, Eduardo and {Shaw}, Richard A. and {Shipsey}, Ian and {Sick}, Jonathan and {Silvestri}, Nicole and {Slater}, Colin T. and {Smith}, J. Allyn and {Smith}, R. Chris and {Sobhani}, Shahram and {Soldahl}, Christine and {Storrie-Lombardi}, Lisa and {Stover}, Edward and {Strauss}, Michael A. and {Street}, Rachel A. and {Stubbs}, Christopher W. and {Sullivan}, Ian S. and {Sweeney}, Donald and {Swinbank}, John D. and {Szalay}, Alexander and {Takacs}, Peter and {Tether}, Stephen A. and {Thaler}, Jon J. and {Thayer}, John Gregg and {Thomas}, Sandrine and {Thornton}, Adam J. and {Thukral}, Vaikunth and {Tice}, Jeffrey and {Trilling}, David E. and {Turri}, Max and {Van Berg}, Richard and {Vanden Berk}, Daniel and {Vetter}, Kurt and {Virieux}, Francoise and {Vucina}, Tomislav and {Wahl}, William and {Walkowicz}, Lucianne and {Walsh}, Brian and {Walter}, Christopher W. and {Wang}, Daniel L. and {Wang}, Shin-Yawn and {Warner}, Michael and {Wiecha}, Oliver and {Willman}, Beth and {Winters}, Scott E. and {Wittman}, David and {Wolff}, Sidney C. and {Wood-Vasey}, W. Michael and {Wu}, Xiuqin and {Xin}, Bo and {Yoachim}, Peter and {Zhan}, Hu},
        title = "{LSST: From Science Drivers to Reference Design and Anticipated Data Products}",
      journal = {\apj},
         year = 2019,
        month = mar,
       volume = {873},
       number = {2},
          eid = {111},
        pages = {111},
          doi = {10.3847/1538-4357/ab042c},
archivePrefix = {arXiv},
       eprint = {0805.2366},
 primaryClass = {astro-ph},
       adsurl = {https://ui.adsabs.harvard.edu/abs/2019ApJ...873..111I}
}

@ARTICLE{takei_2022,
       author = {{Takei}, Yuki and {Tsuna}, Daichi and {Kuriyama}, Naoto and {Ko}, Takatoshi and {Shigeyama}, Toshikazu},
        title = "{CHIPS: Complete History of Interaction-powered Supernovae}",
      journal = {\apj},
         year = 2022,
        month = apr,
       volume = {929},
       number = {2},
          eid = {177},
        pages = {177},
          doi = {10.3847/1538-4357/ac60fe},
archivePrefix = {arXiv},
       eprint = {2109.05871},
 primaryClass = {astro-ph.HE},
       adsurl = {https://ui.adsabs.harvard.edu/abs/2022ApJ...929..177T}
}

@ARTICLE{iwata_2025,
       author = {{Iwata}, Yuhei and {Akimoto}, Masanori and {Matsuoka}, Tomoki and {Maeda}, Keiichi and {Yonekura}, Yoshinori and {Tominaga}, Nozomu and {Moriya}, Takashi J. and {Fujisawa}, Kenta and {Niinuma}, Kotaro and {Yoon}, Sung-Chul and {Lee}, Jae-Joon and {Jung}, Taehyun and {Byun}, Do-Young},
        title = "{Radio Follow-up Observations of SN 2023ixf by Japanese and Korean Very Long Baseline Interferometers}",
      journal = {\apj},
         year = 2025,
        month = jan,
       volume = {978},
       number = {2},
          eid = {138},
        pages = {138},
          doi = {10.3847/1538-4357/ad9a62},
archivePrefix = {arXiv},
       eprint = {2411.07542},
 primaryClass = {astro-ph.HE},
       adsurl = {https://ui.adsabs.harvard.edu/abs/2025ApJ...978..138I}
}

@ARTICLE{nayana_2025,
       author = {{Nayana}, A.~J. and {Margutti}, Raffaella and {Wiston}, Eli and {Chornock}, Ryan and {Campana}, Sergio and {Laskar}, Tanmoy and {Murase}, Kohta and {Krips}, Melanie and {Migliori}, Giulia and {Tsuna}, Daichi and {Alexander}, Kate D. and {Chandra}, Poonam and {Bietenholz}, Michael and {Berger}, Edo and {Chevalier}, Roger A. and {De Colle}, Fabio and {Dessart}, Luc and {Diesing}, Rebecca and {Grefenstette}, Brian W. and {Jacobson-Gal{\'a}n}, Wynn V. and {Maeda}, Keiichi and {Marcote}, Benito and {Matthews}, Daisy and {Milisavljevic}, Dan and {Ray}, Alak K. and {Reguitti}, Andrea and {Polzin}, Ava},
        title = "{Dinosaur in a Haystack: X-Ray View of the Entrails of SN 2023ixf and the Radio Afterglow of Its Interaction with the Medium Spawned by the Progenitor Star (Paper I)}",
      journal = {\apj},
         year = 2025,
        month = may,
       volume = {985},
       number = {1},
          eid = {51},
        pages = {51},
          doi = {10.3847/1538-4357/adc2fb},
archivePrefix = {arXiv},
       eprint = {2411.02647},
 primaryClass = {astro-ph.HE},
       adsurl = {https://ui.adsabs.harvard.edu/abs/2025ApJ...985...51N}
}

@ARTICLE{rest_2025,
       author = {{Rest}, S. and {Rest}, A. and {Kilpatrick}, C.~D. and {Jencson}, J.~E. and {von Coelln}, S. and {Strolger}, L. and {Smartt}, S. and {Anderson}, J.~P. and {Clocchiatti}, A. and {Coulter}, D.~A. and {Denneau}, L. and {Gomez}, S. and {Heinze}, A. and {Ridden-Harper}, R. and {Smith}, K.~W. and {Stalder}, B. and {Tonry}, J.~L. and {Wang}, Q. and {Zenati}, Y.},
        title = "{ATClean: A Novel Method for Detecting Low-luminosity Transients and Application to Pre-explosion Counterparts from SN 2023ixf}",
      journal = {\apj},
         year = 2025,
        month = feb,
       volume = {979},
       number = {2},
          eid = {114},
        pages = {114},
          doi = {10.3847/1538-4357/ad973d},
archivePrefix = {arXiv},
       eprint = {2405.03747},
 primaryClass = {astro-ph.HE},
       adsurl = {https://ui.adsabs.harvard.edu/abs/2025ApJ...979..114R}
}

@INPROCEEDINGS{offner_2023,
       author = {{Offner}, S.~S.~R. and {Moe}, M. and {Kratter}, K.~M. and {Sadavoy}, S.~I. and {Jensen}, E.~L.~N. and {Tobin}, J.~J.},
        title = "{The Origin and Evolution of Multiple Star Systems}",
    booktitle = {Protostars and Planets VII},
         year = 2023,
       editor = {{Inutsuka}, S. and {Aikawa}, Y. and {Muto}, T. and {Tomida}, K. and {Tamura}, M.},
       series = {Astronomical Society of the Pacific Conference Series},
       volume = {534},
        month = jul,
        pages = {275},
          doi = {10.48550/arXiv.2203.10066},
archivePrefix = {arXiv},
       eprint = {2203.10066},
 primaryClass = {astro-ph.SR},
       adsurl = {https://ui.adsabs.harvard.edu/abs/2023ASPC..534..275O}
}

@ARTICLE{sana_2025,
       author = {{Sana}, H. and {Shenar}, T. and {Bodensteiner}, J. and {Britavskiy}, N. and {Langer}, N. and {Lennon}, D.~J. and {Mahy}, L. and {Mandel}, I. and {de Mink}, S.~E. and {Patrick}, L.~R. and {Villase{\~n}or}, J.~I. and {Dirickx}, M. and {Abdul-Masih}, M. and {Almeida}, L.~A. and {Backs}, F. and {Berlanas}, S.~R. and {Bernini-Peron}, M. and {Bowman}, D.~M. and {Bronner}, V.~A. and {Crowther}, P.~A. and {Deshmukh}, K. and {Evans}, C.~J. and {Fabry}, M. and {Gieles}, M. and {Gilkis}, A. and {Gonz{\'a}lez-Tor{\`a}}, G. and {Gr{\"a}fener}, G. and {G{\"o}tberg}, Y. and {Hawcroft}, C. and {H{\'e}nault-Brunet}, V. and {Herrero}, A. and {Holgado}, G. and {Izzard}, R.~G. and {de Koter}, A. and {Janssens}, S. and {Johnston}, C. and {Josiek}, J. and {Justham}, S. and {Kalari}, V.~M. and {Klencki}, J. and {Kub{\'a}t}, J. and {Kub{\'a}tov{\'a}}, B. and {Lefever}, R.~R. and {van Loon}, J. Th. and {Ludwig}, B. and {Mackey}, J. and {Ma{\'\i}z Apell{\'a}niz}, J. and {Maravelias}, G. and {Marchant}, P. and {Mazeh}, T. and {Menon}, A. and {Moe}, M. and {Najarro}, F. and {Oskinova}, L.~M. and {Ovadia}, R. and {Pauli}, D. and {Pawlak}, M. and {Ramachandran}, V. and {Renzo}, M. and {Rocha}, D.~F. and {Sander}, A.~A.~C. and {Schneider}, F.~R.~N. and {Schootemeijer}, A. and {Sch{\"o}sser}, E.~C. and {Sch{\"u}rmann}, C. and {Sen}, K. and {Shahaf}, S. and {Sim{\'o}n-D{\'\i}az}, S. and {van Son}, L.~A.~C. and {Stoop}, M. and {Toonen}, S. and {Tramper}, F. and {Valli}, R. and {Vigna-G{\'o}mez}, A. and {Vink}, J.~S. and {Wang}, C. and {Willcox}, R.},
        title = "{A high fraction of close massive binary stars at low metallicity}",
      journal = {Nature Astronomy},
         year = 2025,
        month = sep,
       volume = {9},
        pages = {1337-1346},
          doi = {10.1038/s41550-025-02610-x},
archivePrefix = {arXiv},
       eprint = {2509.12488},
 primaryClass = {astro-ph.SR},
       adsurl = {https://ui.adsabs.harvard.edu/abs/2025NatAs...9.1337S}
}

@ARTICLE{sana_2013,
       author = {{Sana}, H. and {de Koter}, A. and {de Mink}, S.~E. and {Dunstall}, P.~R. and {Evans}, C.~J. and {H{\'e}nault-Brunet}, V. and {Ma{\'\i}z Apell{\'a}niz}, J. and {Ram{\'\i}rez-Agudelo}, O.~H. and {Taylor}, W.~D. and {Walborn}, N.~R. and {Clark}, J.~S. and {Crowther}, P.~A. and {Herrero}, A. and {Gieles}, M. and {Langer}, N. and {Lennon}, D.~J. and {Vink}, J.~S.},
        title = "{The VLT-FLAMES Tarantula Survey. VIII. Multiplicity properties of the O-type star population}",
      journal = {\aap},
         year = 2013,
        month = feb,
       volume = {550},
          eid = {A107},
        pages = {A107},
          doi = {10.1051/0004-6361/201219621},
archivePrefix = {arXiv},
       eprint = {1209.4638},
 primaryClass = {astro-ph.SR},
       adsurl = {https://ui.adsabs.harvard.edu/abs/2013A&A...550A.107S}
}

@ARTICLE{shenar_2026,
       author = {{Shenar}, T. and {Bodensteiner}, J. and {Abdul-Masih}, M. and {Backs}, F. and {Berlanas}, S.~R. and {Bestenlehner}, J.~M. and {Bobrick}, A. and {Bowman}, D.~M. and {Britavskiy}, N. and {Crowther}, P.~A. and {Deshmukh}, K. and {Fabry}, M. and {Gieles}, M. and {Gilkis}, A. and {Gonz{\'a}lez-Tor{\`a}}, G. and {Gr{\"a}fener}, G. and {Gull}, M. and {G{\"o}tberg}, Y. and {Hawcroft}, C. and {H{\'e}nault-Brunet}, V. and {Herrero}, A. and {Holgado}, G. and {Izzard}, R.~G. and {Janssens}, S. and {Jin}, H. and {Kalari}, V.~M. and {Katabi}, Z. and {de Koter}, A. and {Klencki}, J. and {Kub{\'a}tov{\'a}}, B. and {Kub{\'a}t}, J. and {Langer}, N. and {Lechien}, T. and {Lennon}, D. and {Lefever}, R. and {Ludwig}, B. and {Mahy}, L. and {Ma{\'\i}z Apell{\'a}niz}, J. and {Mandel}, I. and {Mang Rom{\'a}n}, A. and {Maravelias}, G. and {Marchant}, P. and {Menon}, A. and {Najarro}, F. and {O'Grady}, A. and {Oskinova}, L. and {Ovadia}, R. and {Patrick}, L.~R. and {Pauli}, D. and {Pawlak}, M. and {Bernini Peron}, M. and {Picco}, A. and {Ramachandran}, V. and {Renzo}, M. and {Rocha}, D.~F. and {Sana}, H. and {Sander}, A.~A. and {Sayada}, T. and {Schootemeijer}, A. and {Schneider}, F.~R.~N. and {Seeburger}, R. and {Sen}, K. and {Shahaf}, S. and {Sim{\'o}n-D{\'\i}az}, S. and {Stoop}, M. and {Toonen}, S. and {Tramper}, F. and {Valli}, R. and {Van Daele}, P. and {van Loon}, J.~T. and {Son}, L. v. and {Vigna-G{\'o}mez}, A. and {Villase{\~n}or}, J.~I. and {Vink}, J.~S. and {Wang}, C. and {Xu}, X.-T.},
        title = "{Multiplicity of Massive Stars at Low Metallicity: Early Results from the BLOeM Campaign}",
      journal = {The Messenger},
         year = 2026,
        month = mar,
       volume = {196},
        pages = {15-19},
          doi = {10.18727/0722-6691/5417},
archivePrefix = {arXiv},
       eprint = {2607.27395},
 primaryClass = {physics.gen-ph},
       adsurl = {https://ui.adsabs.harvard.edu/abs/2026Msngr.196...15S}
}

@INCOLLECTION{gal_yam_2017,
       author = {{Gal-Yam}, Avishay},
        title = "{Observational and Physical Classification of Supernovae}",
    booktitle = {Handbook of Supernovae},
         year = 2017,
       editor = {{Alsabti}, Athem W. and {Murdin}, Paul},
        pages = {195},
          doi = {10.1007/978-3-319-21846-5_35},
       adsurl = {https://ui.adsabs.harvard.edu/abs/2017hsn..book..195G}
}

@ARTICLE{dong_2023,
       author = {{Dong}, Yize and {Sand}, David J. and {Valenti}, Stefano and {Bostroem}, K. Azalee and {Andrews}, Jennifer E. and {Hosseinzadeh}, Griffin and {Hoang}, Emily and {Janzen}, Daryl and {Jencson}, Jacob E. and {Lundquist}, Michael and {Meza Retamal}, Nicolas E. and {Pearson}, Jeniveve and {Shrestha}, Manisha and {Haislip}, Joshua and {Kouprianov}, Vladimir and {Reichart}, Daniel E.},
        title = "{A Comprehensive Optical Search for Pre-explosion Outbursts from the Quiescent Progenitor of SN 2023ixf}",
      journal = {\apj},
         year = 2023,
        month = nov,
       volume = {957},
       number = {1},
          eid = {28},
        pages = {28},
          doi = {10.3847/1538-4357/acef18},
archivePrefix = {arXiv},
       eprint = {2307.02539},
 primaryClass = {astro-ph.HE},
       adsurl = {https://ui.adsabs.harvard.edu/abs/2023ApJ...957...28D}
}

@ARTICLE{smith_2011,
       author = {{Smith}, Nathan and {Li}, Weidong and {Filippenko}, Alexei V. and {Chornock}, Ryan},
        title = "{Observed fractions of core-collapse supernova types and initial masses of their single and binary progenitor stars}",
      journal = {\mnras},
         year = 2011,
        month = apr,
       volume = {412},
       number = {3},
        pages = {1522-1538},
          doi = {10.1111/j.1365-2966.2011.17229.x},
archivePrefix = {arXiv},
       eprint = {1006.3899},
 primaryClass = {astro-ph.HE},
       adsurl = {https://ui.adsabs.harvard.edu/abs/2011MNRAS.412.1522S}
}

@ARTICLE{smartt_2009,
       author = {{Smartt}, S.~J. and {Eldridge}, J.~J. and {Crockett}, R.~M. and {Maund}, J.~R.},
        title = "{The death of massive stars - I. Observational constraints on the progenitors of Type II-P supernovae}",
      journal = {\mnras},
         year = 2009,
        month = may,
       volume = {395},
       number = {3},
        pages = {1409-1437},
          doi = {10.1111/j.1365-2966.2009.14506.x},
archivePrefix = {arXiv},
       eprint = {0809.0403},
 primaryClass = {astro-ph},
       adsurl = {https://ui.adsabs.harvard.edu/abs/2009MNRAS.395.1409S}
}

@INCOLLECTION{smith_2017,
       author = {{Smith}, Nathan},
        title = "{Interacting Supernovae: Types IIn and Ibn}",
    booktitle = {Handbook of Supernovae},
         year = 2017,
       editor = {{Alsabti}, Athem W. and {Murdin}, Paul},
        pages = {403},
          doi = {10.1007/978-3-319-21846-5_38},
       adsurl = {https://ui.adsabs.harvard.edu/abs/2017hsn..book..403S}
}

@ARTICLE{filippenko_1997,
       author = {{Filippenko}, Alexei V.},
        title = "{Optical Spectra of Supernovae}",
      journal = {\araa},
         year = 1997,
        month = jan,
       volume = {35},
        pages = {309-355},
          doi = {10.1146/annurev.astro.35.1.309},
       adsurl = {https://ui.adsabs.harvard.edu/abs/1997ARA&A..35..309F}
}

@ARTICLE{kilpatrick_2018,
       author = {{Kilpatrick}, Charles D. and {Foley}, Ryan J.},
        title = "{The dusty progenitor star of the Type II supernova 2017eaw}",
      journal = {\mnras},
         year = 2018,
        month = dec,
       volume = {481},
       number = {2},
        pages = {2536-2547},
          doi = {10.1093/mnras/sty2435},
archivePrefix = {arXiv},
       eprint = {1806.00348},
 primaryClass = {astro-ph.SR},
       adsurl = {https://ui.adsabs.harvard.edu/abs/2018MNRAS.481.2536K}
}

@ARTICLE{jencson_2023,
       author = {{Jencson}, Jacob E. and {Pearson}, Jeniveve and {Beasor}, Emma R. and {Lau}, Ryan M. and {Andrews}, Jennifer E. and {Bostroem}, K. Azalee and {Dong}, Yize and {Engesser}, Michael and {Gomez}, Sebastian and {Guolo}, Muryel and {Hoang}, Emily and {Hosseinzadeh}, Griffin and {Jha}, Saurabh W. and {Karambelkar}, Viraj and {Kasliwal}, Mansi M. and {Lundquist}, Michael and {Meza Retamal}, Nicolas E. and {Rest}, Armin and {Sand}, David J. and {Shahbandeh}, Melissa and {Shrestha}, Manisha and {Smith}, Nathan and {Strader}, Jay and {Valenti}, Stefano and {Wang}, Qinan and {Zenati}, Yossef},
        title = "{A Luminous Red Supergiant and Dusty Long-period Variable Progenitor for SN 2023ixf}",
      journal = {\apjl},
         year = 2023,
        month = aug,
       volume = {952},
       number = {2},
          eid = {L30},
        pages = {L30},
          doi = {10.3847/2041-8213/ace618},
archivePrefix = {arXiv},
       eprint = {2306.08678},
 primaryClass = {astro-ph.SR},
       adsurl = {https://ui.adsabs.harvard.edu/abs/2023ApJ...952L..30J}
}

@ARTICLE{schlegel_1990,
       author = {{Schlegel}, Eric M.},
        title = "{A new subclass of type II supernovae ?}",
      journal = {\mnras},
         year = 1990,
        month = may,
       volume = {244},
        pages = {269-271},
       adsurl = {https://ui.adsabs.harvard.edu/abs/1990MNRAS.244..269S}
}

@ARTICLE{kilpatrick_2023,
       author = {{Kilpatrick}, Charles D. and {Izzo}, Luca and {Bentley}, Rory O. and {Chambers}, Kenneth C. and {Coulter}, David A. and {Drout}, Maria R. and {de Boer}, Thomas and {Foley}, Ryan J. and {Gall}, Christa and {Halford}, Melissa R. and {Jones}, David O. and {Langeroodi}, Danial and {Lin}, Chien-Cheng and {Magnier}, Eugene A. and {McGill}, Peter and {O'Grady}, Anna J.~G. and {Pan}, Yen-Chen and {Ramirez-Ruiz}, Enrico and {Rest}, Armin and {Swift}, Jonathan J. and {Tinyanont}, Samaporn and {Villar}, V. Ashley and {Wainscoat}, Richard J. and {Wasserman}, Amanda Rose and {Yadavalli}, S. Karthik and {Yang}, Grace},
        title = "{Type II-P supernova progenitor star initial masses and SN 2020jfo: direct detection, light-curve properties, nebular spectroscopy, and local environment}",
      journal = {\mnras},
         year = 2023,
        month = sep,
       volume = {524},
       number = {2},
        pages = {2161-2185},
          doi = {10.1093/mnras/stad1954},
archivePrefix = {arXiv},
       eprint = {2307.00550},
 primaryClass = {astro-ph.SR},
       adsurl = {https://ui.adsabs.harvard.edu/abs/2023MNRAS.524.2161K}
}

@ARTICLE{meynet_2015,
       author = {{Meynet}, G. and {Chomienne}, V. and {Ekstr{\"o}m}, S. and {Georgy}, C. and {Granada}, A. and {Groh}, J. and {Maeder}, A. and {Eggenberger}, P. and {Levesque}, E. and {Massey}, P.},
        title = "{Impact of mass-loss on the evolution and pre-supernova properties of red supergiants}",
      journal = {\aap},
         year = 2015,
        month = mar,
       volume = {575},
          eid = {A60},
        pages = {A60},
          doi = {10.1051/0004-6361/201424671},
archivePrefix = {arXiv},
       eprint = {1410.8721},
 primaryClass = {astro-ph.SR},
       adsurl = {https://ui.adsabs.harvard.edu/abs/2015A&A...575A..60M}
}

@ARTICLE{muller_2016,
       author = {{M{\"u}ller}, Bernhard and {Heger}, Alexander and {Liptai}, David and {Cameron}, Joshua B.},
        title = "{A simple approach to the supernova progenitor-explosion connection}",
      journal = {\mnras},
         year = 2016,
        month = jul,
       volume = {460},
       number = {1},
        pages = {742-764},
          doi = {10.1093/mnras/stw1083},
archivePrefix = {arXiv},
       eprint = {1602.05956},
 primaryClass = {astro-ph.SR},
       adsurl = {https://ui.adsabs.harvard.edu/abs/2016MNRAS.460..742M}
}

@ARTICLE{gossan_2020,
       author = {{Gossan}, Sarah E. and {Fuller}, Jim and {Roberts}, Luke F.},
        title = "{Wave heating from proto-neutron star convection and the core-collapse supernova explosion mechanism}",
      journal = {\mnras},
         year = 2020,
        month = feb,
       volume = {491},
       number = {4},
        pages = {5376-5391},
          doi = {10.1093/mnras/stz3243},
archivePrefix = {arXiv},
       eprint = {1910.07599},
 primaryClass = {astro-ph.HE},
       adsurl = {https://ui.adsabs.harvard.edu/abs/2020MNRAS.491.5376G}
}

@ARTICLE{van_dyk_2023,
       author = {{Van Dyk}, Schuyler D. and {Bostroem}, K. Azalee and {Zheng}, WeiKang and {Brink}, Thomas G. and {Fox}, Ori D. and {Andrews}, Jennifer E. and {Filippenko}, Alexei V. and {Dong}, Yize and {Hoang}, Emily and {Hosseinzadeh}, Griffin and {Janzen}, Daryl and {Jencson}, Jacob E. and {Lundquist}, Michael J. and {Meza}, Nicolas and {Milisavljevic}, Dan and {Pearson}, Jeniveve and {Sand}, David J. and {Shrestha}, Manisha and {Valenti}, Stefano and {Howell}, D. Andrew},
        title = "{Identifying the SN 2022acko progenitor with JWST}",
      journal = {\mnras},
         year = 2023,
        month = sep,
       volume = {524},
       number = {2},
        pages = {2186-2194},
          doi = {10.1093/mnras/stad2001},
archivePrefix = {arXiv},
       eprint = {2302.00274},
 primaryClass = {astro-ph.HE},
       adsurl = {https://ui.adsabs.harvard.edu/abs/2023MNRAS.524.2186V}
}

@ARTICLE{tonry_2018,
       author = {{Tonry}, J.~L. and {Denneau}, L. and {Heinze}, A.~N. and {Stalder}, B. and {Smith}, K.~W. and {Smartt}, S.~J. and {Stubbs}, C.~W. and {Weiland}, H.~J. and {Rest}, A.},
        title = "{ATLAS: A High-cadence All-sky Survey System}",
      journal = {\pasp},
         year = 2018,
        month = jun,
       volume = {130},
       number = {988},
        pages = {064505},
          doi = {10.1088/1538-3873/aabadf},
archivePrefix = {arXiv},
       eprint = {1802.00879},
 primaryClass = {astro-ph.IM},
       adsurl = {https://ui.adsabs.harvard.edu/abs/2018PASP..130f4505T}
}

@ARTICLE{shingles_2021,
       author = {{Shingles}, L. and {Smith}, K.~W. and {Young}, D.~R. and {Smartt}, S.~J. and {Tonry}, J. and {Denneau}, L. and {Heinze}, A. and {Weiland}, H. and {Flewelling}, H. and {Stalder}, B. and {Clocchiatti}, A. and {F{\"o}rster}, F. and {Pignata}, G. and {Rest}, A. and {Anderson}, J. and {Stubbs}, C. and {Erasmus}, N.},
        title = "{Release of the ATLAS Forced Photometry server for public use}",
      journal = {Transient Name Server AstroNote},
         year = 2021,
        month = jan,
       volume = {7},
        pages = {1-7},
       adsurl = {https://ui.adsabs.harvard.edu/abs/2021TNSAN...7....1S}
}

@ARTICLE{zhai_2024,
       author = {{Zhai}, Q. and {Li}, L. and {Zhang}, J. and {Wang}, X.},
        title = "{LiONS Transient Classification Report for 2024-04-11}",
      journal = {Transient Name Server Classification Report},
         year = 2024,
        month = apr,
       volume = {2024-1031},
        pages = {1},
       adsurl = {https://ui.adsabs.harvard.edu/abs/2024TNSCR1031....1Z}
}

@ARTICLE{hoogendam_2024,
       author = {{Hoogendam}, W. and {Auchettl}, K. and {Tucker}, M. and {Ashall}, C. and {Shappee}, B. and {Huber}, M.},
        title = "{Early Spectrum of SN 2024ggi from SCAT shows a Type II SN with Flash Ionized Features}",
      journal = {Transient Name Server AstroNote},
         year = 2024,
        month = apr,
       volume = {103},
        pages = {1},
       adsurl = {https://ui.adsabs.harvard.edu/abs/2024TNSAN.103....1H}
}

@ARTICLE{jacobson_galan_2022,
       author = {{Jacobson-Gal{\'a}n}, W.~V. and {Dessart}, L. and {Jones}, D.~O. and {Margutti}, R. and {Coppejans}, D.~L. and {Dimitriadis}, G. and {Foley}, R.~J. and {Kilpatrick}, C.~D. and {Matthews}, D.~J. and {Rest}, S. and {Terreran}, G. and {Aleo}, P.~D. and {Auchettl}, K. and {Blanchard}, P.~K. and {Coulter}, D.~A. and {Davis}, K.~W. and {de Boer}, T.~J.~L. and {DeMarchi}, L. and {Drout}, M.~R. and {Earl}, N. and {Gagliano}, A. and {Gall}, C. and {Hjorth}, J. and {Huber}, M.~E. and {Ibik}, A.~L. and {Milisavljevic}, D. and {Pan}, Y. -C. and {Rest}, A. and {Ridden-Harper}, R. and {Rojas-Bravo}, C. and {Siebert}, M.~R. and {Smith}, K.~W. and {Taggart}, K. and {Tinyanont}, S. and {Wang}, Q. and {Zenati}, Y.},
        title = "{Final Moments. I. Precursor Emission, Envelope Inflation, and Enhanced Mass Loss Preceding the Luminous Type II Supernova 2020tlf}",
      journal = {\apj},
         year = 2022,
        month = jan,
       volume = {924},
       number = {1},
          eid = {15},
        pages = {15},
          doi = {10.3847/1538-4357/ac3f3a},
archivePrefix = {arXiv},
       eprint = {2109.12136},
 primaryClass = {astro-ph.HE},
       adsurl = {https://ui.adsabs.harvard.edu/abs/2022ApJ...924...15J}
}

@ARTICLE{geron_2026,
       author = {{G{\'e}ron}, Tobias and {Drout}, Maria R. and {Jacobson-Gal{\'a}n}, W.~V. and {Kilpatrick}, C.~D.},
        title = "{DETECT: A Pipeline to Quantify Detection Thresholds in Rubin for Nearby Targets Embedded in Bright Host Galaxies}",
      journal = {\aj},
         year = 2026,
        month = jul,
       volume = {172},
       number = {1},
          eid = {45},
        pages = {45},
          doi = {10.3847/1538-3881/ae75ad},
       adsurl = {https://ui.adsabs.harvard.edu/abs/2026AJ....172...45G}
}

@ARTICLE{yaron_2017,
       author = {{Yaron}, O. and {Perley}, D.~A. and {Gal-Yam}, A. and {Groh}, J.~H. and {Horesh}, A. and {Ofek}, E.~O. and {Kulkarni}, S.~R. and {Sollerman}, J. and {Fransson}, C. and {Rubin}, A. and {Szabo}, P. and {Sapir}, N. and {Taddia}, F. and {Cenko}, S.~B. and {Valenti}, S. and {Arcavi}, I. and {Howell}, D.~A. and {Kasliwal}, M.~M. and {Vreeswijk}, P.~M. and {Khazov}, D. and {Fox}, O.~D. and {Cao}, Y. and {Gnat}, O. and {Kelly}, P.~L. and {Nugent}, P.~E. and {Filippenko}, A.~V. and {Laher}, R.~R. and {Wozniak}, P.~R. and {Lee}, W.~H. and {Rebbapragada}, U.~D. and {Maguire}, K. and {Sullivan}, M. and {Soumagnac}, M.~T.},
        title = "{Confined dense circumstellar material surrounding a regular type II supernova}",
      journal = {Nature Physics},
         year = 2017,
        month = feb,
       volume = {13},
       number = {5},
        pages = {510-517},
          doi = {10.1038/nphys4025},
archivePrefix = {arXiv},
       eprint = {1701.02596},
 primaryClass = {astro-ph.HE},
       adsurl = {https://ui.adsabs.harvard.edu/abs/2017NatPh..13..510Y}
}

@ARTICLE{morozova_2018,
       author = {{Morozova}, Viktoriya and {Piro}, Anthony L. and {Valenti}, Stefano},
        title = "{Measuring the Progenitor Masses and Dense Circumstellar Material of Type II Supernovae}",
      journal = {\apj},
         year = 2018,
        month = may,
       volume = {858},
       number = {1},
          eid = {15},
        pages = {15},
          doi = {10.3847/1538-4357/aab9a6},
archivePrefix = {arXiv},
       eprint = {1709.04928},
 primaryClass = {astro-ph.HE},
       adsurl = {https://ui.adsabs.harvard.edu/abs/2018ApJ...858...15M}
}

@ARTICLE{gagliano_2025,
       author = {{Gagliano}, A. and {Berger}, E. and {Villar}, V.~A. and {Hiramatsu}, D. and {Kessler}, R. and {Matsumoto}, T. and {Gilkis}, A. and {Laplace}, E.},
        title = "{Finding the Fuse: Prospects for the Detection and Characterization of Hydrogen-rich Core-collapse Supernova Precursor Emission with the LSST}",
      journal = {\apj},
         year = 2025,
        month = jan,
       volume = {978},
       number = {1},
          eid = {110},
        pages = {110},
          doi = {10.3847/1538-4357/ad9748},
archivePrefix = {arXiv},
       eprint = {2408.13314},
 primaryClass = {astro-ph.HE},
       adsurl = {https://ui.adsabs.harvard.edu/abs/2025ApJ...978..110G}
}

@ARTICLE{davies_2022,
       author = {{Davies}, Ben and {Plez}, Bertrand and {Petrault}, Mike},
        title = "{Explosion imminent: the appearance of red supergiants at the point of core-collapse}",
      journal = {\mnras},
         year = 2022,
        month = nov,
       volume = {517},
       number = {1},
        pages = {1483-1490},
          doi = {10.1093/mnras/stac2427},
archivePrefix = {arXiv},
       eprint = {2208.10883},
 primaryClass = {astro-ph.SR},
       adsurl = {https://ui.adsabs.harvard.edu/abs/2022MNRAS.517.1483D}
}

@ARTICLE{jiang_2025,
       author = {{Jiang}, Nan and {Moon}, Dae-Sik and {Ni}, Yuan Qi and {Drout}, Maria R. and {Park}, Hong Soo and {Gonz{\'a}lez-Gait{\'a}n}, Santiago and {Kim}, Sang Chul and {Lee}, Youngdae and {Chang}, Ernest},
        title = "{Infant Core-collapse Supernovae with Circumstellar Interactions from KMTNet I: Luminous Transitional Case of KSP-SN-2022c}",
      journal = {arXiv e-prints},
         year = 2025,
        month = mar,
          eid = {arXiv:2503.23074},
        pages = {arXiv:2503.23074},
          doi = {10.48550/arXiv.2503.23074},
archivePrefix = {arXiv},
       eprint = {2503.23074},
 primaryClass = {astro-ph.HE},
       adsurl = {https://ui.adsabs.harvard.edu/abs/2025arXiv250323074J}
}

@ARTICLE{dong_2024,
       author = {{Dong}, Yize and {Tsuna}, Daichi and {Valenti}, Stefano and {Sand}, David J. and {Andrews}, Jennifer E. and {Bostroem}, K. Azalee and {Hosseinzadeh}, Griffin and {Hoang}, Emily and {Jha}, Saurabh W. and {Janzen}, Daryl and {Jencson}, Jacob E. and {Lundquist}, Michael and {Mehta}, Darshana and {Ravi}, Aravind P. and {Meza Retamal}, Nicolas E. and {Pearson}, Jeniveve and {Shrestha}, Manisha and {Bonanos}, Alceste Z. and {Howell}, D. Andrew and {Smith}, Nathan and {Farah}, Joseph and {Hiramatsu}, Daichi and {Itagaki}, Koichi and {McCully}, Curtis and {Newsome}, Megan and {Padilla Gonzalez}, Estefania and {Paraskeva}, Emmanouela and {Pellegrino}, Craig and {Terreran}, Giacomo and {Haislip}, Joshua and {Kouprianov}, Vladimir and {Reichart}, Daniel E.},
        title = "{SN2023fyq: A Type Ibn Supernova with Long-standing Precursor Activity Due to Binary Interaction}",
      journal = {\apj},
         year = 2024,
        month = dec,
       volume = {977},
       number = {2},
          eid = {254},
        pages = {254},
          doi = {10.3847/1538-4357/ad8de6},
archivePrefix = {arXiv},
       eprint = {2405.04583},
 primaryClass = {astro-ph.HE},
       adsurl = {https://ui.adsabs.harvard.edu/abs/2024ApJ...977..254D}
}

@ARTICLE{quataert_2012,
       author = {{Quataert}, E. and {Shiode}, J.},
        title = "{Wave-driven mass loss in the last year of stellar evolution: setting the stage for the most luminous core-collapse supernovae}",
      journal = {\mnras},
         year = 2012,
        month = jun,
       volume = {423},
       number = {1},
        pages = {L92-L96},
          doi = {10.1111/j.1745-3933.2012.01264.x},
archivePrefix = {arXiv},
       eprint = {1202.5036},
 primaryClass = {astro-ph.SR},
       adsurl = {https://ui.adsabs.harvard.edu/abs/2012MNRAS.423L..92Q}
}

@ARTICLE{shiode_2014,
       author = {{Shiode}, Joshua H. and {Quataert}, Eliot},
        title = "{Setting the Stage for Circumstellar Interaction in Core-Collapse Supernovae. II. Wave-driven Mass Loss in Supernova Progenitors}",
      journal = {\apj},
         year = 2014,
        month = jan,
       volume = {780},
       number = {1},
          eid = {96},
        pages = {96},
          doi = {10.1088/0004-637X/780/1/96},
archivePrefix = {arXiv},
       eprint = {1308.5978},
 primaryClass = {astro-ph.SR},
       adsurl = {https://ui.adsabs.harvard.edu/abs/2014ApJ...780...96S}
}

@ARTICLE{takei_2024,
       author = {{Takei}, Yuki and {Tsuna}, Daichi and {Ko}, Takatoshi and {Shigeyama}, Toshikazu},
        title = "{Simulating Hydrogen-poor Interaction-powered Supernovae with CHIPS}",
      journal = {\apj},
         year = 2024,
        month = jan,
       volume = {961},
       number = {1},
          eid = {67},
        pages = {67},
          doi = {10.3847/1538-4357/ad0da4},
archivePrefix = {arXiv},
       eprint = {2308.10785},
 primaryClass = {astro-ph.HE},
       adsurl = {https://ui.adsabs.harvard.edu/abs/2024ApJ...961...67T}
}

@ARTICLE{kuriyama_2020,
       author = {{Kuriyama}, Naoto and {Shigeyama}, Toshikazu},
        title = "{Radiation hydrodynamical simulations of eruptive mass loss from progenitors of Type Ibn/IIn supernovae}",
      journal = {\aap},
         year = 2020,
        month = mar,
       volume = {635},
          eid = {A127},
        pages = {A127},
          doi = {10.1051/0004-6361/201937226},
archivePrefix = {arXiv},
       eprint = {1912.09738},
 primaryClass = {astro-ph.SR},
       adsurl = {https://ui.adsabs.harvard.edu/abs/2020A&A...635A.127K}
}

@ARTICLE{jacobson_galan_2025,
       author = {{Jacobson-Gal{\'a}n}, W.~V. and {Gonzalez}, S. and {Patel}, S. and {Dessart}, L. and {Jones}, D.~O. and {Coppejans}, D.~L. and {Dimitriadis}, G. and {Foley}, R.~J. and {Kilpatrick}, C.~D. and {Matthews}, D.~J. and {Rest}, S. and {Terreran}, G. and {Aleo}, P.~D. and {Auchettl}, K. and {Blanchard}, P.~K. and {Coulter}, D.~A. and {Davis}, K.~W. and {de Boer}, T.~J.~L. and {DeMarchi}, L. and {Drout}, M.~R. and {Earl}, N. and {Gagliano}, A. and {Gall}, C. and {Hjorth}, J. and {Huber}, M.~E. and {Ibik}, A.~L. and {Milisavljevic}, D. and {Pan}, Y. -C. and {Rest}, A. and {Ridden-Harper}, R. and {Rojas-Bravo}, C. and {Siebert}, M.~R. and {Smith}, K.~W. and {Taggart}, K. and {Tinyanont}, S. and {Wang}, Q. and {Zenati}, Y.},
        title = "{An Updated Detection Pipeline for Precursor Emission in Type II Supernova 2020tlf}",
      journal = {Research Notes of the American Astronomical Society},
         year = 2025,
        month = jan,
       volume = {9},
       number = {1},
          eid = {5},
        pages = {5},
          doi = {10.3847/2515-5172/ada367},
       adsurl = {https://ui.adsabs.harvard.edu/abs/2025RNAAS...9....5J}
}

@ARTICLE{jacobson_galan_2023,
       author = {{Jacobson-Gal{\'a}n}, W.~V. and {Dessart}, L. and {Margutti}, R. and {Chornock}, R. and {Foley}, R.~J. and {Kilpatrick}, C.~D. and {Jones}, D.~O. and {Taggart}, K. and {Angus}, C.~R. and {Bhattacharjee}, S. and {Braff}, L.~A. and {Brethauer}, D. and {Burgasser}, A.~J. and {Cao}, F. and {Carlile}, C.~M. and {Chambers}, K.~C. and {Coulter}, D.~A. and {Dominguez-Ruiz}, E. and {Dickinson}, C.~B. and {de Boer}, T. and {Gagliano}, A. and {Gall}, C. and {Gao}, H. and {Gates}, E.~L. and {Gomez}, S. and {Guolo}, M. and {Halford}, M.~R.~J. and {Hjorth}, J. and {Huber}, M.~E. and {Johnson}, M.~N. and {Karpoor}, P.~R. and {Laskar}, T. and {LeBaron}, N. and {Li}, Z. and {Lin}, Y. and {Loch}, S.~D. and {Lynam}, P.~D. and {Magnier}, E.~A. and {Maloney}, P. and {Matthews}, D.~J. and {McDonald}, M. and {Miao}, H. -Y. and {Milisavljevic}, D. and {Pan}, Y. -C. and {Pradyumna}, S. and {Ransome}, C.~L. and {Rees}, J.~M. and {Rest}, A. and {Rojas-Bravo}, C. and {Sandford}, N.~R. and {Ascencio}, L. Sandoval and {Sanjaripour}, S. and {Savino}, A. and {Sears}, H. and {Sharei}, N. and {Smartt}, S.~J. and {Softich}, E.~R. and {Theissen}, C.~A. and {Tinyanont}, S. and {Tohfa}, H. and {Villar}, V.~A. and {Wang}, Q. and {Wainscoat}, R.~J. and {Westerling}, A.~L. and {Wiston}, E. and {Wozniak}, M.~A. and {Yadavalli}, S.~K. and {Zenati}, Y.},
        title = "{SN 2023ixf in Messier 101: Photo-ionization of Dense, Close-in Circumstellar Material in a Nearby Type II Supernova}",
      journal = {\apjl},
         year = 2023,
        month = sep,
       volume = {954},
       number = {2},
          eid = {L42},
        pages = {L42},
          doi = {10.3847/2041-8213/acf2ec},
archivePrefix = {arXiv},
       eprint = {2306.04721},
 primaryClass = {astro-ph.HE},
       adsurl = {https://ui.adsabs.harvard.edu/abs/2023ApJ...954L..42J}
}

@ARTICLE{jacobson_galan_2024,
       author = {{Jacobson-Gal{\'a}n}, W.~V. and {Davis}, K.~W. and {Kilpatrick}, C.~D. and {Dessart}, L. and {Margutti}, R. and {Chornock}, R. and {Foley}, R.~J. and {Arunachalam}, P. and {Auchettl}, K. and {Bom}, C.~R. and {Cartier}, R. and {Coulter}, D.~A. and {Dimitriadis}, G. and {Dickinson}, D. and {Drout}, M.~R. and {Gagliano}, A.~T. and {Gall}, C. and {Garretson}, B. and {Izzo}, L. and {Jones}, D.~O. and {LeBaron}, N. and {Miao}, H. -Y. and {Milisavljevic}, D. and {Pan}, Y. -C. and {Rest}, A. and {Rojas-Bravo}, C. and {Santos}, A. and {Sears}, H. and {Subrayan}, B.~M. and {Taggart}, K. and {Tinyanont}, S.},
        title = "{SN 2024ggi in NGC 3621: Rising Ionization in a Nearby, Circumstellar-material-interacting Type II Supernova}",
      journal = {\apj},
         year = 2024,
        month = sep,
       volume = {972},
       number = {2},
          eid = {177},
        pages = {177},
          doi = {10.3847/1538-4357/ad5c64},
archivePrefix = {arXiv},
       eprint = {2404.19006},
 primaryClass = {astro-ph.HE},
       adsurl = {https://ui.adsabs.harvard.edu/abs/2024ApJ...972..177J}
}

@ARTICLE{bruch_2021,
       author = {{Bruch}, Rachel J. and {Gal-Yam}, Avishay and {Schulze}, Steve and {Yaron}, Ofer and {Yang}, Yi and {Soumagnac}, Maayane and {Rigault}, Mickael and {Strotjohann}, Nora L. and {Ofek}, Eran and {Sollerman}, Jesper and {Masci}, Frank J. and {Barbarino}, Cristina and {Ho}, Anna Y.~Q. and {Fremling}, Christoffer and {Perley}, Daniel and {Nordin}, Jakob and {Cenko}, S. Bradley and {Adams}, S. and {Adreoni}, Igor and {Bellm}, Eric C. and {Blagorodnova}, Nadia and {Bulla}, Mattia and {Burdge}, Kevin and {De}, Kishalay and {Dhawan}, Suhail and {Drake}, Andrew J. and {Duev}, Dmitry A. and {Dugas}, Alison and {Graham}, Matthew and {Graham}, Melissa L. and {Irani}, Ido and {Jencson}, Jacob and {Karamehmetoglu}, Emir and {Kasliwal}, Mansi and {Kim}, Young-Lo and {Kulkarni}, Shrinivas and {Kupfer}, Thomas and {Liang}, Jingyi and {Mahabal}, Ashish and {Miller}, A.~A. and {Prince}, Thomas A. and {Riddle}, Reed and {Sharma}, Y. and {Smith}, Roger and {Taddia}, Francesco and {Taggart}, Kirsty and {Walters}, Richard and {Yan}, Lin},
        title = "{A Large Fraction of Hydrogen-rich Supernova Progenitors Experience Elevated Mass Loss Shortly Prior to Explosion}",
      journal = {\apj},
         year = 2021,
        month = may,
       volume = {912},
       number = {1},
          eid = {46},
        pages = {46},
          doi = {10.3847/1538-4357/abef05},
archivePrefix = {arXiv},
       eprint = {2008.09986},
 primaryClass = {astro-ph.HE},
       adsurl = {https://ui.adsabs.harvard.edu/abs/2021ApJ...912...46B}
}

@ARTICLE{bruch_2023,
       author = {{Bruch}, Rachel J. and {Gal-Yam}, Avishay and {Yaron}, Ofer and {Chen}, Ping and {Strotjohann}, Nora L. and {Irani}, Ido and {Zimmerman}, Erez and {Schulze}, Steve and {Yang}, Yi and {Kim}, Young-Lo and {Bulla}, Mattia and {Sollerman}, Jesper and {Rigault}, Mickael and {Ofek}, Eran and {Soumagnac}, Maayane and {Masci}, Frank J. and {Fremling}, Christoffer and {Perley}, Daniel and {Nordin}, Jakob and {Cenko}, S. Bradley and {Ho}, Anna Y.~Q. and {Adams}, S. and {Adreoni}, Igor and {Bellm}, Eric C. and {Blagorodnova}, Nadia and {Burdge}, Kevin and {De}, Kishalay and {Dekany}, Richard G. and {Dhawan}, Suhail and {Drake}, Andrew J. and {Duev}, Dmitry A. and {Graham}, Matthew and {Graham}, Melissa L. and {Jencson}, Jacob and {Karamehmetoglu}, Emir and {Kasliwal}, Mansi M. and {Kulkarni}, Shrinivas and {Miller}, A.~A. and {Neill}, James D. and {Prince}, Thomas A. and {Riddle}, Reed and {Rusholme}, Benjamin and {Sharma}, Y. and {Smith}, Roger and {Sravan}, Niharika and {Taggart}, Kirsty and {Walters}, Richard and {Yan}, Lin},
        title = "{The Prevalence and Influence of Circumstellar Material around Hydrogen-rich Supernova Progenitors}",
      journal = {\apj},
         year = 2023,
        month = aug,
       volume = {952},
       number = {2},
          eid = {119},
        pages = {119},
          doi = {10.3847/1538-4357/acd8be},
archivePrefix = {arXiv},
       eprint = {2212.03313},
 primaryClass = {astro-ph.HE},
       adsurl = {https://ui.adsabs.harvard.edu/abs/2023ApJ...952..119B}
}

@ARTICLE{bellm_2019,
       author = {{Bellm}, Eric C. and {Kulkarni}, Shrinivas R. and {Graham}, Matthew J. and {Dekany}, Richard and {Smith}, Roger M. and {Riddle}, Reed and {Masci}, Frank J. and {Helou}, George and {Prince}, Thomas A. and {Adams}, Scott M. and {Barbarino}, C. and {Barlow}, Tom and {Bauer}, James and {Beck}, Ron and {Belicki}, Justin and {Biswas}, Rahul and {Blagorodnova}, Nadejda and {Bodewits}, Dennis and {Bolin}, Bryce and {Brinnel}, Valery and {Brooke}, Tim and {Bue}, Brian and {Bulla}, Mattia and {Burruss}, Rick and {Cenko}, S. Bradley and {Chang}, Chan-Kao and {Connolly}, Andrew and {Coughlin}, Michael and {Cromer}, John and {Cunningham}, Virginia and {De}, Kishalay and {Delacroix}, Alex and {Desai}, Vandana and {Duev}, Dmitry A. and {Eadie}, Gwendolyn and {Farnham}, Tony L. and {Feeney}, Michael and {Feindt}, Ulrich and {Flynn}, David and {Franckowiak}, Anna and {Frederick}, S. and {Fremling}, C. and {Gal-Yam}, Avishay and {Gezari}, Suvi and {Giomi}, Matteo and {Goldstein}, Daniel A. and {Golkhou}, V. Zach and {Goobar}, Ariel and {Groom}, Steven and {Hacopians}, Eugean and {Hale}, David and {Henning}, John and {Ho}, Anna Y.~Q. and {Hover}, David and {Howell}, Justin and {Hung}, Tiara and {Huppenkothen}, Daniela and {Imel}, David and {Ip}, Wing-Huen and {Ivezi{\'c}}, {\v{Z}}eljko and {Jackson}, Edward and {Jones}, Lynne and {Juric}, Mario and {Kasliwal}, Mansi M. and {Kaspi}, S. and {Kaye}, Stephen and {Kelley}, Michael S.~P. and {Kowalski}, Marek and {Kramer}, Emily and {Kupfer}, Thomas and {Landry}, Walter and {Laher}, Russ R. and {Lee}, Chien-De and {Lin}, Hsing Wen and {Lin}, Zhong-Yi and {Lunnan}, Ragnhild and {Giomi}, Matteo and {Mahabal}, Ashish and {Mao}, Peter and {Miller}, Adam A. and {Monkewitz}, Serge and {Murphy}, Patrick and {Ngeow}, Chow-Choong and {Nordin}, Jakob and {Nugent}, Peter and {Ofek}, Eran and {Patterson}, Maria T. and {Penprase}, Bryan and {Porter}, Michael and {Rauch}, Ludwig and {Rebbapragada}, Umaa and {Reiley}, Dan and {Rigault}, Mickael and {Rodriguez}, Hector and {van Roestel}, Jan and {Rusholme}, Ben and {van Santen}, Jakob and {Schulze}, S. and {Shupe}, David L. and {Singer}, Leo P. and {Soumagnac}, Maayane T. and {Stein}, Robert and {Surace}, Jason and {Sollerman}, Jesper and {Szkody}, Paula and {Taddia}, F. and {Terek}, Scott and {Van Sistine}, Angela and {van Velzen}, Sjoert and {Vestrand}, W. Thomas and {Walters}, Richard and {Ward}, Charlotte and {Ye}, Quan-Zhi and {Yu}, Po-Chieh and {Yan}, Lin and {Zolkower}, Jeffry},
        title = "{The Zwicky Transient Facility: System Overview, Performance, and First Results}",
      journal = {\pasp},
         year = 2019,
        month = jan,
       volume = {131},
       number = {995},
        pages = {018002},
          doi = {10.1088/1538-3873/aaecbe},
archivePrefix = {arXiv},
       eprint = {1902.01932},
 primaryClass = {astro-ph.IM},
       adsurl = {https://ui.adsabs.harvard.edu/abs/2019PASP..131a8002B}
}

@ARTICLE{goto_2025,
       author = {{Goto}, Sota and {Yamanaka}, Masayuki and {Nagayama}, Takahiro and {Maeda}, Keiichi and {Kawabata}, Miho and {Sahu}, D.~K. and {Singh}, Avinash and {Gangopadhyay}, Anjasha and {Dkuniya}, Naveen and {Misra}, Kuntal and {Dubey}, Monalisa and {Ailawadhi}, Bhuvya},
        title = "{SN 2023vbg: A Type IIn Supernova Resembling SN 2009ip, with a Long-Duration Precursor and Early-Time Bump}",
      journal = {arXiv e-prints},
         year = 2025,
        month = apr,
          eid = {arXiv:2504.15988},
        pages = {arXiv:2504.15988},
          doi = {10.48550/arXiv.2504.15988},
archivePrefix = {arXiv},
       eprint = {2504.15988},
 primaryClass = {astro-ph.HE},
       adsurl = {https://ui.adsabs.harvard.edu/abs/2025arXiv250415988G}
}

@ARTICLE{fraser_2013,
       author = {{Fraser}, M. and {Magee}, M. and {Kotak}, R. and {Smartt}, S.~J. and {Smith}, K.~W. and {Polshaw}, J. and {Drake}, A.~J. and {Boles}, T. and {Lee}, C. -H. and {Burgett}, W.~S. and {Chambers}, K.~C. and {Draper}, P.~W. and {Flewelling}, H. and {Hodapp}, K.~W. and {Kaiser}, N. and {Kudritzki}, R. -P. and {Magnier}, E.~A. and {Price}, P.~A. and {Tonry}, J.~L. and {Wainscoat}, R.~J. and {Waters}, C.},
        title = "{Detection of an Outburst One Year Prior to the Explosion of SN 2011ht}",
      journal = {\apjl},
         year = 2013,
        month = dec,
       volume = {779},
       number = {1},
          eid = {L8},
        pages = {L8},
          doi = {10.1088/2041-8205/779/1/L8},
archivePrefix = {arXiv},
       eprint = {1309.4695},
 primaryClass = {astro-ph.SR},
       adsurl = {https://ui.adsabs.harvard.edu/abs/2013ApJ...779L...8F}
}

@ARTICLE{law_2009,
       author = {{Law}, Nicholas M. and {Kulkarni}, Shrinivas R. and {Dekany}, Richard G. and {Ofek}, Eran O. and {Quimby}, Robert M. and {Nugent}, Peter E. and {Surace}, Jason and {Grillmair}, Carl C. and {Bloom}, Joshua S. and {Kasliwal}, Mansi M. and {Bildsten}, Lars and {Brown}, Tim and {Cenko}, S. Bradley and {Ciardi}, David and {Croner}, Ernest and {Djorgovski}, S. George and {van Eyken}, Julian and {Filippenko}, Alexei V. and {Fox}, Derek B. and {Gal-Yam}, Avishay and {Hale}, David and {Hamam}, Nouhad and {Helou}, George and {Henning}, John and {Howell}, D. Andrew and {Jacobsen}, Janet and {Laher}, Russ and {Mattingly}, Sean and {McKenna}, Dan and {Pickles}, Andrew and {Poznanski}, Dovi and {Rahmer}, Gustavo and {Rau}, Arne and {Rosing}, Wayne and {Shara}, Michael and {Smith}, Roger and {Starr}, Dan and {Sullivan}, Mark and {Velur}, Viswa and {Walters}, Richard and {Zolkower}, Jeff},
        title = "{The Palomar Transient Factory: System Overview, Performance, and First Results}",
      journal = {\pasp},
         year = 2009,
        month = dec,
       volume = {121},
       number = {886},
        pages = {1395},
          doi = {10.1086/648598},
archivePrefix = {arXiv},
       eprint = {0906.5350},
 primaryClass = {astro-ph.IM},
       adsurl = {https://ui.adsabs.harvard.edu/abs/2009PASP..121.1395L}
}

@ARTICLE{ofek_2014,
       author = {{Ofek}, Eran O. and {Sullivan}, Mark and {Shaviv}, Nir J. and {Steinbok}, Aviram and {Arcavi}, Iair and {Gal-Yam}, Avishay and {Tal}, David and {Kulkarni}, Shrinivas R. and {Nugent}, Peter E. and {Ben-Ami}, Sagi and {Kasliwal}, Mansi M. and {Cenko}, S. Bradley and {Laher}, Russ and {Surace}, Jason and {Bloom}, Joshua S. and {Filippenko}, Alexei V. and {Silverman}, Jeffrey M. and {Yaron}, Ofer},
        title = "{Precursors Prior to Type IIn Supernova Explosions are Common: Precursor Rates, Properties, and Correlations}",
      journal = {\apj},
         year = 2014,
        month = jul,
       volume = {789},
       number = {2},
          eid = {104},
        pages = {104},
          doi = {10.1088/0004-637X/789/2/104},
archivePrefix = {arXiv},
       eprint = {1401.5468},
 primaryClass = {astro-ph.HE},
       adsurl = {https://ui.adsabs.harvard.edu/abs/2014ApJ...789..104O}
}

@ARTICLE{ofek_2013,
       author = {{Ofek}, E.~O. and {Sullivan}, M. and {Cenko}, S.~B. and {Kasliwal}, M.~M. and {Gal-Yam}, A. and {Kulkarni}, S.~R. and {Arcavi}, I. and {Bildsten}, L. and {Bloom}, J.~S. and {Horesh}, A. and {Howell}, D.~A. and {Filippenko}, A.~V. and {Laher}, R. and {Murray}, D. and {Nakar}, E. and {Nugent}, P.~E. and {Silverman}, J.~M. and {Shaviv}, N.~J. and {Surace}, J. and {Yaron}, O.},
        title = "{An outburst from a massive star 40 days before a supernova explosion}",
      journal = {\nat},
         year = 2013,
        month = feb,
       volume = {494},
       number = {7435},
        pages = {65-67},
          doi = {10.1038/nature11877},
archivePrefix = {arXiv},
       eprint = {1302.2633},
 primaryClass = {astro-ph.HE},
       adsurl = {https://ui.adsabs.harvard.edu/abs/2013Natur.494...65O}
}

@Article{numpy_2020,
    title         = {Array programming with {NumPy}},
     author        = {Charles R. Harris and K. Jarrod Millman and St{\'{e}}fan J.
         van der Walt and Ralf Gommers and Pauli Virtanen and David
         Cournapeau and Eric Wieser and Julian Taylor and Sebastian
         Berg and Nathaniel J. Smith and Robert Kern and Matti Picus
         and Stephan Hoyer and Marten H. van Kerkwijk and Matthew
         Brett and Allan Haldane and Jaime Fern{\'{a}}ndez del
         R{\'{i}}o and Mark Wiebe and Pearu Peterson and Pierre
         G{\'{e}}rard-Marchant and Kevin Sheppard and Tyler Reddy and
         Warren Weckesser and Hameer Abbasi and Christoph Gohlke and
         Travis E. Oliphant},
    year          = {2020},
     month         = sep,
     journal       = {Nature},
     volume        = {585},
     number        = {7825},
     pages         = {357--362},
     doi           = {10.1038/s41586-020-2649-2},
     publisher     = {Springer Science and Business Media {LLC}},
     url           = {https://doi.org/10.1038/s41586-020-2649-2}
}

@ARTICLE{scipy_2020,
  author  = {Virtanen, Pauli and Gommers, Ralf and Oliphant, Travis E. and
            Haberland, Matt and Reddy, Tyler and Cournapeau, David and
            Burovski, Evgeni and Peterson, Pearu and Weckesser, Warren and
            Bright, Jonathan and {van der Walt}, St{\'e}fan J. and
            Brett, Matthew and Wilson, Joshua and Millman, K. Jarrod and
            Mayorov, Nikolay and Nelson, Andrew R. J. and Jones, Eric and
            Kern, Robert and Larson, Eric and Carey, C J and
            Polat, {\.I}lhan and Feng, Yu and Moore, Eric W. and
            {VanderPlas}, Jake and Laxalde, Denis and Perktold, Josef and
            Cimrman, Robert and Henriksen, Ian and Quintero, E. A. and
            Harris, Charles R. and Archibald, Anne M. and
            Ribeiro, Ant{\^o}nio H. and Pedregosa, Fabian and
            {van Mulbregt}, Paul and {SciPy 1.0 Contributors}},
  title   = {{{SciPy} 1.0: Fundamental Algorithms for Scientific
            Computing in Python}},
  journal = {Nature Methods},
  year    = {2020},
  volume  = {17},
  pages   = {261--272},
  adsurl  = {https://rdcu.be/b08Wh},
  doi     = {10.1038/s41592-019-0686-2},
}

@ARTICLE{johnson_2018,
       author = {{Johnson}, Samson A. and {Kochanek}, C.~S. and {Adams}, S.~M.},
        title = "{The quiescent progenitors of four Type II-P/L supernovae}",
      journal = {\mnras},
         year = 2018,
        month = oct,
       volume = {480},
       number = {2},
        pages = {1696-1704},
          doi = {10.1093/mnras/sty1966},
archivePrefix = {arXiv},
       eprint = {1712.03957},
 primaryClass = {astro-ph.SR},
       adsurl = {https://ui.adsabs.harvard.edu/abs/2018MNRAS.480.1696J}
}

@ARTICLE{kochanek_2017,
       author = {{Kochanek}, C.~S. and {Fraser}, M. and {Adams}, S.~M. and {Sukhbold}, T. and {Prieto}, J.~L. and {M{\"u}ller}, T. and {Bock}, G. and {Brown}, J.~S. and {Dong}, Subo and {Holoien}, T.~W. -S. and {Khan}, R. and {Shappee}, B.~J. and {Stanek}, K.~Z.},
        title = "{Supernova progenitors, their variability and the Type IIP Supernova ASASSN-16fq in M66}",
      journal = {\mnras},
         year = 2017,
        month = may,
       volume = {467},
       number = {3},
        pages = {3347-3360},
          doi = {10.1093/mnras/stx291},
archivePrefix = {arXiv},
       eprint = {1609.00022},
 primaryClass = {astro-ph.SR},
       adsurl = {https://ui.adsabs.harvard.edu/abs/2017MNRAS.467.3347K}
}

@ARTICLE{kiewe_2012,
       author = {{Kiewe}, Michael and {Gal-Yam}, Avishay and {Arcavi}, Iair and {Leonard}, Douglas C. and {Emilio Enriquez}, J. and {Cenko}, S. Bradley and {Fox}, Derek B. and {Moon}, Dae-Sik and {Sand}, David J. and {Soderberg}, Alicia M. and {CCCP}, The},
        title = "{Caltech Core-Collapse Project (CCCP) Observations of Type IIn Supernovae: Typical Properties and Implications for Their Progenitor Stars}",
      journal = {\apj},
         year = 2012,
        month = jan,
       volume = {744},
       number = {1},
          eid = {10},
        pages = {10},
          doi = {10.1088/0004-637X/744/1/10},
archivePrefix = {arXiv},
       eprint = {1010.2689},
 primaryClass = {astro-ph.CO},
       adsurl = {https://ui.adsabs.harvard.edu/abs/2012ApJ...744...10K}
}

@ARTICLE{strotjohann_2021,
       author = {{Strotjohann}, Nora L. and {Ofek}, Eran O. and {Gal-Yam}, Avishay and {Bruch}, Rachel and {Schulze}, Steve and {Shaviv}, Nir and {Sollerman}, Jesper and {Filippenko}, Alexei V. and {Yaron}, Ofer and {Fremling}, Christoffer and {Nordin}, Jakob and {Kool}, Erik C. and {Perley}, Dan A. and {Ho}, Anna Y.~Q. and {Yang}, Yi and {Yao}, Yuhan and {Soumagnac}, Maayane T. and {Graham}, Melissa L. and {Barbarino}, Cristina and {Tartaglia}, Leonardo and {De}, Kishalay and {Goldstein}, Daniel A. and {Cook}, David O. and {Brink}, Thomas G. and {Taggart}, Kirsty and {Yan}, Lin and {Lunnan}, Ragnhild and {Kasliwal}, Mansi and {Kulkarni}, Shri R. and {Nugent}, Peter E. and {Masci}, Frank J. and {Rosnet}, Philippe and {Adams}, Scott M. and {Andreoni}, Igor and {Bagdasaryan}, Ashot and {Bellm}, Eric C. and {Burdge}, Kevin and {Duev}, Dmitry A. and {Dugas}, Alison and {Frederick}, Sara and {Goldwasser}, Samantha and {Hankins}, Matthew and {Irani}, Ido and {Karambelkar}, Viraj and {Kupfer}, Thomas and {Liang}, Jingyi and {Neill}, James D. and {Porter}, Michael and {Riddle}, Reed L. and {Sharma}, Yashvi and {Short}, Phil and {Taddia}, Francesco and {Tzanidakis}, Anastasios and {van Roestel}, Jan and {Walters}, Richard and {Zhuang}, Zhuyun},
        title = "{Bright, Months-long Stellar Outbursts Announce the Explosion of Interaction-powered Supernovae}",
      journal = {\apj},
         year = 2021,
        month = feb,
       volume = {907},
       number = {2},
          eid = {99},
        pages = {99},
          doi = {10.3847/1538-4357/abd032},
archivePrefix = {arXiv},
       eprint = {2010.11196},
 primaryClass = {astro-ph.HE},
       adsurl = {https://ui.adsabs.harvard.edu/abs/2021ApJ...907...99S}
}

@ARTICLE{werner_2004,
       author = {{Werner}, M.~W. and {Roellig}, T.~L. and {Low}, F.~J. and {Rieke}, G.~H. and {Rieke}, M. and {Hoffmann}, W.~F. and {Young}, E. and {Houck}, J.~R. and {Brandl}, B. and {Fazio}, G.~G. and {Hora}, J.~L. and {Gehrz}, R.~D. and {Helou}, G. and {Soifer}, B.~T. and {Stauffer}, J. and {Keene}, J. and {Eisenhardt}, P. and {Gallagher}, D. and {Gautier}, T.~N. and {Irace}, W. and {Lawrence}, C.~R. and {Simmons}, L. and {Van Cleve}, J.~E. and {Jura}, M. and {Wright}, E.~L. and {Cruikshank}, D.~P.},
        title = "{The Spitzer Space Telescope Mission}",
      journal = {\apjs},
         year = 2004,
        month = sep,
       volume = {154},
       number = {1},
        pages = {1-9},
          doi = {10.1086/422992},
archivePrefix = {arXiv},
       eprint = {astro-ph/0406223},
 primaryClass = {astro-ph},
       adsurl = {https://ui.adsabs.harvard.edu/abs/2004ApJS..154....1W}
}

@ARTICLE{holtzman_1995,
       author = {{Holtzman}, Jon A. and {Hester}, J. Jeff and {Casertano}, Stefano and {Trauger}, John T. and {Watson}, Alan M. and {Ballester}, Gilda E. and {Burrows}, Christopher J. and {Clarke}, John T. and {Crisp}, David and {Evans}, Robin W. and {Gallagher}, III, John S. and {Griffiths}, Richard E. and {Hoessel}, John G. and {Matthews}, Lynn D. and {Mould}, Jeremy R. and {Scowen}, Paul A. and {Stapelfeldt}, Karl R. and {Westphal}, James A.},
        title = "{The Performance and Calibration of WFPC2 on the Hubble Space Telescope}",
      journal = {\pasp},
         year = 1995,
        month = feb,
       volume = {107},
        pages = {156},
          doi = {10.1086/133533},
       adsurl = {https://ui.adsabs.harvard.edu/abs/1995PASP..107..156H}
}

@Article{matplotlib_2007,
  Author    = {Hunter, J. D.},
  Title     = {Matplotlib: A 2D graphics environment},
  Journal   = {Computing in Science \& Engineering},
  Volume    = {9},
  Number    = {3},
  Pages     = {90--95},
  publisher = {IEEE COMPUTER SOC},
  doi       = {10.1109/MCSE.2007.55},
  year      = 2007
}

@ARTICLE{saha_2006,
       author = {{Saha}, A. and {Thim}, F. and {Tammann}, G.~A. and {Reindl}, B. and {Sandage}, A.},
        title = "{Cepheid Distances to SNe Ia Host Galaxies Based on a Revised Photometric Zero Point of the HST WFPC2 and New PL Relations and Metallicity Corrections}",
      journal = {\apjs},
         year = 2006,
        month = jul,
       volume = {165},
       number = {1},
        pages = {108-137},
          doi = {10.1086/503800},
archivePrefix = {arXiv},
       eprint = {astro-ph/0602572},
 primaryClass = {astro-ph},
       adsurl = {https://ui.adsabs.harvard.edu/abs/2006ApJS..165..108S}
}

@ARTICLE{chen_2024,
       author = {{Chen}, Xinlei and {Kumar}, Brajesh and {Er}, Xinzhong and {Guo}, Helong and {Yang}, Yuan-Pei and {Lin}, Weikang and {Fang}, Yuan and {Du}, Guowang and {Liu}, Chenxu and {Zhao}, Jiewei and {Zhang}, Tianyu and {Bao}, Yuxi and {Zou}, Xingzhu and {Pan}, Yu and {Wang}, Yu and {Zhu}, Xufeng and {Chatterjee}, Kaushik and {Liu}, Xiangkun and {Liu}, Dezi and {Lagioia}, Edoardo P. and {Rangwal}, Geeta and {Zhong}, Shiyan and {Zhang}, Jinghua and {Lian}, Jianhui and {Cai}, Yongzhi and {Zhang}, Yangwei and {Liu}, Xiaowei},
        title = "{Early-phase Simultaneous Multiband Observations of the Type II Supernova SN 2024ggi with Mephisto}",
      journal = {\apjl},
         year = 2024,
        month = aug,
       volume = {971},
       number = {1},
          eid = {L2},
        pages = {L2},
          doi = {10.3847/2041-8213/ad62f7},
archivePrefix = {arXiv},
       eprint = {2405.07964},
 primaryClass = {astro-ph.HE},
       adsurl = {https://ui.adsabs.harvard.edu/abs/2024ApJ...971L...2C}
}

@ARTICLE{chen_2024b,
       author = {{Chen}, Ting-Wan and {Yang}, Sheng and {Srivastav}, Shubham and {Moriya}, Takashi J. and {Smartt}, Stephen J. and {Rest}, Sofia and {Rest}, Armin and {Lin}, Hsing Wen and {Miao}, Hao-Yu and {Cheng}, Yu-Chi and {Aryan}, Amar and {Cheng}, Chia-Yu and {Fraser}, Morgan and {Huang}, Li-Ching and {Lee}, Meng-Han and {Lai}, Cheng-Han and {Liu}, Yu Hsuan and {Sankar. K}, Aiswarya and {Smith}, Ken W. and {Stevance}, Heloise F. and {Wang}, Ze-Ning and {Anderson}, Joseph P. and {Angus}, Charlotte R. and {de Boer}, Thomas and {Chambers}, Kenneth and {Duan}, Hao-Yuan and {Erasmus}, Nicolas and {Gao}, Hua and {Herman}, Joanna and {Hou}, Wei-Jie and {Hsiao}, Hsiang-Yao and {Huber}, Mark E. and {Lin}, Chien-Cheng and {Lin}, Hung-Chin and {Magnier}, Eugene A. and {Kit Man}, Ka and {Moore}, Thomas and {Ngeow}, Chow-Choong and {Nicholl}, Matt and {Ou}, Po-Sheng and {Pignata}, Giuliano and {Shiau}, Yu-Chien and {Silvester Sommer}, Julian and {Tonry}, John L. and {Wang}, Xiao-Feng and {Young}, David R. and {Yeh}, You-Ting and {Zhang}, Jujia},
        title = "{Discovery and Extensive Follow-Up of SN 2024ggi, a nearby type IIP supernova in NGC 3621}",
      journal = {arXiv e-prints},
         year = 2024,
        month = jun,
          eid = {arXiv:2406.09270},
        pages = {arXiv:2406.09270},
          doi = {10.48550/arXiv.2406.09270},
archivePrefix = {arXiv},
       eprint = {2406.09270},
 primaryClass = {astro-ph.HE},
       adsurl = {https://ui.adsabs.harvard.edu/abs/2024arXiv240609270C}
}

@ARTICLE{hong_2024,
       author = {{Hong}, Xinyi and {Sun}, Ning-Chen and {Niu}, Zexi and {Wu}, Junjie and {Xi}, Qiang and {Liu}, Jifeng},
        title = "{Constraining the Progenitor of the Nearby Type II-P SN 2024ggi with Environmental Analysis}",
      journal = {\apjl},
         year = 2024,
        month = dec,
       volume = {977},
       number = {2},
          eid = {L50},
        pages = {L50},
          doi = {10.3847/2041-8213/ad99da},
archivePrefix = {arXiv},
       eprint = {2411.14685},
 primaryClass = {astro-ph.SR},
       adsurl = {https://ui.adsabs.harvard.edu/abs/2024ApJ...977L..50H}
}

@ARTICLE{antoniadis_2024,
       author = {{Antoniadis}, K. and {Bonanos}, A.~Z. and {de Wit}, S. and {Zapartas}, E. and {Munoz-Sanchez}, G. and {Maravelias}, G.},
        title = "{Establishing a mass-loss rate relation for red supergiants in the Large Magellanic Cloud}",
      journal = {\aap},
         year = 2024,
        month = jun,
       volume = {686},
          eid = {A88},
        pages = {A88},
          doi = {10.1051/0004-6361/202449383},
archivePrefix = {arXiv},
       eprint = {2401.15163},
 primaryClass = {astro-ph.SR},
       adsurl = {https://ui.adsabs.harvard.edu/abs/2024A&A...686A..88A}
}

@ARTICLE{lutovinov_2024,
       author = {{Lutovinov}, A.~A. and {Semena}, A.~N. and {Mereminskiy}, I.~A. and {Sazonov}, S. Yu. and {Molkov}, S.~V. and {Tkachenko}, A. Yu. and {Arefiev}, V.~A.},
        title = "{SRG/ART-XC detects SN2024ggi in X-rays}",
      journal = {The Astronomer's Telegram},
         year = 2024,
        month = apr,
       volume = {16586},
        pages = {1},
       adsurl = {https://ui.adsabs.harvard.edu/abs/2024ATel16586....1L}
}

@ARTICLE{yoon_2010,
       author = {{Yoon}, Sung-Chul and {Cantiello}, Matteo},
        title = "{Evolution of Massive Stars with Pulsation-driven Superwinds During the Red Supergiant Phase}",
      journal = {\apjl},
         year = 2010,
        month = jul,
       volume = {717},
       number = {1},
        pages = {L62-L65},
          doi = {10.1088/2041-8205/717/1/L62},
archivePrefix = {arXiv},
       eprint = {1005.4925},
 primaryClass = {astro-ph.SR},
       adsurl = {https://ui.adsabs.harvard.edu/abs/2010ApJ...717L..62Y}
}

@ARTICLE{jacobson_galan_2024b,
       author = {{Jacobson-Gal{\'a}n}, W.~V. and {Dessart}, L. and {Davis}, K.~W. and {Kilpatrick}, C.~D. and {Margutti}, R. and {Foley}, R.~J. and {Chornock}, R. and {Terreran}, G. and {Hiramatsu}, D. and {Newsome}, M. and {Padilla Gonzalez}, E. and {Pellegrino}, C. and {Howell}, D.~A. and {Filippenko}, A.~V. and {Anderson}, J.~P. and {Angus}, C.~R. and {Auchettl}, K. and {Bostroem}, K.~A. and {Brink}, T.~G. and {Cartier}, R. and {Coulter}, D.~A. and {de Boer}, T. and {Drout}, M.~R. and {Earl}, N. and {Ertini}, K. and {Farah}, J.~R. and {Farias}, D. and {Gall}, C. and {Gao}, H. and {Gerlach}, M.~A. and {Guo}, F. and {Haynie}, A. and {Hosseinzadeh}, G. and {Ibik}, A.~L. and {Jha}, S.~W. and {Jones}, D.~O. and {Langeroodi}, D. and {LeBaron}, N. and {Magnier}, E.~A. and {Piro}, A.~L. and {Raimundo}, S.~I. and {Rest}, A. and {Rest}, S. and {Rich}, R. Michael and {Rojas-Bravo}, C. and {Sears}, H. and {Taggart}, K. and {Villar}, V.~A. and {Wainscoat}, R.~J. and {Wang}, X. -F. and {Wasserman}, A.~R. and {Yan}, S. and {Yang}, Y. and {Zhang}, J. and {Zheng}, W.},
        title = "{Final Moments. II. Observational Properties and Physical Modeling of Circumstellar-material-interacting Type II Supernovae}",
      journal = {\apj},
         year = 2024,
        month = aug,
       volume = {970},
       number = {2},
          eid = {189},
        pages = {189},
          doi = {10.3847/1538-4357/ad4a2a},
archivePrefix = {arXiv},
       eprint = {2403.02382},
 primaryClass = {astro-ph.HE},
       adsurl = {https://ui.adsabs.harvard.edu/abs/2024ApJ...970..189J}
}

@ARTICLE{dessart_2010,
       author = {{Dessart}, Luc and {Livne}, Eli and {Waldman}, Roni},
        title = "{Shock-heating of stellar envelopes: a possible common mechanism at the origin of explosions and eruptions in massive stars}",
      journal = {\mnras},
         year = 2010,
        month = jul,
       volume = {405},
       number = {4},
        pages = {2113-2131},
          doi = {10.1111/j.1365-2966.2010.16626.x},
archivePrefix = {arXiv},
       eprint = {0910.3655},
 primaryClass = {astro-ph.SR},
       adsurl = {https://ui.adsabs.harvard.edu/abs/2010MNRAS.405.2113D}
}

@ARTICLE{tsang_2022,
       author = {{Tsang}, Benny T. -H. and {Kasen}, Daniel and {Bildsten}, Lars},
        title = "{3D Hydrodynamics of Pre-supernova Outbursts in Convective Red Supergiant Envelopes}",
      journal = {\apj},
         year = 2022,
        month = sep,
       volume = {936},
       number = {1},
          eid = {28},
        pages = {28},
          doi = {10.3847/1538-4357/ac83bc},
archivePrefix = {arXiv},
       eprint = {2207.13090},
 primaryClass = {astro-ph.SR},
       adsurl = {https://ui.adsabs.harvard.edu/abs/2022ApJ...936...28T}
}

@ARTICLE{dessart_2017,
       author = {{Dessart}, Luc and {Hillier}, D. John and {Audit}, Edouard},
        title = "{Explosion of red-supergiant stars: Influence of the atmospheric structure on shock breakout and early-time supernova radiation}",
      journal = {\aap},
         year = 2017,
        month = sep,
       volume = {605},
          eid = {A83},
        pages = {A83},
          doi = {10.1051/0004-6361/201730942},
archivePrefix = {arXiv},
       eprint = {1704.01697},
 primaryClass = {astro-ph.SR},
       adsurl = {https://ui.adsabs.harvard.edu/abs/2017A&A...605A..83D}
}

@ARTICLE{fuller_2024,
       author = {{Fuller}, Jim and {Tsuna}, Daichi},
        title = "{Boil-off of red supergiants: mass loss and type II-P supernovae}",
      journal = {The Open Journal of Astrophysics},
         year = 2024,
        month = jun,
       volume = {7},
          eid = {47},
        pages = {47},
          doi = {10.33232/001c.120130},
archivePrefix = {arXiv},
       eprint = {2405.21049},
 primaryClass = {astro-ph.SR},
       adsurl = {https://ui.adsabs.harvard.edu/abs/2024OJAp....7E..47F}
}

@ARTICLE{bianco_2022,
       author = {{Bianco}, Federica B. and {Ivezi{\'c}}, {\v{Z}}eljko and {Jones}, R. Lynne and {Graham}, Melissa L. and {Marshall}, Phil and {Saha}, Abhijit and {Strauss}, Michael A. and {Yoachim}, Peter and {Ribeiro}, Tiago and {Anguita}, Timo and {Bauer}, A.~E. and {Bauer}, Franz E. and {Bellm}, Eric C. and {Blum}, Robert D. and {Brandt}, William N. and {Brough}, Sarah and {Catelan}, M{\'a}rcio and {Clarkson}, William I. and {Connolly}, Andrew J. and {Gawiser}, Eric and {Gizis}, John E. and {Hlo{\v{z}}ek}, Ren{\'e}e and {Kaviraj}, Sugata and {Liu}, Charles T. and {Lochner}, Michelle and {Mahabal}, Ashish A. and {Mandelbaum}, Rachel and {McGehee}, Peregrine and {Neilsen}, Jr., Eric H. and {Olsen}, Knut A.~G. and {Peiris}, Hiranya V. and {Rhodes}, Jason and {Richards}, Gordon T. and {Ridgway}, Stephen and {Schwamb}, Megan E. and {Scolnic}, Dan and {Shemmer}, Ohad and {Slater}, Colin T. and {Slosar}, An{\v{z}}e and {Smartt}, Stephen J. and {Strader}, Jay and {Street}, Rachel and {Trilling}, David E. and {Verma}, Aprajita and {Vivas}, A.~K. and {Wechsler}, Risa H. and {Willman}, Beth},
        title = "{Optimization of the Observing Cadence for the Rubin Observatory Legacy Survey of Space and Time: A Pioneering Process of Community-focused Experimental Design}",
      journal = {\apjs},
         year = 2022,
        month = jan,
       volume = {258},
       number = {1},
          eid = {1},
        pages = {1},
          doi = {10.3847/1538-4365/ac3e72},
archivePrefix = {arXiv},
       eprint = {2108.01683},
 primaryClass = {astro-ph.IM},
       adsurl = {https://ui.adsabs.harvard.edu/abs/2022ApJS..258....1B}
}

@ARTICLE{eldridge_2018,
       author = {{Eldridge}, J.~J. and {Xiao}, L. and {Stanway}, E.~R. and {Rodrigues}, N. and {Guo}, N. -Y.},
        title = "{Supernova lightCURVE POPulation Synthesis I: Including interacting binaries is key to understanding the diversity of type II supernova lightcurves}",
      journal = {\pasa},
         year = 2018,
        month = dec,
       volume = {35},
          eid = {e049},
        pages = {e049},
          doi = {10.1017/pasa.2018.47},
archivePrefix = {arXiv},
       eprint = {1811.00282},
 primaryClass = {astro-ph.SR},
       adsurl = {https://ui.adsabs.harvard.edu/abs/2018PASA...35...49E}
}

@ARTICLE{matsuoka_2024,
       author = {{Matsuoka}, Tomoki and {Sawada}, Ryo},
        title = "{Binary Interaction Can Yield a Diversity of Circumstellar Media around Type II Supernova Progenitors}",
      journal = {\apj},
         year = 2024,
        month = mar,
       volume = {963},
       number = {2},
          eid = {105},
        pages = {105},
          doi = {10.3847/1538-4357/ad1829},
archivePrefix = {arXiv},
       eprint = {2307.00727},
 primaryClass = {astro-ph.SR},
       adsurl = {https://ui.adsabs.harvard.edu/abs/2024ApJ...963..105M}
}

@ARTICLE{pessi_2024,
       author = {{Pessi}, Thallis and {Cartier}, R{\'e}gis and {Hueichapan}, Emilio and {de Brito Silva}, Danielle and {Prieto}, Jose L. and {Mu{\~n}oz}, Ricardo R. and {Medina}, Gustavo E. and {Diaz}, Paula and {Li}, Ting S.},
        title = "{Early emission lines in SN 2024ggi revealed by high-resolution spectroscopy}",
      journal = {\aap},
         year = 2024,
        month = aug,
       volume = {688},
          eid = {L28},
        pages = {L28},
          doi = {10.1051/0004-6361/202450608},
archivePrefix = {arXiv},
       eprint = {2405.02274},
 primaryClass = {astro-ph.HE},
       adsurl = {https://ui.adsabs.harvard.edu/abs/2024A&A...688L..28P}
}

@ARTICLE{shrestha_2024,
       author = {{Shrestha}, Manisha and {Bostroem}, K. Azalee and {Sand}, David J. and {Hosseinzadeh}, Griffin and {Andrews}, Jennifer E. and {Dong}, Yize and {Hoang}, Emily and {Janzen}, Daryl and {Pearson}, Jeniveve and {Jencson}, Jacob E. and {Lundquist}, M.~J. and {Mehta}, Darshana and {Ravi}, Aravind P. and {Meza Retamal}, Nicol{\'a}s and {Valenti}, Stefano and {Brown}, Peter J. and {Jha}, Saurabh W. and {Macrie}, Colin and {Hsu}, Brian and {Farah}, Joseph and {Howell}, D. Andrew and {McCully}, Curtis and {Newsome}, Megan and {Padilla Gonzalez}, Estefania and {Pellegrino}, Craig and {Terreran}, Giacomo and {Kwok}, Lindsey and {Smith}, Nathan and {Schwab}, Michaela and {Martas}, Aidan and {Munoz}, Ricardo R. and {Medina}, Gustavo E. and {Li}, Ting S. and {Diaz}, Paula and {Hiramatsu}, Daichi and {Tucker}, Brad E. and {Wheeler}, J.~C. and {Wang}, Xiaofeng and {Zhai}, Qian and {Zhang}, Jujia and {Gangopadhyay}, Anjasha and {Yang}, Yi and {Guti{\'e}rrez}, Claudia P.},
        title = "{Extended Shock Breakout and Early Circumstellar Interaction in SN 2024ggi}",
      journal = {\apjl},
         year = 2024,
        month = sep,
       volume = {972},
       number = {1},
          eid = {L15},
        pages = {L15},
          doi = {10.3847/2041-8213/ad6907},
archivePrefix = {arXiv},
       eprint = {2405.18490},
 primaryClass = {astro-ph.HE},
       adsurl = {https://ui.adsabs.harvard.edu/abs/2024ApJ...972L..15S}
}

@ARTICLE{li_2011,
       author = {{Li}, Weidong and {Leaman}, Jesse and {Chornock}, Ryan and {Filippenko}, Alexei V. and {Poznanski}, Dovi and {Ganeshalingam}, Mohan and {Wang}, Xiaofeng and {Modjaz}, Maryam and {Jha}, Saurabh and {Foley}, Ryan J. and {Smith}, Nathan},
        title = "{Nearby supernova rates from the Lick Observatory Supernova Search - II. The observed luminosity functions and fractions of supernovae in a complete sample}",
      journal = {\mnras},
         year = 2011,
        month = apr,
       volume = {412},
       number = {3},
        pages = {1441-1472},
          doi = {10.1111/j.1365-2966.2011.18160.x},
archivePrefix = {arXiv},
       eprint = {1006.4612},
 primaryClass = {astro-ph.SR},
       adsurl = {https://ui.adsabs.harvard.edu/abs/2011MNRAS.412.1441L}
}

@ARTICLE{tonry_2024,
       author = {{Tonry}, J. and {Denneau}, L. and {Weiland}, H. and {Lawrence}, A. and {Siverd}, R. and {Erasmus}, N. and {Koorts}, W. and {Jordan}, A. and {Suc}, V. and {Smartt}, S.~J. and {Smith}, K.~W. and {Young}, D.~R. and {Nicholl}, M. and {Fulton}, M. and {McCollum}, M. and {Moore}, T. and {Weston}, J. and {Sheng}, X. and {Ramsden}, P. and {Angus}, C.~R. and {Aamer}, A. and {Shingles}, L. and {Srivastav}, S. and {Gillanders}, J. and {Rhodes}, L. and {Andersson}, A. and {Stevance}, H. and {Rest}, A. and {Chen}, T.~W. and {Stubbs}, C. and {Sommer}, J.},
        title = "{ATLAS Transient Discovery Report for 2024-04-11}",
      journal = {Transient Name Server Discovery Report},
         year = 2024,
        month = apr,
       volume = {2024-1020},
        pages = {1},
       adsurl = {https://ui.adsabs.harvard.edu/abs/2024TNSTR1020....1T}
}

@ARTICLE{wyrzykowski_2025,
       author = {{Wyrzykowski}, L. and {Mikolajczyk}, P. and {Kotysz}, K. and {Zielinski}, P. and {Hambsch}, J. and {Bronikowski}, M.},
        title = "{Photometric follow-up of SN2024ggi with BHTOM.space global telescope network}",
      journal = {Transient Name Server AstroNote},
         year = 2025,
        month = jan,
       volume = {22},
        pages = {1},
       adsurl = {https://ui.adsabs.harvard.edu/abs/2025TNSAN..22....1W}
}

@ARTICLE{marti_devesa_2024,
       author = {{Marti-Devesa}, G. and {Fermi-LAT Collaboration}},
        title = "{Fermi-LAT gamma-ray observations of SN 2024ggi}",
      journal = {The Astronomer's Telegram},
         year = 2024,
        month = apr,
       volume = {16601},
        pages = {1},
       adsurl = {https://ui.adsabs.harvard.edu/abs/2024ATel16601....1M}
}

@ARTICLE{chandra_2024,
       author = {{Chandra}, Poonam and {Maeda}, Keiichi and {Nayana}, A.~J. and {Chevalier}, Roger A. and {Ryder}, Stuart and {Ho}, Anna Y.~Q. and {Ray}, Alak K.},
        title = "{Centimeter-wavelength upper limits on SN 2024ggi with the JVLA and the uGMRT}",
      journal = {The Astronomer's Telegram},
         year = 2024,
        month = may,
       volume = {16612},
        pages = {1},
       adsurl = {https://ui.adsabs.harvard.edu/abs/2024ATel16612....1C}
}

@ARTICLE{hu_2025,
       author = {{Hu}, Maokai and {Ao}, Yiping and {Yang}, Yi and {Hu}, Lei and {Li}, Fulin and {Wang}, Lifan and {Wang}, Xiaofeng},
        title = "{Early-time Millimeter Observations of the Nearby Type II SN 2024ggi}",
      journal = {\apjl},
         year = 2025,
        month = jan,
       volume = {978},
       number = {2},
          eid = {L27},
        pages = {L27},
          doi = {10.3847/2041-8213/ada1cd},
archivePrefix = {arXiv},
       eprint = {2412.11389},
 primaryClass = {astro-ph.SR},
       adsurl = {https://ui.adsabs.harvard.edu/abs/2025ApJ...978L..27H}
}

@ARTICLE{ryder_2024,
       author = {{Ryder}, Stuart and {Maeda}, Keiichi and {Chandra}, Poonam and {Alsaberi}, Rami and {Kotak}, Rubina},
        title = "{Radio detection of SN 2024ggi}",
      journal = {The Astronomer's Telegram},
         year = 2024,
        month = may,
       volume = {16616},
        pages = {1},
       adsurl = {https://ui.adsabs.harvard.edu/abs/2024ATel16616....1R}
}

@ARTICLE{margutti_2024,
       author = {{Margutti}, Raffaella and {Grefenstette}, Brian},
        title = "{NuSTAR detection of SN2024ggi at 2 days post discovery}",
      journal = {The Astronomer's Telegram},
         year = 2024,
        month = apr,
       volume = {16587},
        pages = {1},
       adsurl = {https://ui.adsabs.harvard.edu/abs/2024ATel16587....1M}
}

@ARTICLE{zhang_2024b,
       author = {{Zhang}, J. and {Li}, C.~K. and {Cheng}, H.~Q. and {Wu}, Q.~Y. and {Jia}, S.~M. and {Chen}, Y. and {Cui}, W.~W. and {Feng}, H. and {Guan}, J. and {Han}, D.~W. and {Li}, W. and {Liu}, C.~Z. and {Lu}, F.~J. and {Song}, L.~M. and {Wang}, J. and {Xu}, J.~J. and {Zhang}, S.~N. and {Zhao}, H.~S. and {Zhao}, X.~F. and {Jin}, C.~C. and {Ling}, Z.~X. and {Liu}, H.~Y. and {Liu}, M.~J. and {Liu}, Y. and {Li}, D.~Y. and {Sun}, H. and {Yuan}, W. and {Zhang}, C. and {Zhang}, W.~D. and {Li}, R.~Z. and {Wang}, Y. and {Zhou}, H. and {Nandra}, K. and {Rau}, A. and {Friedrich}, P. and {Meidinger}, N. and {Burwitz}, V. and {Kuulkers}, E. and {Santovincenzo}, A. and {O'Brien}, P. and {Cordier}, B. and {Wang}, X.~F. and {Li}, W.~X.},
        title = "{SN 2024ggi: detection of X-ray emission by EP-FXT}",
      journal = {The Astronomer's Telegram},
         year = 2024,
        month = apr,
       volume = {16588},
        pages = {1},
       adsurl = {https://ui.adsabs.harvard.edu/abs/2024ATel16588....1Z}
}

@ARTICLE{weiler_2007,
       author = {{Weiler}, Kurt W. and {Williams}, Christopher L. and {Panagia}, Nino and {Stockdale}, Christopher J. and {Kelley}, Matthew T. and {Sramek}, Richard A. and {Van Dyk}, Schuyler D. and {Marcaide}, J.~M.},
        title = "{Long-Term Radio Monitoring of SN 1993J}",
      journal = {\apj},
         year = 2007,
        month = dec,
       volume = {671},
       number = {2},
        pages = {1959-1980},
          doi = {10.1086/523258},
archivePrefix = {arXiv},
       eprint = {0709.1136},
 primaryClass = {astro-ph},
       adsurl = {https://ui.adsabs.harvard.edu/abs/2007ApJ...671.1959W}
}

@ARTICLE{zhang_2024,
       author = {{Zhang}, Jujia and {Dessart}, Luc and {Wang}, Xiaofeng and {Zhai}, Qian and {Yang}, Yi and {Li}, Liping and {Lin}, Han and {Valerin}, Giorgio and {Cai}, Yongzhi and {Guo}, Zhen and {Wang}, Lingzhi and {Zhao}, Zeyi and {Wang}, Zhenyu and {Yan}, Shengyu},
        title = "{Probing the Shock Breakout Signal of SN 2024ggi from the Transformation of Early Flash Spectroscopy}",
      journal = {\apjl},
         year = 2024,
        month = jul,
       volume = {970},
       number = {1},
          eid = {L18},
        pages = {L18},
          doi = {10.3847/2041-8213/ad5da4},
archivePrefix = {arXiv},
       eprint = {2406.07806},
 primaryClass = {astro-ph.HE},
       adsurl = {https://ui.adsabs.harvard.edu/abs/2024ApJ...970L..18Z}
}

@ARTICLE{smith_2014,
       author = {{Smith}, Nathan},
        title = "{Mass Loss: Its Effect on the Evolution and Fate of High-Mass Stars}",
      journal = {\araa},
         year = 2014,
        month = aug,
       volume = {52},
        pages = {487-528},
          doi = {10.1146/annurev-astro-081913-040025},
archivePrefix = {arXiv},
       eprint = {1402.1237},
 primaryClass = {astro-ph.SR},
       adsurl = {https://ui.adsabs.harvard.edu/abs/2014ARA&A..52..487S}
}

@ARTICLE{chevalier_2012,
       author = {{Chevalier}, Roger A.},
        title = "{Common Envelope Evolution Leading to Supernovae with Dense Interaction}",
      journal = {\apjl},
         year = 2012,
        month = jun,
       volume = {752},
       number = {1},
          eid = {L2},
        pages = {L2},
          doi = {10.1088/2041-8205/752/1/L2},
archivePrefix = {arXiv},
       eprint = {1204.3300},
 primaryClass = {astro-ph.HE},
       adsurl = {https://ui.adsabs.harvard.edu/abs/2012ApJ...752L...2C}
}

@ARTICLE{woosley_2002,
       author = {{Woosley}, S.~E. and {Heger}, A. and {Weaver}, T.~A.},
        title = "{The evolution and explosion of massive stars}",
      journal = {Reviews of Modern Physics},
         year = 2002,
        month = nov,
       volume = {74},
       number = {4},
        pages = {1015-1071},
          doi = {10.1103/RevModPhys.74.1015},
       adsurl = {https://ui.adsabs.harvard.edu/abs/2002RvMP...74.1015W}
}

@ARTICLE{heger_2003,
       author = {{Heger}, A. and {Fryer}, C.~L. and {Woosley}, S.~E. and {Langer}, N. and {Hartmann}, D.~H.},
        title = "{How Massive Single Stars End Their Life}",
      journal = {\apj},
         year = 2003,
        month = jul,
       volume = {591},
       number = {1},
        pages = {288-300},
          doi = {10.1086/375341},
archivePrefix = {arXiv},
       eprint = {astro-ph/0212469},
 primaryClass = {astro-ph},
       adsurl = {https://ui.adsabs.harvard.edu/abs/2003ApJ...591..288H}
}

@ARTICLE{morozova_2017,
       author = {{Morozova}, Viktoriya and {Piro}, Anthony L. and {Valenti}, Stefano},
        title = "{Unifying Type II Supernova Light Curves with Dense Circumstellar Material}",
      journal = {\apj},
         year = 2017,
        month = mar,
       volume = {838},
       number = {1},
          eid = {28},
        pages = {28},
          doi = {10.3847/1538-4357/aa6251},
archivePrefix = {arXiv},
       eprint = {1610.08054},
 primaryClass = {astro-ph.HE},
       adsurl = {https://ui.adsabs.harvard.edu/abs/2017ApJ...838...28M}
}

@ARTICLE{jencson_2016,
       author = {{Jencson}, J.~E. and {Prieto}, J.~L. and {Kochanek}, C.~S. and {Shappee}, B.~J. and {Stanek}, K.~Z. and {Pogge}, R.~W.},
        title = "{Optical observations of the luminous Type IIn Supernova 2010jl for over 900 d}",
      journal = {\mnras},
         year = 2016,
        month = mar,
       volume = {456},
       number = {3},
        pages = {2622-2635},
          doi = {10.1093/mnras/stv2795},
archivePrefix = {arXiv},
       eprint = {1505.01186},
 primaryClass = {astro-ph.HE},
       adsurl = {https://ui.adsabs.harvard.edu/abs/2016MNRAS.456.2622J}
}

@ARTICLE{bauer_2008,
       author = {{Bauer}, F.~E. and {Dwarkadas}, V.~V. and {Brandt}, W.~N. and {Immler}, S. and {Smartt}, S. and {Bartel}, N. and {Bietenholz}, M.~F.},
        title = "{Supernova 1996cr: SN 1987A's Wild Cousin?}",
      journal = {\apj},
         year = 2008,
        month = dec,
       volume = {688},
       number = {2},
        pages = {1210-1234},
          doi = {10.1086/589761},
archivePrefix = {arXiv},
       eprint = {0804.3597},
 primaryClass = {astro-ph},
       adsurl = {https://ui.adsabs.harvard.edu/abs/2008ApJ...688.1210B}
}

@ARTICLE{williams_2002,
       author = {{Williams}, Christopher L. and {Panagia}, Nino and {Van Dyk}, Schuyler D. and {Lacey}, Christina K. and {Weiler}, Kurt W. and {Sramek}, Richard A.},
        title = "{Radio Emission from SN 1988Z and Very Massive Star Evolution}",
      journal = {\apj},
         year = 2002,
        month = dec,
       volume = {581},
       number = {1},
        pages = {396-403},
          doi = {10.1086/344087},
archivePrefix = {arXiv},
       eprint = {astro-ph/0208190},
 primaryClass = {astro-ph},
       adsurl = {https://ui.adsabs.harvard.edu/abs/2002ApJ...581..396W}
}

@ARTICLE{ryder_2016,
       author = {{Ryder}, S.~D. and {Kotak}, R. and {Smith}, I.~A. and {Tingay}, S.~J. and {Kool}, E.~C. and {Polshaw}, J.},
        title = "{SN 1978K: An evolved supernova outside our Local Group detected at millimetre wavelengths}",
      journal = {\aap},
         year = 2016,
        month = nov,
       volume = {595},
          eid = {L9},
        pages = {L9},
          doi = {10.1051/0004-6361/201629763},
archivePrefix = {arXiv},
       eprint = {1610.03149},
 primaryClass = {astro-ph.HE},
       adsurl = {https://ui.adsabs.harvard.edu/abs/2016A&A...595L...9R}
}

@ARTICLE{cai_2026,
       author = {{Cai}, Y.-Z. and {Pastorello}, A. and {Moriya}, T.~J. and {Wang}, X.-F. and {Reguitti}, A. and {Filippenko}, A.~V. and {Tomasella}, L. and {{\v{C}}otar}, K. and {Siviero}, A. and {Elias-Rosa}, N. and {Brink}, T.~G. and {Valerin}, G. and {Benetti}, S. and {Zhao}, J.-W. and {Peng}, Z.-H. and {Wang}, Z.-Y. and {Altunin}, I. and {Baer-Way}, R. and {Baron}, E. and {Chander}, V. and {Chu}, M. and {deGraw}, A. and {DerKacy}, J.~M. and {Isern}, J. and {Jennings}, C. and {Kotak}, R. and {Li}, L.-P. and {Marziani}, P. and {May}, M. and {Mazzali}, P.~A. and {Morales-Garoffolo}, A. and {Moran}, S. and {Ochner}, P. and {Salmaso}, I. and {Tartaglia}, L. and {Turatto}, M. and {Wu}, H.-Y. and {Xiang}, D.-F. and {Yan}, S.-Y. and {Zhang}, J.-J. and {Zheng}, W.},
        title = "{SN 2019vxm: A luminous and long-lived Type IIn supernova with early flash-ionisation features}",
      journal = {arXiv e-prints},
         year = 2026,
        month = jun,
          eid = {arXiv:2606.30390},
        pages = {arXiv:2606.30390},
          doi = {10.48550/arXiv.2606.30390},
archivePrefix = {arXiv},
       eprint = {2606.30390},
 primaryClass = {astro-ph.SR},
       adsurl = {https://ui.adsabs.harvard.edu/abs/2026arXiv260630390C}
}

@ARTICLE{ertini_2025,
       author = {{Ertini}, K. and {Regna}, T.~A. and {Ferrari}, L. and {Bersten}, M.~C. and {Folatelli}, G. and {Mendez Llorca}, A. and {Fern{\'a}ndez-Laj{\'u}s}, E. and {Ferrero}, G.~A. and {Hueichap{\'a}n D{\'\i}az}, E. and {Cartier}, R. and {Rom{\'a}n Aguilar}, L.~M. and {Putkuri}, C. and {Piccirilli}, M.~P. and {Cellone}, S.~A. and {Moreno}, J. and {Orellana}, M. and {Prieto}, J.~L. and {Gerlach}, M. and {Acosta}, V. and {Ritacco}, M.~J. and {Schujman}, J.~C. and {Vald{\'e}z}, J.},
        title = "{SN 2024ggi: another year, another striking Type II supernova}",
      journal = {arXiv e-prints},
         year = 2025,
        month = mar,
          eid = {arXiv:2503.01577},
        pages = {arXiv:2503.01577},
          doi = {10.48550/arXiv.2503.01577},
archivePrefix = {arXiv},
       eprint = {2503.01577},
 primaryClass = {astro-ph.HE},
       adsurl = {https://ui.adsabs.harvard.edu/abs/2025arXiv250301577E}
}

@ARTICLE{laplace_2026,
       author = {{Laplace}, Eva and {Bronner}, Vincent A. and {Schneider}, Fabian R.~N. and {Podsiadlowski}, Philipp},
        title = "{Pulsations Change the Structures of Massive Stars before Explosion: Interpreting SN 2023ixf and SN 2024ggi}",
      journal = {\apjl},
         year = 2026,
        month = feb,
       volume = {998},
       number = {2},
          eid = {L40},
        pages = {L40},
          doi = {10.3847/2041-8213/ae3d2e},
archivePrefix = {arXiv},
       eprint = {2508.11088},
 primaryClass = {astro-ph.HE},
       adsurl = {https://ui.adsabs.harvard.edu/abs/2026ApJ...998L..40L}
}

@ARTICLE{ferdinand_2026,
       author = {{Ferdinand} and {Jacobson-Gal{\'a}n}, W.~V. and {Kasliwal}, M.~M. and {Zimmerman}, Erez A.},
        title = "{XSNAP: An X-Ray Supernova Analysis Pipeline with Application to the Type II Supernova 2024ggi}",
      journal = {\apj},
         year = 2026,
        month = apr,
       volume = {1001},
       number = {1},
          eid = {26},
        pages = {26},
          doi = {10.3847/1538-4357/ae4ec6},
archivePrefix = {arXiv},
       eprint = {2511.10744},
 primaryClass = {astro-ph.HE},
       adsurl = {https://ui.adsabs.harvard.edu/abs/2026ApJ..1001...26F}
}

@ARTICLE{bostroem_2026,
       author = {{Bostroem}, K. Azalee and {Valenti}, Stefano and {Sand}, David J. and {Pearson}, Jeniveve and {Shrestha}, Manisha and {Andrews}, Jennifer E. and {Dessart}, Luc and {Jacobson-Gal{\'a}n}, W.~V. and {Hsu}, Brian and {Ravi}, Aravind P. and {Andrews}, Moira and {Christy}, Collin and {Dong}, Yize and {Farah}, Joseph and {Franz}, Noah and {Filippenko}, Alexei V. and {Gill}, Kiranjyot and {Hoang}, Emily T. and {Hosseinzadeh}, Griffin and {Howell}, D. Andrew and {Janzen}, Daryl and {Jencson}, Jacob E. and {Jha}, Saurabh W. and {Kwok}, Lindsey A. and {Lundquist}, Michael and {Martas}, Aidan and {McCully}, Curtis and {Mehta}, Darshana and {Newsome}, Megan and {Padilla-Gonzalez}, Estefania and {Ransome}, Conor and {Meza Retamal}, Nicolas E. and {Smith}, Nathan and {Subrayan}, Bhagya M. and {Terreran}, Giacomo},
        title = "{Late-time Hubble Space Telescope Ultraviolet Spectra of SN 2023ixf and SN 2024ggi Show Ongoing Interaction with Circumstellar Material}",
      journal = {\apj},
         year = 2026,
        month = jun,
       volume = {1004},
       number = {1},
          eid = {23},
        pages = {23},
          doi = {10.3847/1538-4357/ae644e},
archivePrefix = {arXiv},
       eprint = {2508.11756},
 primaryClass = {astro-ph.HE},
       adsurl = {https://ui.adsabs.harvard.edu/abs/2026ApJ..1004...23B}
}

@ARTICLE{xiang_2024,
       author = {{Xiang}, Danfeng and {Mo}, Jun and {Wang}, Xiaofeng and {Wang}, Lingzhi and {Zhang}, Jujia and {Lin}, Han and {Chen}, Liyang and {Song}, Cuiying and {Liu}, Liang-Duan and {Wang}, Zhenyu and {Li}, Gaici},
        title = "{The Red Supergiant Progenitor of Type II Supernova 2024ggi}",
      journal = {\apjl},
         year = 2024,
        month = jul,
       volume = {969},
       number = {1},
          eid = {L15},
        pages = {L15},
          doi = {10.3847/2041-8213/ad54b3},
archivePrefix = {arXiv},
       eprint = {2405.07699},
 primaryClass = {astro-ph.HE},
       adsurl = {https://ui.adsabs.harvard.edu/abs/2024ApJ...969L..15X}
}

@ARTICLE{fuller_2017,
       author = {{Fuller}, Jim},
        title = "{Pre-supernova outbursts via wave heating in massive stars - I. Red supergiants}",
      journal = {\mnras},
         year = 2017,
        month = sep,
       volume = {470},
       number = {2},
        pages = {1642-1656},
          doi = {10.1093/mnras/stx1314},
archivePrefix = {arXiv},
       eprint = {1704.08696},
 primaryClass = {astro-ph.SR},
       adsurl = {https://ui.adsabs.harvard.edu/abs/2017MNRAS.470.1642F}
}

@ARTICLE{fuller_2018,
       author = {{Fuller}, Jim and {Ro}, Stephen},
        title = "{Pre-supernova outbursts via wave heating in massive stars - II. Hydrogen-poor stars}",
      journal = {\mnras},
         year = 2018,
        month = may,
       volume = {476},
       number = {2},
        pages = {1853-1868},
          doi = {10.1093/mnras/sty369},
archivePrefix = {arXiv},
       eprint = {1710.04251},
 primaryClass = {astro-ph.SR},
       adsurl = {https://ui.adsabs.harvard.edu/abs/2018MNRAS.476.1853F}
}

@ARTICLE{Hogbom1974,
       author = {{H{\"o}gbom}, J.~A.},
        title = "{Aperture Synthesis with a Non-Regular Distribution of Interferometer Baselines}",
      journal = {\aaps},
         year = 1974,
        month = jun,
       volume = {15},
        pages = {417},
       adsurl = {https://ui.adsabs.harvard.edu/abs/1974A&AS...15..417H}
}

@ARTICLE{Clark1980,
       author = {{Clark}, B.~G.},
        title = "{An efficient implementation of the algorithm 'CLEAN'}",
      journal = {\aap},
         year = 1980,
        month = sep,
       volume = {89},
       number = {3},
        pages = {377},
       adsurl = {https://ui.adsabs.harvard.edu/abs/1980A&A....89..377C}
}

@ARTICLE{Sault1994,
       author = {{Sault}, R.~J. and {Wieringa}, M.~H.},
        title = "{Multi-frequency synthesis techniques in radio interferometric imaging.}",
      journal = {\aaps},
         year = 1994,
        month = dec,
       volume = {108},
        pages = {585-594},
       adsurl = {https://ui.adsabs.harvard.edu/abs/1994A&AS..108..585S}
}

@INPROCEEDINGS{Sault1995,
       author = {{Sault}, R.~J. and {Teuben}, P.~J. and {Wright}, M.~C.~H.},
        title = "{A Retrospective View of MIRIAD}",
    booktitle = {Astronomical Data Analysis Software and Systems IV},
         year = 1995,
       editor = {{Shaw}, R.~A. and {Payne}, H.~E. and {Hayes}, J.~J.~E.},
       series = {Astronomical Society of the Pacific Conference Series},
       volume = {77},
        month = jan,
        pages = {433},
       adsurl = {https://ui.adsabs.harvard.edu/abs/1995ASPC...77..433S}
}

@ARTICLE{Chevalier1998,
       author = {{Chevalier}, Roger A.},
        title = "{Synchrotron Self-Absorption in Radio Supernovae}",
      journal = {\apj},
         year = 1998,
        month = may,
       volume = {499},
       number = {2},
        pages = {810-819},
          doi = {10.1086/305676},
       adsurl = {https://ui.adsabs.harvard.edu/abs/1998ApJ...499..810C}
}

@ARTICLE{Ashton2019,
       author = {{Ashton}, Gregory and {H{\"u}bner}, Moritz and {Lasky}, Paul D. and {Talbot}, Colm and {Ackley}, Kendall and {Biscoveanu}, Sylvia and {Chu}, Qi and {Divakarla}, Atul and {Easter}, Paul J. and {Goncharov}, Boris and {Hernandez Vivanco}, Francisco and {Harms}, Jan and {Lower}, Marcus E. and {Meadors}, Grant D. and {Melchor}, Denyz and {Payne}, Ethan and {Pitkin}, Matthew D. and {Powell}, Jade and {Sarin}, Nikhil and {Smith}, Rory J.~E. and {Thrane}, Eric},
        title = "{BILBY: A User-friendly Bayesian Inference Library for Gravitational-wave Astronomy}",
      journal = {\apjs},
         year = 2019,
        month = apr,
       volume = {241},
       number = {2},
          eid = {27},
        pages = {27},
          doi = {10.3847/1538-4365/ab06fc},
archivePrefix = {arXiv},
       eprint = {1811.02042},
 primaryClass = {astro-ph.IM},
       adsurl = {https://ui.adsabs.harvard.edu/abs/2019ApJS..241...27A}
}

@ARTICLE{Speagle2020,
       author = {{Speagle}, Joshua S.},
        title = "{DYNESTY: a dynamic nested sampling package for estimating Bayesian posteriors and evidences}",
      journal = {\mnras},
         year = 2020,
        month = apr,
       volume = {493},
       number = {3},
        pages = {3132-3158},
          doi = {10.1093/mnras/staa278},
archivePrefix = {arXiv},
       eprint = {1904.02180},
 primaryClass = {astro-ph.IM},
       adsurl = {https://ui.adsabs.harvard.edu/abs/2020MNRAS.493.3132S}
}

@ARTICLE{wang_2026,
	title = {Aspherical Shock Breakout Inferred from the First Light of Supernova 2024ggi},
	author = {Wang, Xiaofeng and Yan, Shengyu and Zhang, Jujia and Song, Cuiying and Hu, Maokai and Yang, Yi and Yuan, Weimin and Aguado, D and Perez-Fournon, Ismael and Wu, Qinyu and Antier, Sarah and Sun, Tianrui and Li, Gaici and Liu, Liangduan and Herman, Eliot and Xiong, Dingrong and Lin, Weili and Li, Wenxiong and Poidevin, F and López-Oramas, A and Nespral, David López Fdez and Wang, Zhenyu and Elias-Rosa, Nancy and Liu, Jialian and Wen, Xudong and Xiang, Danfeng and Wang, Bo and Hussenot-Desenonges, Thomas and Coughlin, Michael and Freeberg, Michael and Hello, Patrice and Kaeouach, Aziz and Karpov, Sergey and Klotz, Alain and Leonini, Simone and Ma\v\{s\}ek, Martin and Odeh, Mohammad and Romanov, Filipp and Damien, Turpin and Zhang, J and Li, C.K. and Cheng, H.-Q. and Liu, Mingjun and Chen, Yong and Jia, S.M. and Zhou, Hao and Castro-Tirado, Alberto},
	DOI = {10.21203/rs.3.rs-8711438/v1},
	journal = {Research Square},
	year = {2026},
	URL = {https://doi.org/10.21203/rs.3.rs-8711438/v1},
}
\bibliographystyle{aasjournal}



\end{document}
